\documentclass[journal,9pt]{IEEEtran}

\usepackage{mathtools}
\usepackage{amssymb, amsthm, amscd, MnSymbol}
\usepackage{physics}

\usepackage[switch]{lineno}
\usepackage[named]{algo}
\usepackage[noend]{algpseudocode}
\usepackage{algorithm}

\usepackage{xurl}
\usepackage{xcolor}

\ifCLASSOPTIONcompsoc
 \usepackage[caption=false,font=normalsize,labelfont=sf,textfont=sf]{subfig}
\else
 \usepackage[caption=false,font=footnotesize]{subfig}
\fi

\usepackage{threeparttable}
\usepackage{array}
\newcolumntype{P}[1]{>{\centering\arraybackslash}p{#1}}

\usepackage{float}
\usepackage{stfloats}
\usepackage[section]{placeins}
\usepackage{balance}

\usepackage{cite}
\usepackage[pdftex]{graphicx}	
\usepackage[none]{hyphenat}
\usepackage{wrapfig} % Allows in-line images

\usepackage{pgfplots}
\pgfplotsset{compat=1.8}

\usepackage{bookmark}
\usepackage{xcolor}

\usepackage{verbatim}

\theoremstyle{remark}

\DeclareMathOperator*{\minimize}{minimize}

\graphicspath{{./figures/}}
\usepackage{cuted,tcolorbox,lipsum}
\definecolor{infocolor}{RGB}{245,245,245}

\begin{document}

\title{An Overview of Advances in Signal Processing Techniques for Classical and Quantum Wideband Synthetic Apertures}

\author{Peter Vouras, Kumar Vijay Mishra, Alexandra Artusio-Glimpse, Samuel Pinilla, Angeliki Xenaki, \\David W. Griffith and Karen Egiazarian% <-this % stops a space
\thanks{P. V. is with the United States Department of Defense, Washington, DC, 20375 USA, e-mail: synthetic\_aperture\_twg@ieee.org.}
\thanks{K. V. M. is with the United States DEVCOM Army Research Laboratory, Adelphi, MD, 20783 USA, e-mail: kvm@ieee.org.}
\thanks{A. A-G. is with the National Institute of Standards and Technology (NIST), Boulder, CO, 80303 USA e-mail: alexandra.artusio-glimpse@nist.gov.}
\thanks{S. P. is with the University of Manchester at Harwell Science and Innovation campus, Oxfordshire, OX110GD, United Kingdom, e-mail: samuel.pinilla@correo.uis.edu.co.}
\thanks{A. X. is with the Centre for Maritime Research and Experimentation, Science and Technology Organization, NATO, La Spezia, 19126, Italy, email: angeliki.xenaki@cmre.nato.int.}
\thanks{D. W. G. is with NIST, Gaithersburg, MD, 20899, USA e-mail: david.griffith@nist.gov.}
\thanks{K. E. is with Faculty of Information Technology and Communication Sciences, Tampere University, Tampere, 33720, Finland, e-mail: karen.egiazarian@noiselessimaging.com.}
\thanks{K. V. M. acknowledges support from the National Academies of Sciences, Engineering, and Medicine via Army Research Laboratory Harry Diamond Distinguished Fellowship.}
}

% make the title area
\maketitle

\begin{abstract}
Rapid developments in synthetic aperture (SA) systems, which generate a larger aperture with greater angular resolution than is inherently possible from the physical dimensions of a single sensor alone, are leading to novel research avenues in several signal processing applications. The SAs may either use a mechanical positioner to move an antenna through space or deploy a distributed network of sensors. With the advent of new hardware technologies, the SAs tend to be denser nowadays. The recent opening of higher frequency bands has led to wide SA bandwidths. In general, new techniques and setups are required to harness the potential of wide SAs in space and bandwidth. Herein, we provide a brief overview of emerging signal processing trends in such spatially and spectrally wideband SA systems. This guide is intended to aid newcomers in navigating the most critical issues in SA analysis and further supports the development of new theories in the field. In particular, we cover the theoretical framework and practical underpinnings of wideband SA radar, channel sounding, sonar, radiometry, and optical applications. Apart from the classical SA applications, we also discuss the quantum electric-field-sensing probes in SAs that are currently undergoing active research but remain at nascent stages of development.
\end{abstract}


\IEEEpeerreviewmaketitle

\section{Introduction}
\label{sec:intro}

%\textbf{1. Introduction} \\
%- Brief history of synthetic aperture concept across different applications\\
Over the past several decades, an array of imaging sensors has been employed to create a single synthetic image by simulating a sensor with a much wider aperture and shallow depth-of-field. This \textit{synthetic aperture} (SA) processing technique has led to a wide variety of cutting-edge applications in radar \cite{soumekh1999synthetic}, sonar \cite{Hayes2009}, radio telescopes \cite{levanda2009synthetic}, channel sounding \cite{Vouras2020}, optics \cite{konda2020fourier}, radiometry \cite{le1999synthetic}, acoustics \cite{fan2021synthetic}, quantum \cite{Holloway2014}, microscopy \cite{ralston2007interferometric} and biomedical applications, including ultrasound \cite{jensen2006synthetic}, magnetic resonance imaging (MRI) \cite{mayer2003synthetic}, magnetometry \cite{herdman2003determination}, and computed tomography (CT) \cite{akiyama1997synthetic}. Compared to a filled hardware aperture capable of the same angular resolution, the SAs offer savings in cost, hardware, and power while also providing a better view of occluded objects, improvement in signal-to-noise ratio (SNR), and enhanced resolution. The SAs may be constructed through the motion of the sensor/object or via a distributed deployment of sensors. Originally invented for radar systems in the 1950s, SAs were first implemented using digital computers in the late 1970s \cite{soumekh1999synthetic}. More advanced techniques were introduced in the late 1980s before widespread adoption in other applications throughout the 1990s \cite{ylitalo1994ultrasound}.

The angle and delay resolution of a metrology system that collects information from the environment by steering a high-gain antenna to different directions in space is determined by the physical size of the antenna and by the instantaneous bandwidth of the transmitted signal. As the size and cost of the sensors have decreased, denser and wider arrays have become feasible. Similarly, with the advent of several remote sensing and communications applications for higher frequency bands such as millimeter-wave \cite{Mishra2019} and Terahertz (THz) \cite{elbir2021terahertz}, SA systems with extremely wide bandwidths are currently being investigated. For example, millimeter-wave SA radar (SAR) is revolutionizing the rapid developments in the automotive industry toward building the next-generation autonomous vehicles \cite{gishkori2019imaging}. In quantum applications, Rydberg sensors are garnering significant interest for wideband receivers \cite{Holloway2014,vouras2023phase}. In SA sonar (SAS), existing algorithms are being adapted for widebeam and wideband systems to discern new properties of sea-bottom scattering \cite{Hayes2009}. In optics, coded diffraction patterns are now used to acquire several snapshots of the scene by changing the spatial configuration \cite{pinilla2022unfolding}.

Novel signal processing techniques are essential for implementations of wideband SA techniques. In wideband SAR, it is necessary to adapt common SAR algorithms based on spotlight Stolt and polar format for wideband processing \cite{carrara2004new}. Other related SAR computed imaging techniques such as autofocusing, Doppler processing, tomographic imaging, 3-D imaging, gridding, and subaperture processing are also significantly different for wideband. Quantum aperture receivers based on Rydberg quantum states sense only the intensity of electric fields thereby requiring novel receiver processing techniques to improve both sensitivity and frequency agility. Rydberg sensors exploit electromagnetically-induced transparency, wherein the intensity of a laser passing through a cloud of Rydberg atoms changes. In SA optics, algorithms for applications such as lensless microscopy, hyperspectral complex-domain imaging and phaseless object identification need to be developed \cite{pinilla2022unfolding}. Currently SAs are less common in MRI; ultrasound and CT SA applications require development of new processing and image quality assessment methods.

\begin{figure*}
\centering
\includegraphics[width=1.0\textwidth]{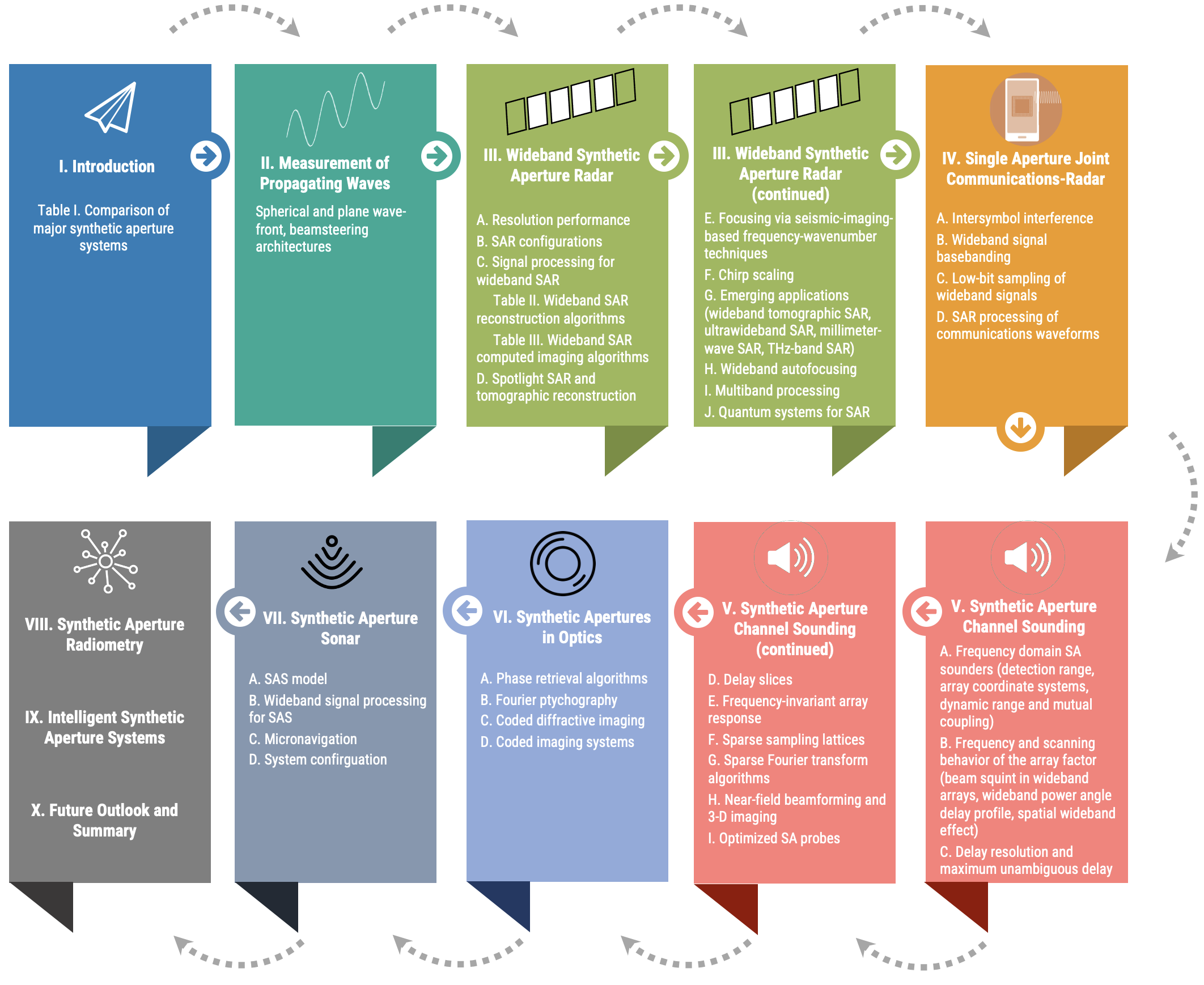}
\caption{Structure of the tutorial.}
\label{fig:outline}
\end{figure*}

In this paper, we provide a tutorial overview of methods to improve the spatial (angular) and delay resolution by synthesizing, respectively, a virtual aperture larger than the physical antenna and a measurement bandwidth greater than the instantaneous signal bandwidth. This concept has been leveraged in several applications, including radar imaging, sonar, optics, and wireless channel sounding as described in the paper.  Fig.~\ref{fig:outline} summarizes the structure of this tutorial.
%-----------------------------------------------------------------------------
  
In Section~\ref{sec:meas_prop} we describe propagating waves and the measurements associated with different phased array architectures.  In Section~\ref{sec:radar}, we provide a background on SAR techniques %. SAR is a powerful imaging technique with all-weather capability and is widely used in both military and civilian applications. After a tutorial background on the principles of SAR and different SAR configurations, 
along with some of the wideband SAR applications, including millimeter-wave SAR, wideband autofocusing, and quantum systems for SAR. Section~\ref{sec:jrc} presents new results for systems that leverage the same antenna aperture, hardware platform, and waveforms to combine radar detection processing with data communications. In Section~\ref{sec:channel_sound}, we discuss another important SA application of channel sounding. In modern 5G/6G communications, channel sounding plays an important role in establishing system performance, especially for single-carrier modulated systems. With multiple-carrier modulations, such as Orthogonal Frequency Division Multiplexing (OFDM), a guard interval is added between symbols, which mitigates the impact of multipath and intersymbol interference (ISI). The SAs have been used in channel sounding to accurately characterize the scattering of electromagnetic fields propagating through a wireless channel. This paper describes SA channel sounders that sample the frequency response of a wireless channel. A brief introduction to possible future paths of research describes time-domain SA sounders that utilize novel quantum sensors to measure the intensity of impinging electric fields. Then, Section~\ref{sec:sonar} explains the use of various lensng techniques in optics to generate wideband apertures. In Section~\ref{sec:sonar}, we introduce and discuss new developments in SAS such as wideband processing, micronavigation, and multiple-input multiple-output (MIMO) systems. Section~\ref{sec:radiometry} describes SA applications in radiometry. New innovations and capabilities enabled through the use of machine learning techniques in SA systems are briefly summarized in Section~\ref{sec:cognitive}. We conclude in Section~\ref{sec:summary}. Table~\ref{tbl:comparison} lists major SA systems by the quantities they measure.

\section{Measurements of Propagating Waves}
\label{sec:meas_prop}
\begin{figure}[htb]
% \begin{minipage}[b]{1.0\linewidth}
 \centering
 \includegraphics[width=1.0\columnwidth]{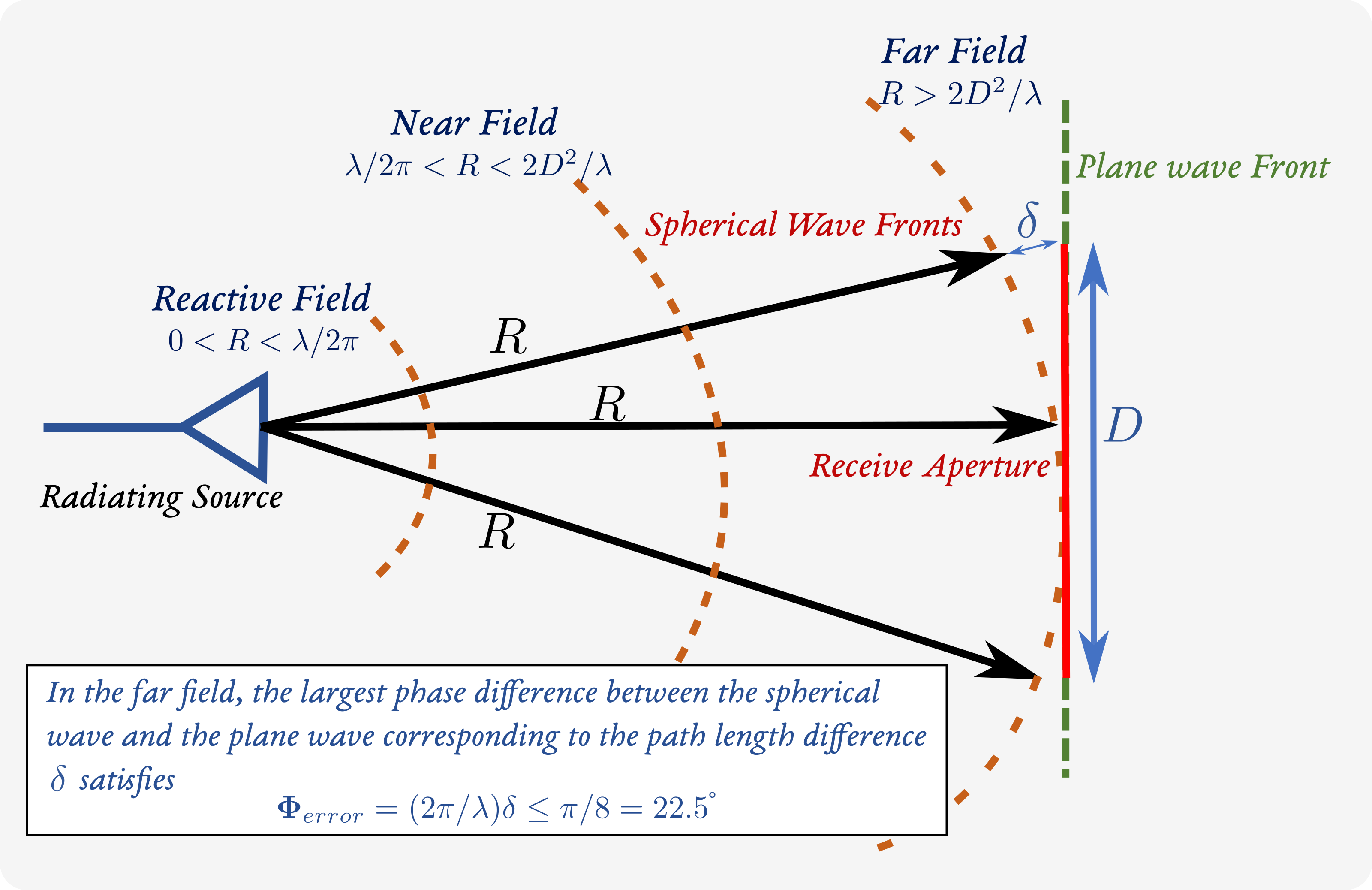}
% \vspace{2.0cm}
 \caption{Spherical propagation of waves.}%\medskip
 \label{fig:spherical}
% \end{minipage}
\end{figure}

Radio-frequency (RF) or acoustic energy radiating from a signal source travels radially outward as shown in Fig. \ref{fig:spherical}.
Beyond the nominal distance of ${2D^2/{\lambda}}$ a plane wave approximation is used to characterize the propagating fields.  Here ${D}$ refers to the largest dimension of the physical antenna and ${\lambda}$ is the operating wavelength.  For a spherical wavefront, the field ${s_{sp}}$ as a function of position ${\mathbf{x}}$ and time ${t}$ is given by,
\begin{equation}
\label{E:wave_sph}
    s_{sp}(\mathbf{x}, t) = e^{-\textrm{j}\frac{2{\pi}}{\lambda}d(\mathbf{x})}e^{\textrm{j}(2{\pi}ft + \phi(t))}
\end{equation}
with ${d(\mathbf{x})}$ equal to the distance between the signal source and the receive location ${\mathbf{x} = [x \quad y \quad z]^{T}}$, ${f}$ is temporal frequency, and ${\phi(t)}$ is a time-varying phase term due to waveform modulation or Doppler shift.  Eqn. (\ref{E:wave_sph}) is valid for any range.  
%----------------------------------------------------------------------------------
 \begin{table*}
 \caption{Comparison of major SA systems} 
 \label{tbl:comparison}  % Give a unique label
 \centering
 \begin{tabular}{p{4.1cm}P{4.1cm}P{4.1cm}P{4.1cm}}
 \hline\hline\noalign{\smallskip}
 %System\tnote{a} & Spatial Aperture Type \tnote{b} & Bandwidth Type
 \\
 Measures Temporal Phase Only &  Measures Temporal and Spatial Phase & Measures Spatial Phase Only & Does Not Measure Any Phase Directly\\
 \noalign{\smallskip}
 \hline\hline\noalign{\smallskip} 
 Single wideband antenna receiving modulated signal & & & \\
 \hline \medskip
 & Synthetic aperture radar (SAR): aperture created along aircraft trajectory, capable of wide instantaneous bandwidth & & \\
 \hline \medskip
 & Inverse synthetic aperture radar (ISAR): aperture created via object motion, capable of wide instantaneous bandwidth & & \\
 \hline \medskip
 & Interferometric synthetic aperture radar (InSAR): aperture created via phase difference from multiple passes, capable of wide instantaneous bandwidth & & \\
 \hline \medskip
 & & Synthetic aperture channel sounder: aperture created using mechanical positioner such as a robot, instantaneously narrowband but can synthesize wide bandwidths & \\
 \hline \medskip
 & Synthetic aperture sonar (SAS): aperture created along ship's trajectory, capable of wide instantaneous bandwidth \\
 \hline \medskip
 & & & Fourier ptychography: high-resolution image synthesized by illuminating object from different angles, uses phase retrieval to recover spatial phase \\
 \hline \medskip
 & Synthetic aperture radiometry: Resolution of large aperture synthesized from thinned array samples, capable of wide instantaneous bandwidth \\
\hline \medskip
Quantum: The use of a single Rydberg probe to receive and demodulate an FM radio signal has been demonstrated & & Quantum: SA constructed via mechanical positioner, has narrow instantaneous bandwidth but measurement is tunable over wide frequency range. Spatial phase is measured after radiating an LO signal that mixes with the carrier signal & Quantum: one version of a SA uses a mechanical positioner to move a Rydberg probe that measures electric field intensity only.  Phase retrieval algorithms are then applied to recover the spatial phase \\
\hline
 \noalign{\smallskip}\hline\noalign{\smallskip}
 \end{tabular}
 \end{table*}
 %----------------------------------------------------------------------------------

In the far-field, the field of a propagating monochromatic plane wave is approximated by 
\begin{equation}
\label{E:wave_prop}
    s(\mathbf{x}, t) = e^{\textrm{j}2{\pi}(-\mathbf{k}^{T}\mathbf{x} + ft) + \textrm{j}\phi(t)}
\end{equation}
where
\begin{align}
    \mathbf{k} &= (1/{\lambda})[\sin\theta \cos\phi \quad \sin\theta \sin\phi \quad \cos\theta ]^T \\ \nonumber
    &\triangleq [k_x \quad k_y \quad k_z]^{T}
\end{align}
is the spatial frequency vector.  The spherical angles ${(\theta,\phi)}$ and other array coordinates are defined further in Section \ref{sec:channel_sound}.  The vector ${\mathbf{k}}$ represents the number of wavelengths per unit distance in each of the three orthogonal spatial directions.  

A continuous 4-D Fourier transform of (\ref{E:wave_prop}) yields the 4-D wavenumber-frequency spectrum ${S(\mathbf{k},\omega)}$,
\begin{equation}
    \label{E:wf-spectrum}
    S(\mathbf{k},\omega) \triangleq \displaystyle\int_{-\infty}^{\infty} \displaystyle\int_{-\infty}^{\infty} s(\mathbf{x},t)e^{-\textrm{j}2{\pi}(f{t} - \mathbf{k}^{T}\mathbf{x})}d{\mathbf{x}}dt
\end{equation}
where ${\omega=2{\pi}f}$.  Any 4-D signal ${s(\mathbf{x},t)}$ can be decomposed into a superposition of propagating plane waves.  The Fourier transform of ${s(\mathbf{x}_{0},t_0)}$ with ${\phi(t)=0}$ yields
\begin{equation}
    S(\mathbf{k},\omega) = \delta(\mathbf{k}-\mathbf{k}_{0})\delta(\omega-\omega_{0})
\end{equation}
which is a 4-D delta function in ${(\mathbf{k},\omega)}$ space at the point ${\mathbf{k}=\mathbf{k}_{0}}$ and ${\omega=\omega_{0}}$.  Thus, each point in ${(\mathbf{k},\omega)}$ space corresponds to a plane wave in ${(\mathbf{x},t)}$ space at the frequency ${\omega}$ and in the direction ${\mathbf{k}}$ \cite{Dudgeon84}. 

\begin{figure}[htb]
% \begin{minipage}[b]{1.0\linewidth}
 \centering
 \includegraphics[width=8.5cm]{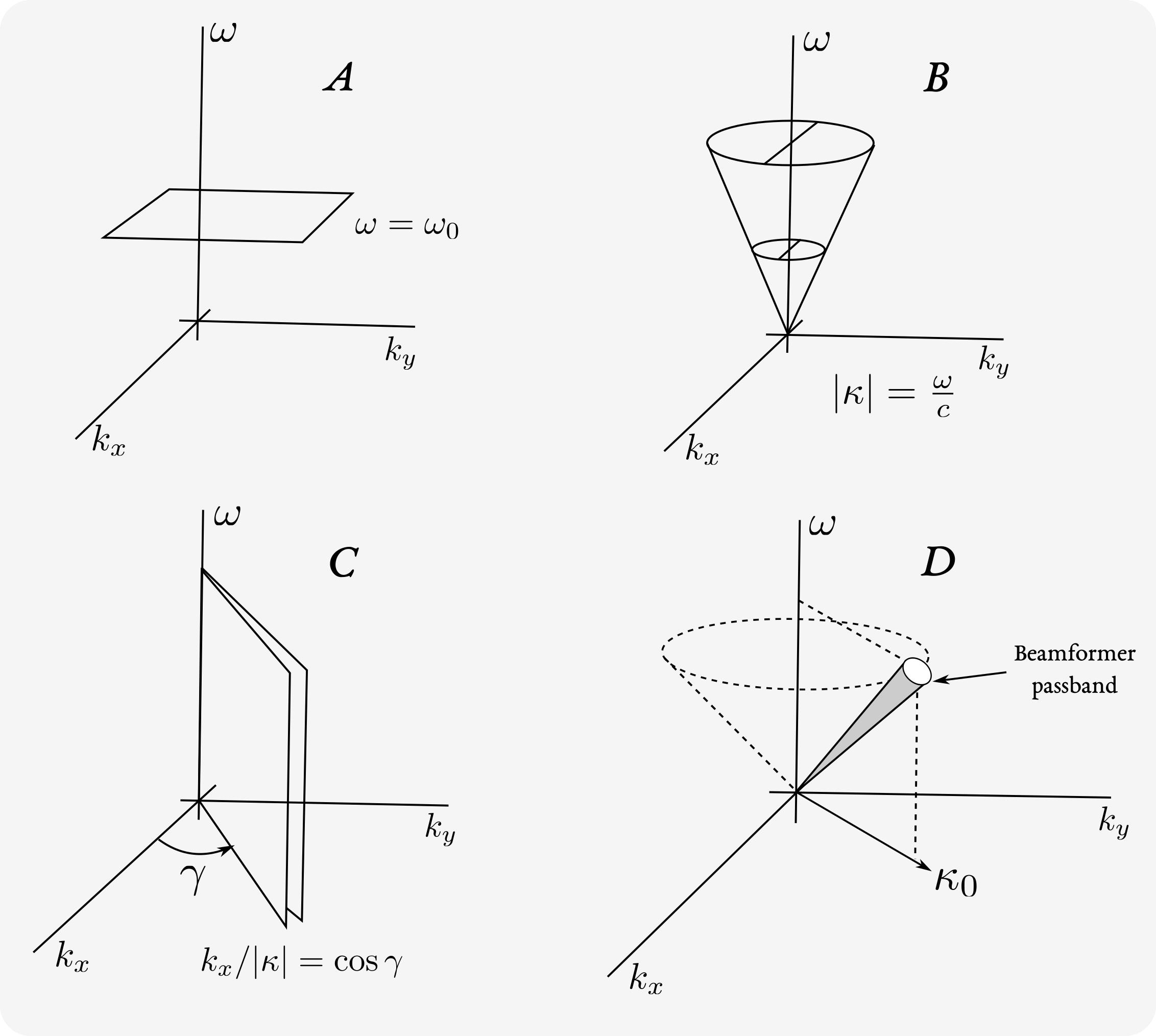}
% \vspace{2.0cm}
 \caption{Wavenumber-frequency representations of wideband propagating signals \cite{Dudgeon84}.} %\medskip
 \label{fig:k-omega}
% \end{minipage}
\end{figure}

In many wideband applications, including beamforming as described in Section \ref{ssec:wideband_beam}, it is often advantageous to describe signals and filtering operations in ${(\mathbf{k},\omega)}$ space.  For example, the illustration in Fig. \ref{fig:k-omega}A represents a signal with constant frequency propagating in all directions.  The ${k_x}$ and ${k_y}$ axes form the horizontal plane and the vertical axis corresponds to frequency ${\omega}$.  The ${k_z}$ axis is omitted for simplicity.  Fig.  \ref{fig:k-omega}B shows that wideband signals with the same propagation velocity ${c}$ lie on a cone in ${(\mathbf{k},\omega)}$ space since ${c=f/\Vert\mathbf{k}\Vert}$.  Wideband signals propagating in the same direction lie on a half-plane that forms an angle ${\gamma}$ with respect to the ${k_x}$ axis determined by the direction of the ${\mathbf{k}}$-vector as shown in Fig. \ref{fig:k-omega}C.  A wideband beamformer acts as a spatial filter that isolates the signals propagating in a particular direction.  All the signals propagating with a speed ${c}$ lie on the surface of the cone given by ${f=c\Vert\mathbf{k}\Vert}$ and the passband of the beamformer is given by the intersection of this cone with the half-plane corresponding to the desired direction vector ${\mathbf{k}_0}$, as illustrated in \ref{fig:k-omega}D.

The fundamental problem facing the designer of a synthetic aperture is how to measure the spatial phase equal to ${-2{\pi}\mathbf{k}^{T}\mathbf{x}}$ in (\ref{E:wave_prop}) and the temporal phase ${\phi(t)}$ simultaneously.  The spatial phase depends on the angle of arrival of a signal and measurements of spatial phase over a large aperture provide high angular resolution.  The temporal phase contains the modulation content of a signal as well as information on the relative Doppler shift between the transmitter and receiver.  The synthetic apertures described in this paper are analyzed according to how they measure spatial and temporal phase in Table \ref{tbl:comparison}.

First, consider the simple case of a phased array antenna that measures spatial and temporal phase simultaneously. 
\begin{figure}[htb]
% \begin{minipage}[b]{1.0\linewidth}
 \centering
 \includegraphics[width=8.5cm]{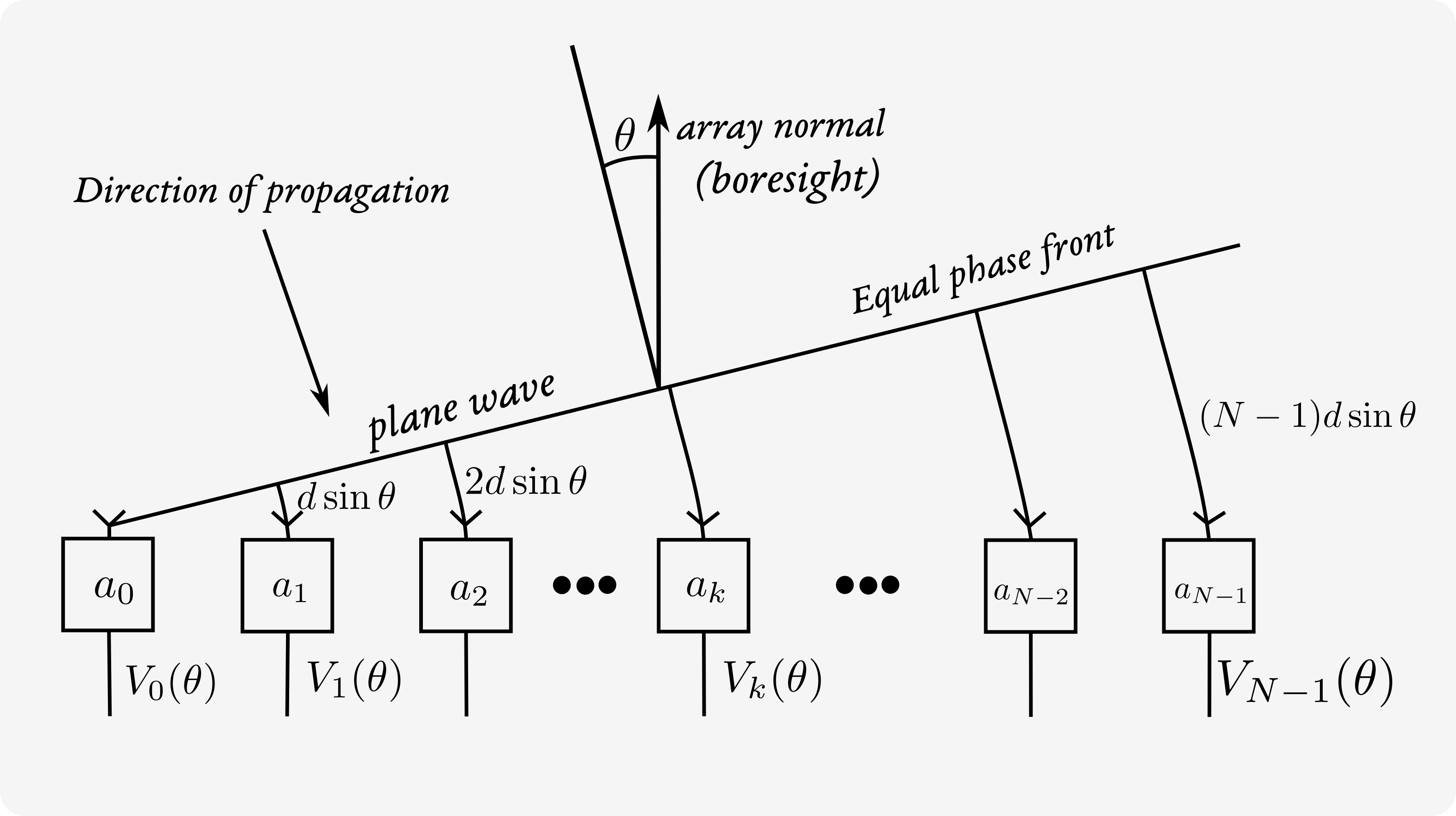}
% \vspace{2.0cm}
 \caption{Planar wavefront propagating across a phased array.}%\medskip
 \label{fig:plane_waves}
% \end{minipage}
\end{figure}
As shown in Fig. \ref{fig:plane_waves} a planar wavefront at some angular offset ${\theta}$ with respect to the array normal (also known as boresight) is incident at the edge element of a linear array.  The wavefront must travel an incremental distance ${d\sin\theta}$, where ${d}$ corresponds to the interelement spacing, before it arrives at the second array element.  Typically, ${d=\lambda/2}$ where ${\lambda}$ corresponds to the operating frequency of the array.  Between the first and last elements, the straight-line wavefront traverses a total incremental distance equal to ${(N-1)d\sin\theta}$, where ${N}$ equals the number of array elements.  In the narrowband case, each increment of distance corresponds to an additional shift of spatial phase equal to ${(2\pi/\lambda)nd\sin\theta}$ for ${n = 0,\ldots,N-1}$.

An analog beamforming network coherently combines the RF outputs of all the array elements to form directional beams in space.  When the array is beamsteered, a phase shifter behind each array element compensates for the spatial phase corresponding to the RF output of each array element.  If all the element outputs are perfectly phase aligned, the beamformer's output is maximum and the array mainbeam is pointed directly towards a signal source.  The temporal phase of the impinging signal is not affected by the analog beamforming operation and is available for demodulation or Doppler processing at the output of the beamformer for every time instant.  The coherent summation of all array elements in the beamfomer improves the received signal power by a factor of ${N^2}$.  Each array element also contributes uncorrelated noise to the summation which increases the total noise power at the beamformer output by a factor of ${N}$.  Thus the total signal-to-noise-ratio (SNR) measured at the output of the array is a factor of ${N}$ greater than the SNR of an individual receive element.

\begin{figure}[t]
% \begin{minipage}[b]{1.0\linewidth}
 \centering
 \includegraphics[width=1.0\columnwidth]{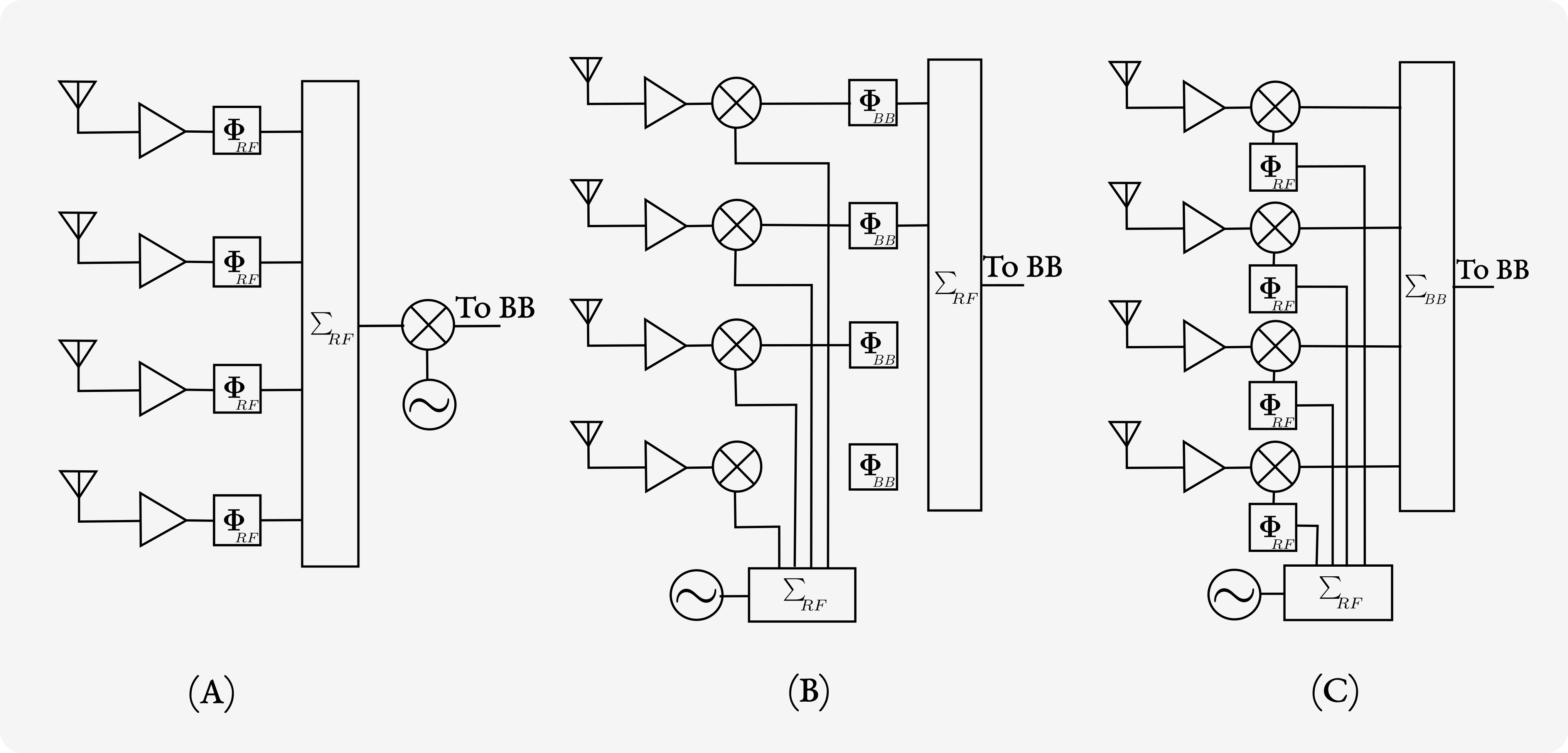}
% \vspace{2.0cm}
 \caption{Beam steering architectures for phased arrays.}%\medskip
 \label{fig:beam_steer}
% \end{minipage}
\end{figure}
The phase shifting operation for steering the beam of a phased array can be performed in one of three ways. As shown in Fig. \ref{fig:beam_steer}A the phase shifter can be placed in the RF path, the local oscillator (LO) path, or at baseband (also called intermediate frequency (IF) phase shifting).  Phase shifting in the RF path requires wide band and low loss phase shifters \cite{Natarajan2007}.  Wide bandwidth is required to accommodate the transmit signal and low loss is necessary to maintain a low receiver noise figure.  After initial amplification using a low-noise amplifier (LNA), the signals are combined at RF which implies the combiner must also be low loss and wideband.  One advantage to this approach is that only one mixer and LO are required.  Disadvantages are that RF phase shifters and power combiners have significant loss which is placed directly in the signal path in this architecture.  Also, it is difficult to achieve many phase quantization levels for high phase resolution with RF phase shifters.

Phase shifting at baseband shown in Fig. \ref{fig:beam_steer}B requires an LNA and a mixer behind each array element.  Furthermore, an identical LO must be fed to each element for downconversion.  Phase shifting is performed on the downconverted signal using IF phase shifters which are very compact and can have wide bandwidth and high resolution while maintaining low power consumption \cite{Marcu2009}. After phase shifting, the individual signals are combined. This architecture is very modular and allows for low power phase shifters and combiners.  Digital signal processing can also be
utilized to implement complex calibrations which are not possible in the RF architecture since the signal generated at baseband has already been combined.  The disadvantage of this architecture is that every element is in effect a full transceiver and the high frequency LO signal must be split and distributed to each one.  The bandwidth of the LO path is no longer determined by the signal's frequency support but by the LO tuning range which could be larger.

The final analog array architecture in Fig. \ref{fig:beam_steer}c implements phase shifting in the LO path \cite{Hashemi2005,Babakhani2006,Natarajan2006}.  This architecture also requires each element to have an LNA and mixer.  However, each mixer is fed an appropriately phase shifted LO signal.  The mixing action transfers the phase of the LO to the downconverted signal after low-pass filtering.  In this architecture high frequency phase shifting is required as in the RF architecture and
LO distribution is required as in the IF architecture.  The only advantage is that signal combination is simpler at baseband than in the IF architecture.  Unfortunately, this topology has the disadvantages of both previous architectures without providing significant performance improvements.

\begin{figure}[t]
% \begin{minipage}[b]{1.0\linewidth}
 \centering
 \includegraphics[width=0.7\columnwidth]{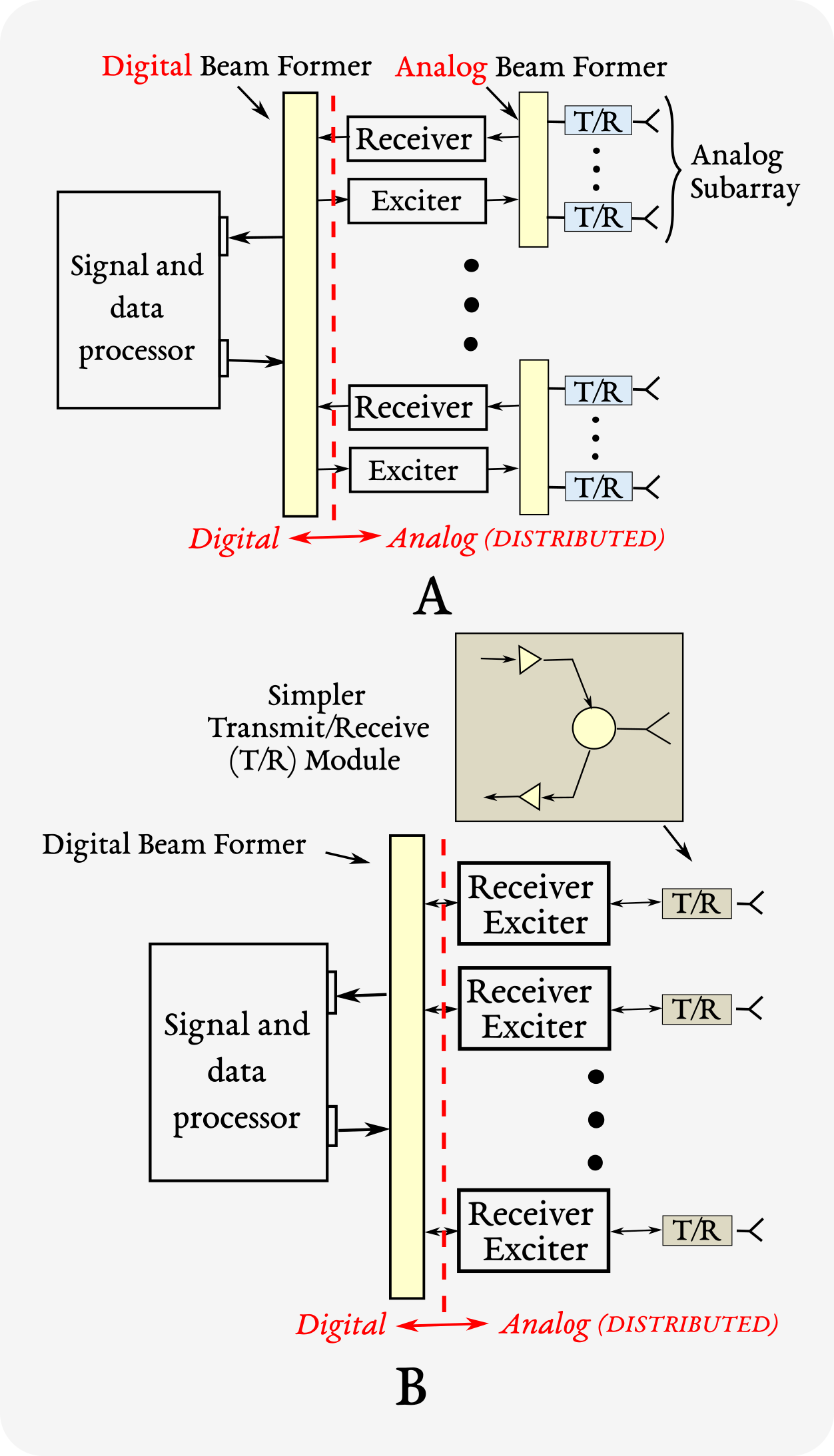}
% \vspace{2.0cm}
 \caption{(a) Subarray-level digital array.  (b) Element-level digital array \cite{Talisa2016}.}%\medskip
 \label{fig:digital_array}
% \end{minipage}
\end{figure}

The most powerful architecture for phased arrays is the digital array.  A fully digital array has a transceiver behind each element and digitizes received energy immediately behind the antenna as shown in Fig. \ref{fig:digital_array}B.  Phase and amplitude beamsteering weights are applied digitally and beamforming is performed using field programmable gate arrays (FPGAs).  Transmitted energy is also generated digitally at each element.  A digital array allows multiple independently-steerable beams to be formed that use the entire aperture and don't suffer a penalty in signal-to-noise-ratio (SNR).  Digital arrays are also easier to calibrate.  Some digital arrays are digitized at the subarray level to reduce the number of transceivers as shown in Fig. \ref{fig:digital_array}A.  In this case, the receivers and waveform generators are placed behind analog-beamformed subarrays \cite{Talisa2016}.

Ultimately the size of a phased array determines the width of the mainbeam and the angular resolution.  To attain the higher spatial resolution that yields superior imaging performance, a large aperture is necessary which is not economical given the complexity of beamsteering, especially at millimeter wave frequencies.  The push towards higher angle resolution motivates the development of synthetic apertures that can measure spatial phase over a larger volume than a hardware array.  The resulting increase in effective aperture area may create difficulties however in simultaneously measuring temporal phase.  We explore these issues in the remainder of the paper and begin our overview of synthetic apertures in Section \ref{sec:radar} by considering radar, which represents the largest commercial application of synthetic aperture techniques.

%\clearpage
\section{Wideband SAR}
\label{sec:radar}
When a radar illuminates an object, conventional processing techniques, such as beamforming and matched filtering, are utilized to obtain downrange resolution along the radar line-of-sight (LoS). If the object is also moving relative to the radar, then the Doppler frequency gradient is used to obtain cross-range resolution that is much finer than the radar's beamwidth. The motion of the object is generated in a variety of ways but ultimately this motion is related to the simplified case of a stationary monostatic radar illuminating a rotating object.  A careful distinction should be made between the terms SAR and inverse SAR (ISAR).  Technically, ISAR refers to a radar configuration with a moving target and a stationary antenna.  SAR refers to a moving antenna and a stationary target.  The analysis for range-doppler imaging is similar in both cases since a stationary target will appear to be rotating if the antenna is moving.
\begin{figure}%[htb]
% \begin{minipage}[b]{1.0\linewidth}
 \centering
 \includegraphics[width=8.5cm]{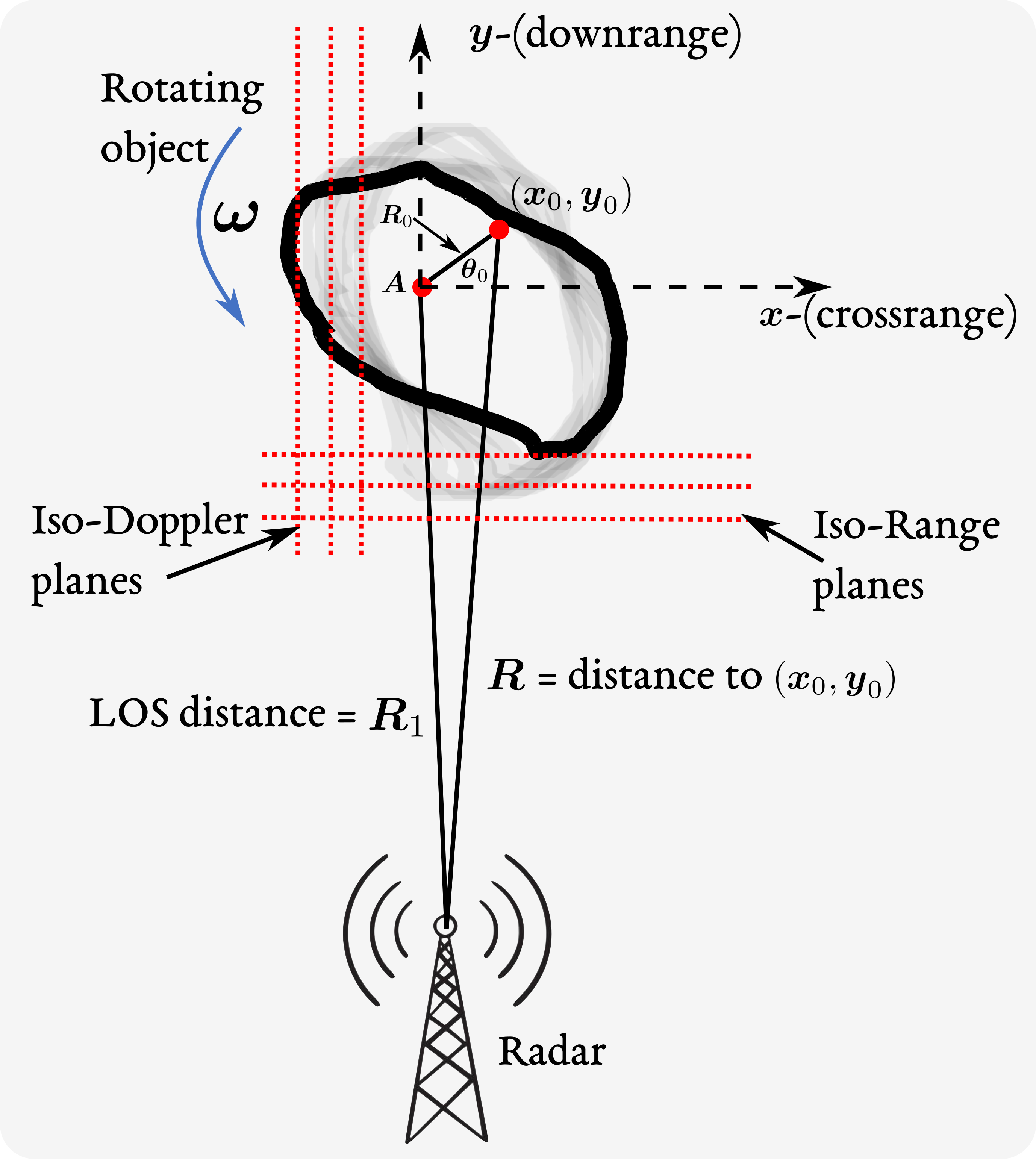}
% \vspace{2.0cm}
 \caption{Range-Doppler imaging through radar using the synthetic aperture principle.  Signal bandwidth and Doppler frequency resolution determine downrange and cross-range resolutions, respectively. The graphic shows a moving target for a stationary antenna. This is the principle of ISAR. In SAR, the SA is created with a moving array sensor and stationary target.}%\medskip
 \label{fig:rangeDopplerImaging}
% \end{minipage}
\end{figure}

In Fig.~\ref{fig:rangeDopplerImaging}, a three-dimensional (3-D) object illuminated by the radar is projected onto the x-y plane with the object rotating about the z-axis with uniform angular velocity. If the object is contained within the main beam of the radar and rotating about the point ${A}$ at ${\omega}$ radians per second with the radar at a LoS distance ${R_{1}}$ from ${A}$, then the distance to a point on the object with initial coordinates ${(R_{0}, \theta_{0}, z_{0})}$ at time ${t=0}$ is 
\begin{equation}
 %\label{E:eqn01}
 R = [R_{0}^2 + R_{1}^2 + 2R_{1}R_{0}\sin(\theta_{0} + {\omega}{t}) + z_{0}^2]^{\frac{1}{2}}.
\end{equation}
If the distance to the object is much larger than the size of the object, i.e. ${R_{1} >> R_{0}, z_{0}}$, then a good approximation is
\begin{equation}
 %\label{E:eqn02}
 R \approx R_{1} + x_{0}\sin(\omega{t}) + y_{0}\cos(\omega{t}),
\end{equation}
and the Doppler frequency of the returned signal is
\begin{equation}
 %\label{E:eqn03}
 f_{d} = \frac{2}{\lambda}\frac{dR}{dt} = \frac{2x_{0}\omega}{\lambda}\cos{\omega{t}} - \frac{2y_{0}\omega}{\lambda}\sin{\omega{t}},
\end{equation}
where ${\lambda}$ is the radar wavelength. If the radar data are processed over a short time interval centered at ${t=0}$, the range to ${(x_{0},y_{0})}$ and the Doppler frequency shift is approximated as
\begin{equation}
 \label{E:eqn04}
 R = R_{1} + y_{0}, \quad f_{d} = \frac{2x_{0}\omega}{\lambda}.
\end{equation}

It follows that the downrange component ${y_{0}}$ of the position of a point scatterer is estimated by analyzing the delay of the radar return and the cross-range component ${x_{0}}$ is obtained by analyzing the Doppler frequency shift \cite{Ausherman1984,Fienup2003,Fienup2001,Fienup1994,Fienup2000,Fienup2000B,Ye1999}. This framework captures the conventional range-Doppler imaging technique used in SAR. An implicit assumption is that the LoS distance ${R_{1}}$ from the radar antenna to the center of the rotating object is a constant and known value. If ${R_{1}}$ is time varying, then the effects of changing range must be compensated for in the signal processing. The parallel lines in Fig.~\ref{fig:rangeDopplerImaging} perpendicular to the radar LoS are surfaces of constant delay or range. The surfaces of constant Doppler are the lines parallel to the plane formed by the LoS and the rotation axis.

\subsection{Resolution Performance}
\label{subsec:resolution}
The downrange resolution ${\Delta{R}}$ of the monostatic radar illuminating a rotating object is determined by the instantaneous bandwidth ${B}$ of the transmitted waveform, ${\Delta{R} = {c}/{2B}}$, where ${c}$ is the speed of light. The factor of two arises because an incremental delay ${\Delta{t} = 1/B}$ corresponds to an incremental downrange distance ${\Delta{R} = c\Delta{t}/2}$. Fine range resolution is achieved with a single pulse and the corresponding processing is termed fast-time processing, meaning that the input data rate is equal to the analog-to-digital converter (ADC) sampling rate.

It follows from \eqref{E:eqn04} that a cross-range resolution ${\Delta{x}}$ is achieved if Doppler frequency is measured with a resolution of
\begin{equation}
 \label{E:eqn05}
 \Delta{f_{d}} = \frac{2\omega{\Delta}x}{\lambda}.
\end{equation}
A resolution of ${\Delta{f_{d}}}$ requires a coherent processing interval of approximately ${T = 1/\Delta{f_{d}}}$. The cross-range resolution is 
\begin{equation}
 \label{E:eqn06}
 \Delta{x} = \frac{\lambda}{2\omega{T}} = \frac{\lambda}{2\Delta{\theta}},
\end{equation}
where ${\Delta{\theta} = \omega{T}}$ is the angle through which the object rotates during the coherent processing interval. Fine cross-range resolution requires multiple pulses and the corresponding processing is often called slow-time processing because the input data rate is equal to the pulse repetition frequency (PRF) of the transmitted waveform.

\subsection{SAR Configurations}
\label{subsec:sar}

\begin{figure}[t]
 %\centering
 \includegraphics[width=1.0\columnwidth]{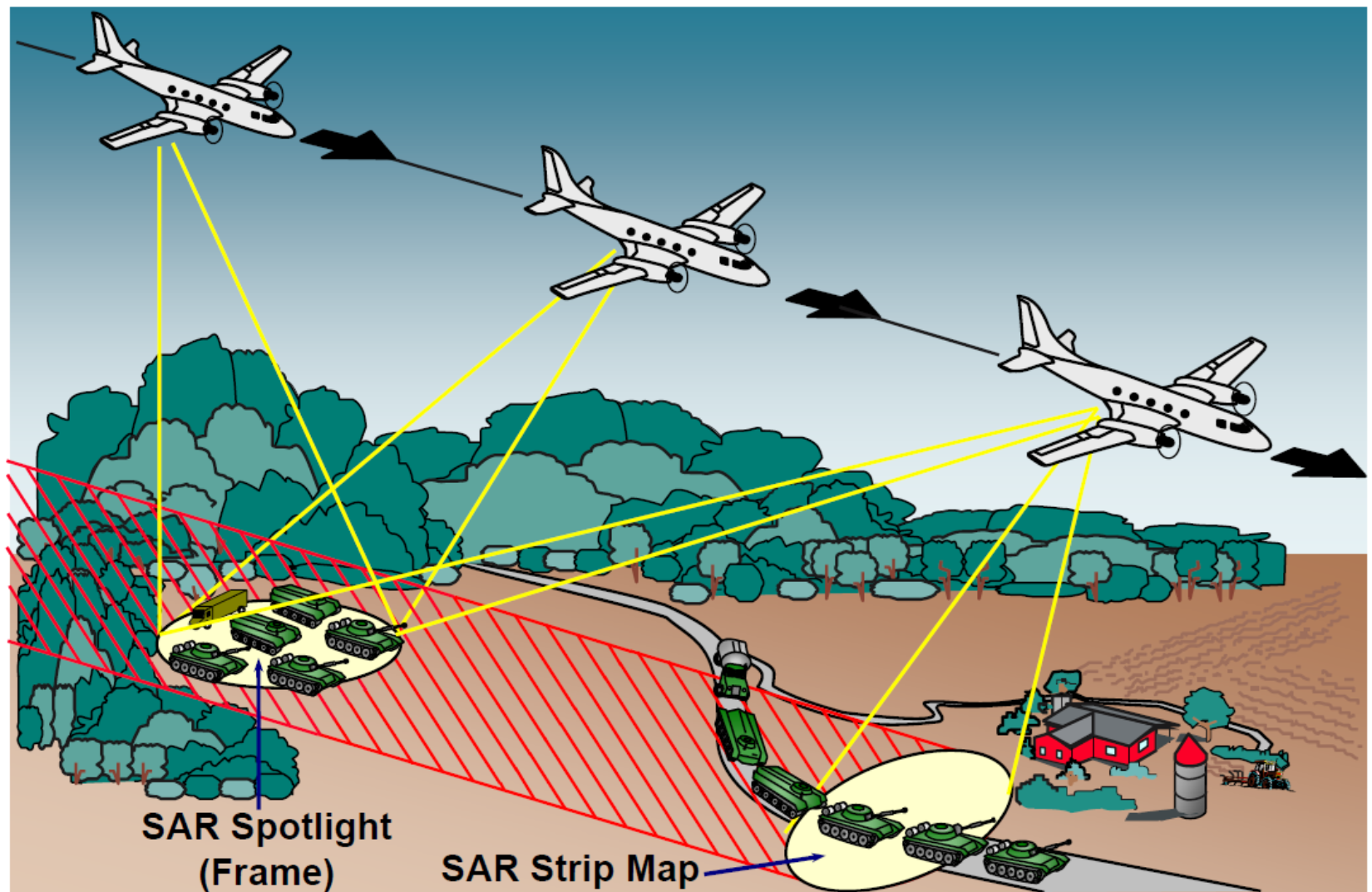}
 \caption{Stripmap and Spotlight SAR Operation~\cite{ODonnell2013}}
 \label{fig:SAR_scans}
\end{figure}

The range-Doppler imaging principle leads to several different SAR configurations \cite{Ender2014}.  Strictly speaking, SAR is a method only useful for improving cross-range resolution.  One common SAR configuration is stripmap SAR where the radar is mounted on an airborne platform and looking down from the side of the aircraft towards terrain.  Another configuration is spotlight SAR where the antenna on the moving aircraft illuminates a fixed area on the ground from a continuously changing aspect angle.  These two modes are illustrated in Fig. \ref{fig:SAR_scans}.

Assume the aircraft is moving in a straight line at a constant altitude with a speed ${V}$ for a duration ${T}$ along the direction perpendicular to the LoS. The length of the SA, ${L = VT}$, is small compared to the range ${R_{1}}$ to the center of the target region, so the angle subtended by the SA is approximately ${\Delta{\theta} \approx L/R_{1} = VT/R_{1}}$.  From the viewpoint of the radar, the scene appears to be rotating with angular velocity ${\omega = V/R_{1}}$. During the duration ${T}$, the total angle through which the scene appears to rotate is ${\Delta{\theta} = \omega{T} = VT/R_{1}}$. A point scatterer in the scene will appear to have a LoS velocity of ${\omega{x}}$ relative to the radar, where ${x}$ is the cross-range distance of the scatterer from the radar LoS. This apparent LoS velocity ${v}$ will create a Doppler frequency of ${f_{d} = 2v/\lambda = 2\omega{x}/\lambda}$.

\begin{figure}
 %\centering
 \includegraphics[width=1.0\columnwidth]{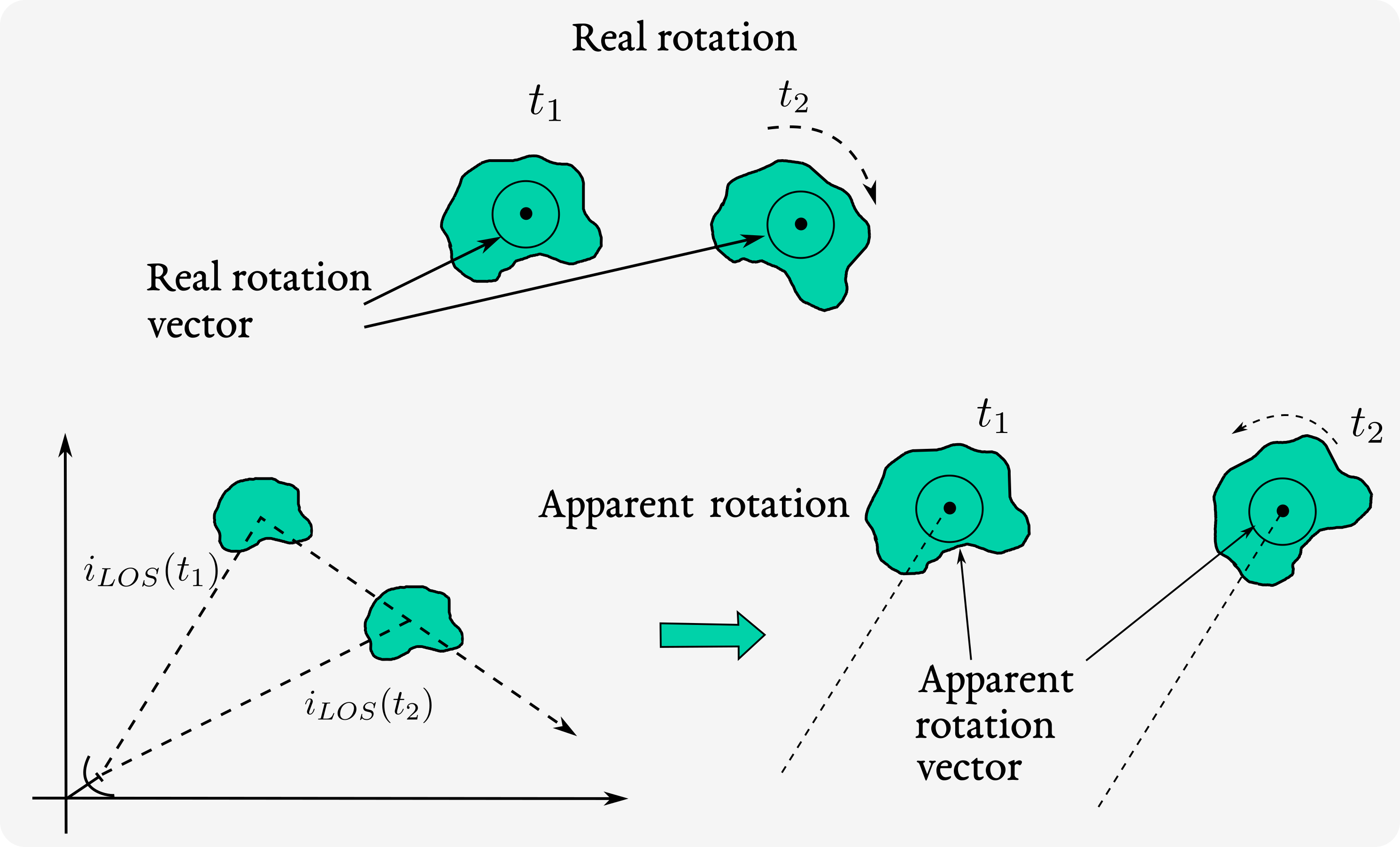}
 \caption{Real vs apparent rotation of an object.  Real rotation refers to the object physically spinning about an axis.  Apparent rotation is induced when the radar LoS to the object changes due to platform motion~\cite{Martorella2010-2}}
 \label{fig:rotate}
\end{figure}

The concept of apparent rotation is illustrated in Fig. \ref{fig:rotate}.  In the case of real rotation the object is physically spinning about the z-axis.  If the object is instead static but the radar LoS to the object is changing due to platform motion, then the object will also appear to be rotating about an axis with the corresponding Doppler frequency~\cite{Martorella2010}.

If the Doppler frequency can be measured with a resolution of ${\Delta{f}_{d}}$, then the corresponding cross-range resolution is
\begin{equation}
 \label{E:eqn07}
 \Delta{x} = \frac{\lambda}{2\omega{T}} = \frac{\lambda}{2\Delta{\theta}} \approx \frac{\lambda{R_{1}}}{2L} = \frac{\lambda{R_{1}}}{2VT}.
\end{equation}
Equations \ref{E:eqn06} and \ref{E:eqn07} yield the same result but are derived using different approaches. Equation \ref{E:eqn06} suggests that cross-range resolution results from the Doppler shifts created by the different apparent LoS velocities of point scatterers in the scene \cite{Werness1990}. Equation \ref{E:eqn07} indicates that cross-range resolution is a result of the larger aperture size as measured by its length ${L}$. Both interpretations are valid and show that synthesizing larger apertures and using coherent processing can increase cross-range and angular resolution.

%------------------------------------------
\begin{figure}[t]
 %\centering
 \includegraphics[width=1.0\columnwidth]{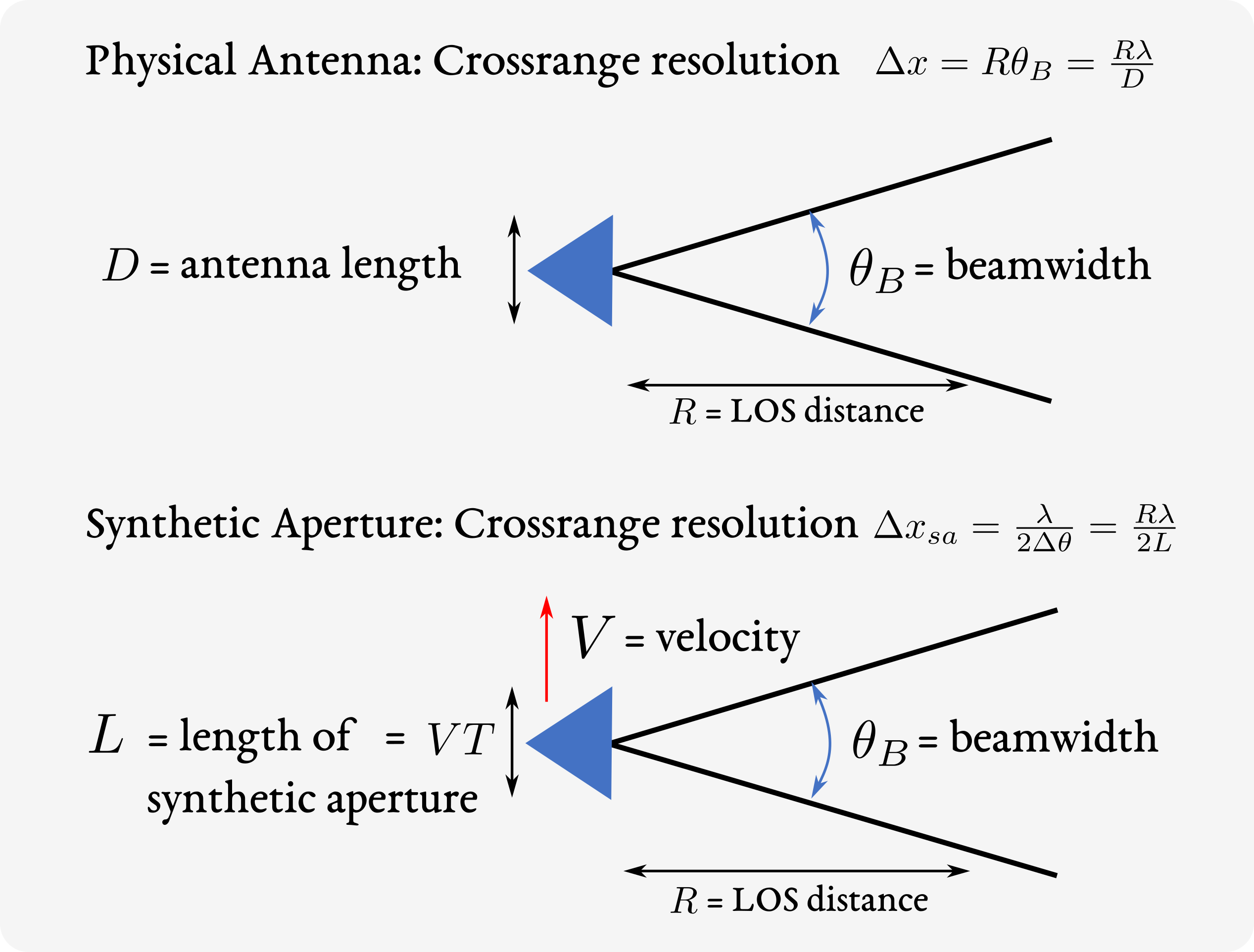}
 \caption{Cross-range resolution of physical antenna (top) and SA (bottom).}
 \label{fig:crossrangeResolution}
\end{figure}
%------------------------------------------
Fig.~\ref{fig:crossrangeResolution} compares the cross range resolution, defined as the distance from the mainlobe peak to the first null in the antenna pattern, for a physical antenna and for a SA of the same size. The beamwidth of the physical antenna is approximately given by ${\theta_{B} \approx \lambda/D}$, where ${D}$ is the cross-range length of the antenna, and in this scenario ${D = L}$. The figure illustrates that for the case of range-Doppler imaging the cross-range resolution that can be achieved by a SA is one-half the cross-range resolution possible using a physical antenna of the same size.

SAR resolution performance will be degraded if point scatterers in the scene move through different range or Doppler resolution cells. Errors caused by range or Doppler bin migration will have to be removed in the signal processing or by shortening the coherent processing interval. Other errors that affect performance include a variable angular rotation rate or a radar LoS that is not orthogonal to the axis of rotation.

The range-Doppler imaging technique is best used when the total variation in aspect angle to the scene is small (${<5^{\circ}}$) and the observation time is short.  In this case, the sampled data collected by the radar is approximately evenly spaced over a rectangular grid and Fast Fourier Transform (FFT) processing can be applied \cite{Martorella2010}.

\subsection{Signal Processing for Wideband SAR}
\label{sec:spwideband}
\begin{table*}
 \label{tbl:reconst}
 \centering
 \caption{Wideband SAR Reconstruction Algorithms}
    \begin{tabular}{lccl}
    \hline Algorithm & Pulse Repetition Interval (PRI) & Spatial Resolution & Remarks \\
    \hline
    \hline
    Spotlight polar format & $\propto$ Antenna size  & $\propto$ Antenna size & Spot size limited by beamwidth; common in UWB SAR \\
    Spotlight Stolt format & $\propto$ Wavelength & $\propto$ Wavelength & No slow-time dwell or scene size limit \\
    Stripmap range-Doppler & $\propto$ Antenna size & $\propto$ Antenna size & Slow-time dwell limited by beamwidth \\
    Stripmap chirp-scaling & $\propto$ Antenna size & $\propto$ Antenna size & No interpolation required; not recommended for squint-mode \\
    \hline
    \end{tabular}
\end{table*}

\begin{table*}
 \label{tbl:ct_comp}
 \centering
 \caption{Wideband SAR Computed Imaging Algorithms}
    \begin{tabular}{llcl}
    \hline Algorithm & Wideband variation & Representative application & cf. \\
    \hline
    \hline
    Autofocusing & Data-driven and inverse problem approaches & mmWave FloSAR & Section~\ref{subsec:autofocus} \\
    Backprojection & Plane-wave Fraunhofer assumption,  spherical Radon transform & General wideband SAR & Section~\ref{sec:spwideband} \\
    Tomographic imaging & Algebraic reconstruction technique (ART) & Asteroid imaging SAR & Section~\ref{subsubsec:tomsar}\\
    Doppler processing & Doppler-induced temporal analysis, Doppler beam sharpening & Automotive MIMO-SAR & Section~\ref{subsubsec:thz}\\
    Subaperture Stolt format & Reduction in FFT size, coherent summing of low-resolution images & Satellite/aircraft-borne SAR & \cite{moreira1990new}\\
    Spectral estimation & Multiple component signal reconstruction & Long-dwell SAR & \cite{fornaro2005three,fornaro2008lmmse} \\
    Gridding & Replacement of Fourier reconstruction, resampling, weighting functions & General wideband SAR & \cite{gorham2007comparison}\\
    Omega-k processing & Parallelized algorithms & General wideband SAR & \cite{yerkes1994implementation}\\
    ISAR & CS-based reconstruction & General wideband ISAR & \cite{hou2015compressed}\\
    InSAR & Split-bandwidth interferometry, multi-frequency processing & Satellite-based interferometry & Section~\ref{subsec:multi}\\
    Multistatic SAR & Equivalent sensor approach & Concealed object imaging SAR & \cite{sheen2001three}\\
    3-D SAR & Multi-pass sensing, volumetric Stolt format & Circular flight SAR & \cite{reigber2000first}\\
    ScanSAR & Multiple beams & Satellite SAR & \cite{claassen1978system} \\
    QSAR & Quantum illumination & QTMS radar & Section~\ref{ssec:quantumSAR} \\
    \hline
    \end{tabular}
\end{table*}
SAR signal processing encompasses a diverse and huge set of algorithms that vary as per the processing domain, wave-field model, and various wave-field approximations. We summarize some of the major algorithms in Table~\ref{tbl:reconst} with a focus on wideband SAR. In addition, some of these algorithms may be used in conjunction with computed imaging algorithms depending on the application. Table~\ref{tbl:ct_comp} summarizes the unique wideband aspect of these allied techniques along with representative applications discussed later. It is difficult to explain each one of these algorithms in detail here. Therefore, we provide suitable references to the reader. In the sequel, we give a high-level overview of SAR processing based on the back-projection algorithm.

We begin by defining phrases and terms that leverage the concept of signal coherence.  For example, a coherent pulse train implies that the receiver knows the initial phase of each pulse.  A coherently processed pulse train implies that the processing (e.g. an FFT) makes use of the known phases.  Coherent signals are signals with a known phase relationship.  For example, transmit and receive signals referenced to the same LO are coherent.  Coherently integrated data refers to the complex sum over the spatial, spectral, or temporal domains of coherent data with measured amplitude and phase.  For example, the data may be samples of a coherent pulse train or coherent spatial samples of the signal received across a planar synthetic aperture.  In some radar systems a transmitted pulse may have random initial phase.  If this phase is recorded and used as a reference for the received signal, then the system is known as coherent on receive.  Coherent on receive systems know the phase of the previous pulse but the pulses themselves are not coherent because there is no phase relationship between them.

Section \ref{subsec:resolution} highlighted the importance of spatial resolution in SAR systems.  Equally important for the system designer is achieving adequate SNR on a target.  The simple approach of transmitting a short-duration, high-bandwidth pulse to attain better range resolution sacrifices energy on the target.  The standard solution is to use pulse compression waveforms, such as a Linear Frequency Modulated (LFM) chirp or a Binary Phase Modulated signal, that illuminate the target with a long pulse.  A long pulse can have the same bandwidth and delay resolution as a short pulse if it is modulated in frequency or phase.  A chirp signal is described by 
\begin{equation}
    s(t) = A\text{rect}\left(\frac{t}{T}\right)e^{\textrm{j}2{\pi}f_{c}t + \textrm{j}{\pi}Kt^2}.
\end{equation}
The parameter ${f_c}$ is the center frequency, ${A}$ is the amplitude, and the chirp rate ${K = B/T}$ is given by the ratio of the bandwidth to the pulse duration.  The bandwidth ${B = f_{max} - f_{low}}$ is equal to the difference between the upper and lower frequencies of the chirp.  Often, the time-bandwidth product ${TBP = BT = T/T_c}$ is used to describe chirps.  The ${TBP}$ is also equal to the ratio of the uncompressed pulse duration to the compressed pulse width, ${T_c}$.  Higher values of ${TBP}$ result in a narrower mainlobe after matched filtering and higher delay resolution.

A matched filter maximizes the peak SNR at its output for any signal in white noise \cite{Skolnik2008}.  The frequency response of the matched filter is given by
\begin{equation}
    H( f ) = S^{*}(f)e^{-\textrm{j}2{\pi}fT},
\end{equation}
where ${S(f)}$ is the Fourier transform of the signal ${s(t)}$.  The signal received by the radar will be a delayed and attenuated copy of the transmitted signal.  If matched filtering is performed at baseband, then ${f_c = 0}$ and the complex signal after downconversion will be
\begin{equation}
    s(t) = A\text{rect}\left(\frac{t}{T}\right)e^{\textrm{j}{\pi}Kt^2}.
\end{equation}
The matched filter impulse response is
\begin{equation}
    h(t) = A\text{rect}\left(\frac{t}{T}\right)e^{-\textrm{j}{\pi}Kt^2}
\end{equation}
and the magnitude squared output of the matched filter ${\Psi(\tau,f_{d})}$ as a function of delay and Doppler shift is \cite{Levanon2004}
\begin{align}
\label{eqn:mf}
    \Psi(\tau,f_{d}) = \left|\left(1-\frac{|t|}{T}\right)\frac{\sin\left({\pi}T(Kt+f_{d})\left(1-\frac{|t|}{T}\right)\right)}{{\pi}T(Kt+f_{d})\left(1-\frac{|t|}{T}\right)} \right|^2, \quad |t| \leq T.
\end{align}

The detection performance of a radar waveform in terms of measuring a target's velocity and range is often analyzed using the concept of an ambiguity function. The ambiguity function represents the output of a matched filter for all possible target delays and Doppler shifts.  In general, the ambiguity function ${\Psi(\tau,f_{d})}$ for a transmit waveform with complex envelope ${u(t)}$ is defined as the squared magnitude of the autocorrelation function ${\chi(\tau,f_{d})}$,
\begin{equation}
 \Psi(\tau, f_{d}) = |\chi(\tau,f_{d})|^{2},
\end{equation}
where ${\tau}$ represents relative time delay and ${f_{d}}$ is Doppler shift. The autocorrelation function ${\chi(\tau,f_{d})}$ for ${u(t)}$ is defined as,
\begin{equation}
 \chi(\tau,f_{d}) = \int\limits_{-\infty}^{\infty}u(t)u^{*}(t+\tau)e^{\mathrm{j}2{\pi}f_{d}t}dt.
\end{equation}
The value of the ambiguity function at the origin is equal to ${(2E)^{2}}$ where ${E}$ is the energy of the bandpass signal corresponding to ${u(t)}$, and the volume under the ambiguity function is also equal to ${(2E)^{2}}$.

The impulse response of a filter matched to a waveform ${u(t)}$ is given by
\begin{equation}
 h(t) = u^{*}(-t).
\end{equation}
The output ${y(t)}$ of a matched filter to an input signal ${s(t) = u(t)e^{\mathrm{j}2{\pi}f_{d}t}}$ with zero time delay and Doppler shift ${f_{d}}$ is given by the convolution of ${s(t)}$ with the matched filter impulse response ${h_{mf}(t)}$,
\begin{equation}
 y(t) = \int\limits_{-\infty}^{\infty}u(t')u^{*}(t'-t)e^{\mathrm{j}2{\pi}f_{d}t'}dt'.
\end{equation}
Comparing this result with the definition of the autocorrelation function shows that the matched filter response can be expressed as,
\begin{equation}
 y(t) = \chi(-t, f_{d}).
\end{equation}
Thus, the matched filter output for a target with Doppler frequency ${f_{d}}$ is a time-reversed version of the autocorrelation function.

The ambiguity function computed from the magnitude squared autocorrelation of the baseband radar waveform ${u(t)}$ can be used to describe the resolution performance of the waveform. For example, assume ${u(t)}$ is normalized to have unit energy,
\begin{equation}
 \int\limits_{-\infty}^{\infty}|u(t)|^{2}dt = 1,
\end{equation}
and two targets are located in the same angular direction and with equal radar cross sections. If one target is located at the origin of the delay-doppler plane with zero Doppler and zero relative time delay, then the value of the ambiguity function is unity, ${\Psi(0, 0)}$ = 1. If a second target is located at a slightly different Doppler frequency ${f_{d}}$ and delay offset ${\tau}$, then it is not resolvable at locations in the delay-Doppler plane that place the peak value of ${\Psi(\tau, f_{d})}$ within the mainlobe of the reference target at ${\Psi(0,0)}$. 

\begin{figure}
\centering
\includegraphics[width=0.50\columnwidth]{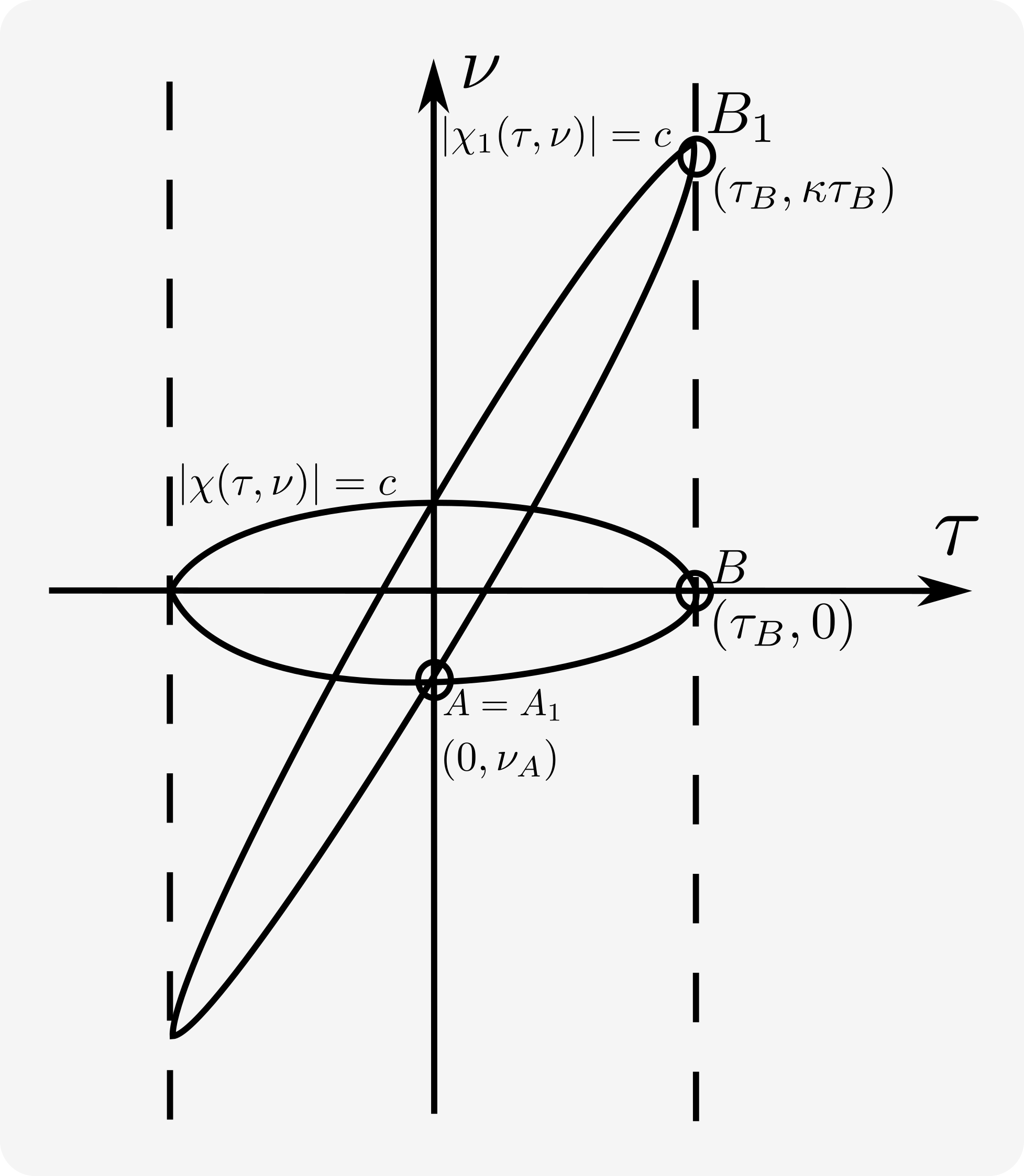}
\caption[example]{\label{fig:LFM_ambiguity} 
Ambiguity diagrams of LFM waveform and single unmodulated CW pulse \cite{Levanon2004}}.
\end{figure} 
In Fig. \ref{fig:LFM_ambiguity}, a constant-value contour of the ambiguity function versus delay and Doppler, denoted as ${|\chi_{1}(\tau,\nu)| = c}$, for an LFM up-chirp is shown as the slanted ridge.  Also shown in the plot as a horizontal ellipsoid is the ambiguity function for a single pulse of a continuous-wave (CW) sinusoid with the same duration as the chirp.  The diagram shows the improved delay resolution of the chirp waveform since the LFM ridge has a narrower width.  The slope of the LFM ridge reveals that delay and Doppler measurements will be coupled for a frequency modulated waveform.  In other words, a received waveform from a stationary target at the range ${R_{1} = c(t_{0}+\Delta{t})/2}$, where ${t_{0}}$ corresponds to the origin in the diagram, ${c}$ is equal to the speed of light and ${\Delta{t}< T}$, will yield the same matched filter output as a moving target at the range ${R=ct_{0}/2}$ with Doppler frequency ${\nu = Bt_{0}/\Delta{t}}$.   

\begin{figure}
%\centering
\includegraphics[width=1.0\columnwidth]{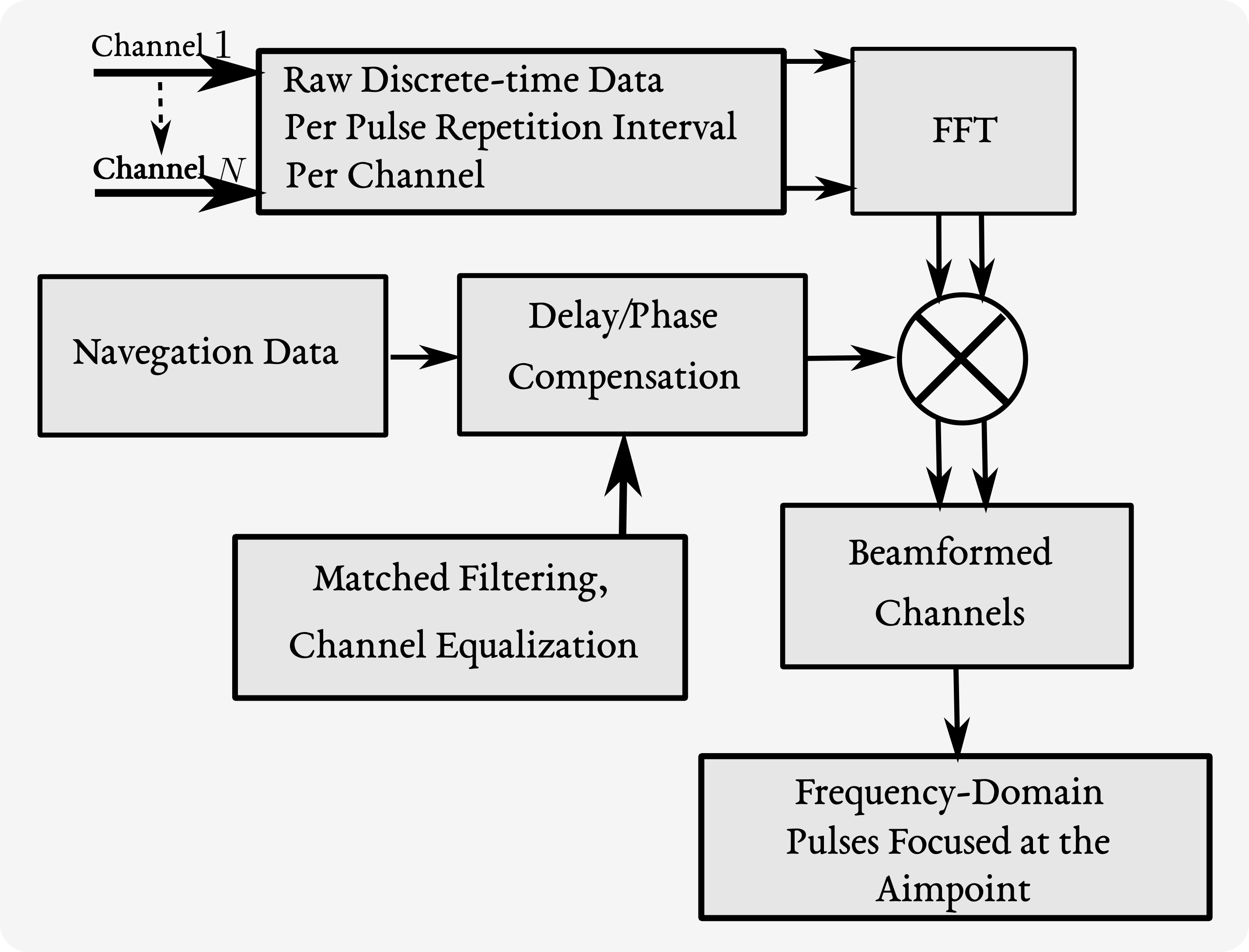}
\caption[example]{\label{fig:SAR_block} 
Notional block diagram of SAR receive chain. \cite{ODonnell2013}}.
\end{figure}
\begin{figure}
%\centering
\includegraphics[width=1.0\columnwidth]{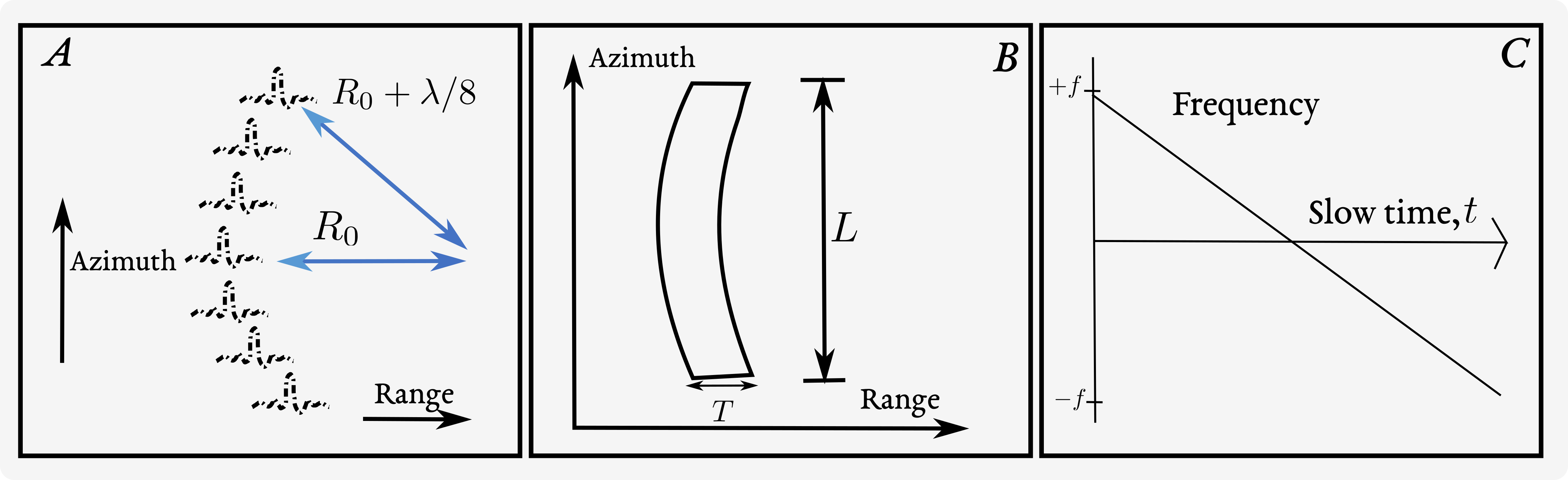}
\caption[example]{\label{fig:SAR_history} 
History of received matched filter outputs along aircraft's flight path. \cite{Rahman2010}}.
\end{figure}

 
Fig. \ref{fig:SAR_block} illustrates the signal processing functions performed on receive for a stripmap SAR.  Over the course of the aircraft's flight the history of matched filter outputs shown in Fig. \ref{fig:SAR_history}A will be received from a single stationary point target in the scene.  As the aircraft flies past the target, the peaks corresponding to the target's azimuth and range will fall along a hyperbolic curve.  Fig. \ref{fig:SAR_history}B illustrates the effective length ${L}$ of the synthetic aperture and the range migration of the target over the interval ${T}$.  The Doppler shift due to the aircraft's motion is shown in Fig. \ref{fig:SAR_history}C.  Initially the plot shows a a positive Doppler shift as the aircraft approaches the target, zero Doppler shift when the aircraft is directly in front of the target, and negative Doppler shift as the aircraft recedes from the target.  The slow-time axis effectively refers to pulse number.  

\begin{figure}[t]
%\centering
\includegraphics[width=1.0\columnwidth]{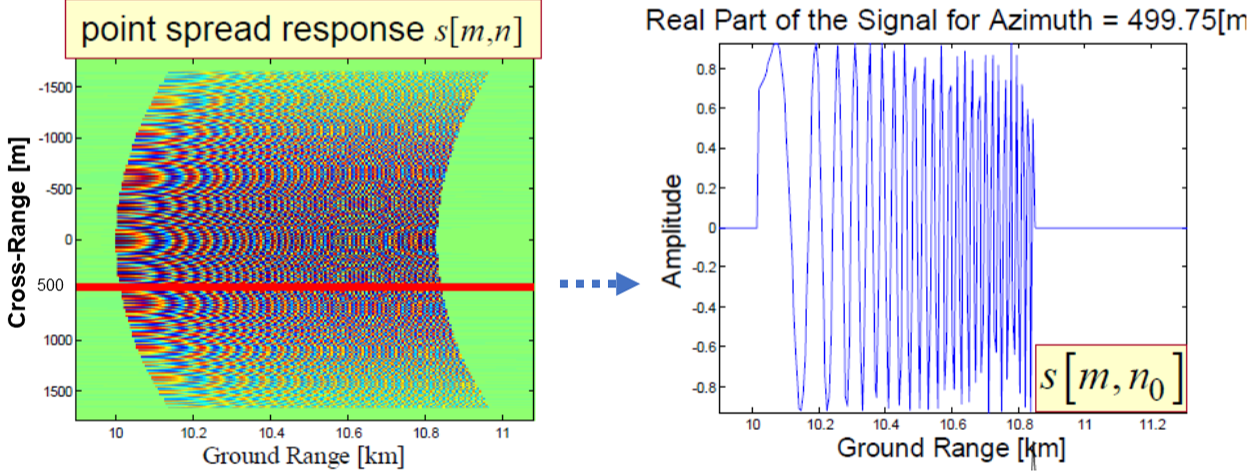}
\caption[example]{\label{fig:horz_slice} 
Horizontal range slice through SAR point spread function. \cite{Richards2010}}.
\end{figure}

\begin{figure}[t]
%\centering
\includegraphics[width=1.0\columnwidth]{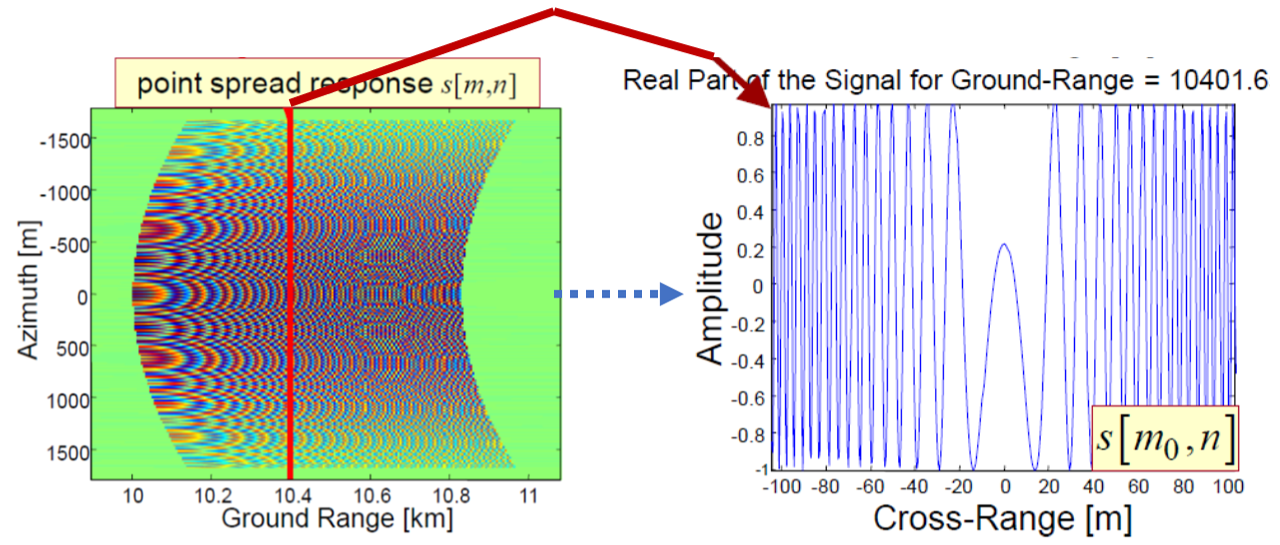}
\caption[example]{\label{fig:vert_slice} 
Vertical cross-range slice through SAR point spread function. \cite{Richards2010}}.
\end{figure}

\begin{figure}[t]
%\centering
\includegraphics[width=1.0\columnwidth]{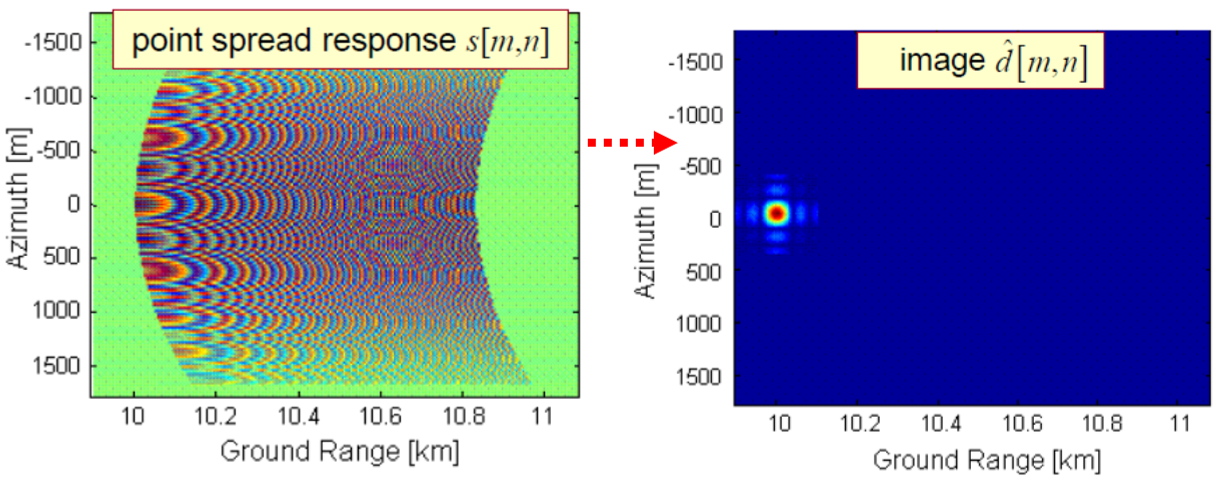}
\caption[example]{\label{fig:focus_image} 
Successful focusing operation results in a single correctly located point corresponding to the target's location.
\cite{Richards2010}}.
\end{figure}

Fig. \ref{fig:horz_slice} illustrates a range cut through the point spread function (PSF) of the stripmap SAR and Fig. \ref{fig:vert_slice} illustrates a cross-range (azimuth) cut through the PSF.  This image represents raw data before matched filtering or pulse compression.  The range and cross-range cuts show that the PSF of the stripmap SAR is essentially a 2-D chirp, curved in range.  The curvature and chirp rates are predictable parameters based on the scene geometry.  The goal of SAR processing is to focus the energy from a single point target smeared across range (often called fast time and corresponding to ADC samples or range bins) and cross-range (also known as slow time and corresponding to pulses).  An example of a focused image is shown in Fig. \ref{fig:focus_image}.  Focusing algorithms essentially implement a matched filter but there are many variations in the exact approach and the resulting accuracy.

\begin{figure}[t]
%\centering
\includegraphics[width=1.0\columnwidth]{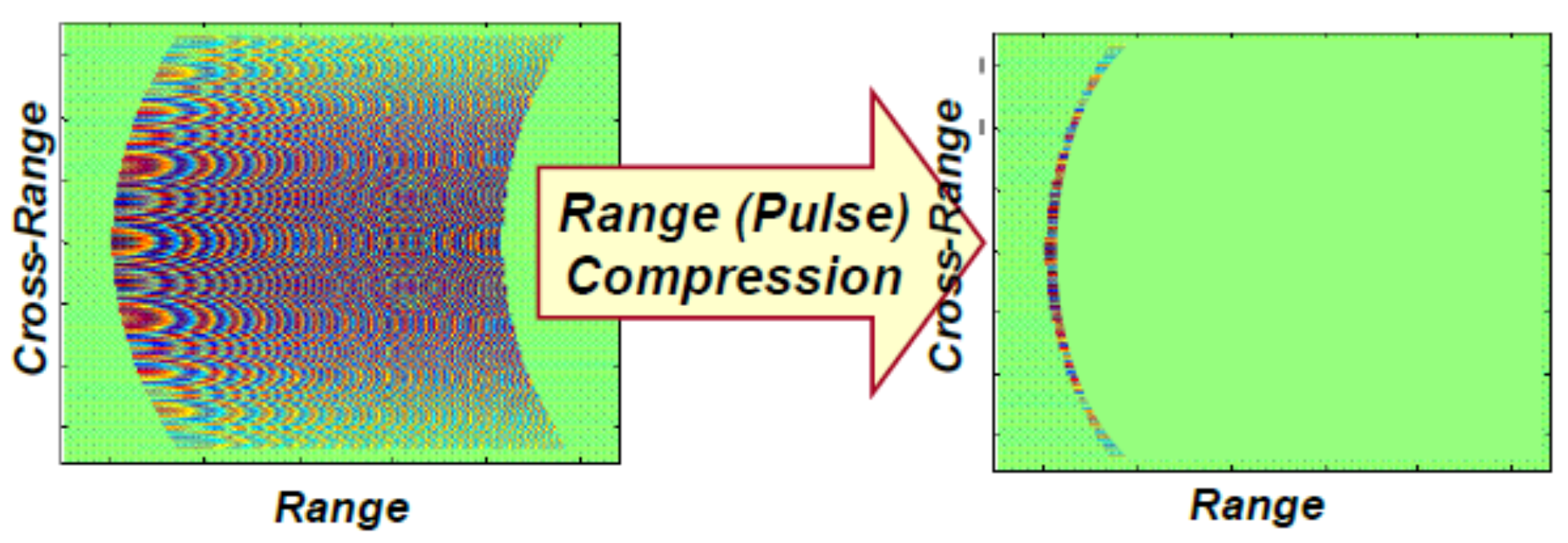}
\caption[example]{\label{fig:backproj_step1} 
Matched filtering of raw SAR data along the range dimension yields a hyperbolic curve.
\cite{Richards2010}}.
\end{figure}

One time-domain algorithm for image formation is called backprojection.  The first step performs pulse compression along the range dimension to yield the result in Fig. \ref{fig:backproj_step1}.
The remaining steps in the backprojection algorithm are summarized in Algorithm \ref{alg:backproj}.
\begin{algorithm}%[H]
\caption{SAR image formation using backprojection}\label{alg:backproj}
\begin{algorithmic}[1]
\Statex \textbf{Input:} Raw discrete-time data output of the radar  ${c[m,n]}$, where the indices ${m}$ and $n$ refer to cross-range and range
\Statex \textbf{Output:} Final image ${b[m_j,n_j]}$ with pixel index ${[m_j,n_j]}$
\State Matched-filter ${c[m,n]}$ along the range dimension to obtain ${x[m,n]}$.
\For{ $\forall m_j$, $n_j$}
\State Compute the PSF ${s[m,n;m_j,n_j]}$ over all indices ${[m,n]}$ for a target located at ${[m_{j},n_{j}]}$
\State ${b[m_j,n_j] \gets \sum\limits_{m=0}^{M-1}\sum\limits_{n=0}^{N-1}s^{*}[m,n;m_j,n_j]x[m,n]}$
\EndFor
\State \Return $b[m_j,n_j]$.
\end{algorithmic}
\end{algorithm}

If no attempt is made to compensate for the phases of the signals received along the synthetic aperture before coherently integrating them, then the effective length of the aperture will be limited for a given target range ${R_0}$.  This case is referred to as an unfocused synthetic aperture.  Then the maximum value of ${L}$ will be equal to the aperture length such that the round-trip distance between the target and the center of the synthetic aperture differs by ${{\lambda}/4}$ from the round-trip distance between the target and the edge of the aperture.  The maximum value of ${L}$ for an unfocused aperture is~\cite{Skolnik2008}
\begin{equation}
    L_{max} = \sqrt{R_0{\lambda}}.
\end{equation}
In the focused case, all the signals received along the aperture from a target at ${R_0}$ are compensated such that they have equal phase before combining them.  The new maximum aperture length is determined by the linear width of the radiated beam at the range ${R_0}$,
\begin{equation}
    L_{max} = \frac{\lambda{R_0}}{D},
\end{equation}
where ${D}$ is the horizontal length of the physical antenna~\cite{Skolnik2008}.

The backprojection algorithm is essentially a time domain implementation of 2-D matched filtering.  However, as shown in Fig. \ref{fig:SAR_block}, it is common to process SAR data in the frequency domain after performing an FFT.  For example, the matched filtering operation can be implemented as ${\text{IFFT}[S(f)H(f)]}$.  The conventional frequency-domain image formation process for stripmap SAR is a beamformer with sidelobe control \cite{Benitz1997}.  The range to the target creates a linear phase ramp of a certain slope versus frequency.  The azimuth (crossrange) location of the target yields a linear phase ramp across the spatial aperture.  After focusing to compensate for the effects of platform motion, signal returns from the desired pixel at location ${(x,y)}$ add coherently (i.e. in phase), while returns from other pixels add incoherently.

For a scatterer at position ${(x,y)}$ the radar measures reflection coefficients over a band of frequencies corresponding to different antenna positions and viewing angles along the aircraft's flight path.  Assume this data is arranged in a matrix ${\mathbf{Z}}$ as in,
\begin{equation}
    \mathbf{Z} = \left[ \begin{array} {ccccc} z_{11} & z_{12} & z_{13} & \ldots & z_{1N} \\
    z_{21} & z_{22} & z_{23} & \ldots & z_{2N} \\
    \vdots  \\
    z_{M1} & z_{M2} & z_{M1} & \ldots & z_{MN} \end{array} \right]
\end{equation}
where the columns correspond to ${N}$ frequencies and the rows correspond to ${M}$ antenna positions along the length of the synthetic aperture.  An optimized ${M}$-by-${N}$ matrix ${\mathbf{W}}$ of weighting coefficients can be computed and applied to the data matrix ${\mathbf{Z}}$ as in,
\begin{equation}
    \mathbf{Y} = \mathbf{Z}\odot\mathbf{W}
\end{equation}
where ${\odot}$ represents the Hadamard product.  The magnitude squared of the sum of the elements in ${\mathbf{Y}}$ represents an estimate of the radar cross section at the location ${(x,y)}$,
\begin{equation}
    \widehat{\sigma}^2(x,y) = {\Bigg\vert}\sum\limits_{m=0}^{M-1}\sum\limits_{n=0}^{N-1}{\left[\mathbf{Y}\right]}_{mn}{\Bigg\vert}^{2}.
\end{equation}

A central assumption in SAR image formation is that the data consists of a superposition of discrete point scatterers in white noise.  The ideal point scatterer has no variation in reflectivity with respect to aspect angle or frequency.  The PSF that results from an ideal point scatterer is a constant amplitude with a phase versus frequency characteristic determined solely by the distance to the object.  This phase delay is ${2\times2{\pi}Rf/c}$ radians where ${R}$ is the distance to the object and ${f}$ is frequency.  The extra factor of 2 accounts for the round trip distance of the radar signal.

Adaptive beamforming techniques compute a unique set of optimized weights that are applied to the SAR data to create each output pixel in the image.  A frequent prerequisite to computing adaptive beamforming weights is forming a sample covariance matrix.  Since the SAR airborne platform typically collects only one look of the scene data, it is common to partition the length of the synthetic aperture into smaller segments, or the frequency extent of the data into sub-bands, so as to build a full-rank covariance matrix.  However, the partitioning process incurs a penalty in terms of resolution.  To mitigate the loss in resolution, the partitioned data subsets may be significantly overlapped.  Although the overlapped data segments are no longer statistically independent, they can still support the formation of a sample covariance matrix.

Another key component of adaptive beamforming algorithms is the steering vector which represents a model of the system's response to a scatterer.  The simplest steering vector corresponds to a point signal source.  In this case, the assumed amplitude of the scatterer is constant with respect to frequency and aspect angle, and the phase characteristic is linear corresponding to a delay.  Each pixel in the output image will have its own associated steering vector.  Steering vectors that correspond to non-ideal scatterers will have an amplitude profile applied to the frequency or azimuth dimension.

The Capon beamforming algorithm is well known because it provides a maximum likelihood estimate of the power spectrum.  The power being estimated in this case is the radar cross section (RCS), ${\sigma^2(x,y)}$, of the scatterer at each pixel.  Capon's beamformer maintains unity gain in the direction of the object at location ${(x,y)}$ and minimizes the energy detected from other interfering signal sources.  The objective function for Capon's beamformer is
\begin{align}
    \sigma^2(x,y) = & \displaystyle \min_{\mathbf{w}} \mathbf{w}^{H}\widehat{\mathbf{R}}\mathbf{w} \\
    \text{such that } & \mathbf{w}^{H}\mathbf{v} = 1,
\end{align}
where ${\mathbf{v}(x,y)}$ is the unit-norm steering vector corresponding to a point scatterer at location ${(x,y)}$ and ${\widehat{\mathbf{R}}}$ is the sample covariance matrix.  With a full-rank covariance matrix the solution to Capon's beamformer is
\begin{equation}
    \sigma^2(x,y) = \frac{1}{\mathbf{v}^{H}{\widehat{\mathbf{R}}^{-1}\mathbf{v}}}.
\end{equation}
To account for the case of SAR data with a rank deficient sample covariance matrix, a norm constraint is placed on ${\mathbf{w}}$ as in,
\begin{equation}
\label{eqn:normcons}
    \displaystyle \min_{\mathbf{w}} \mathbf{w}^{H}\widehat{\mathbf{R}}\mathbf{w} + {\alpha}\mathbf{w}^{H}\mathbf{w}.
\end{equation}
The optimum ${\mathbf{w}}$ for (\ref{eqn:normcons}) adds the scalar ${\alpha>0}$ to the diagonal elements of ${\widehat{\mathbf{R}}}$,
\begin{equation}
    \mathbf{w} \propto (\widehat{\mathbf{R}} + \alpha{\mathbf{I}})^{-1}\mathbf{v}.
\end{equation}

\subsection{Spotlight SAR and Tomographic Reconstruction}
\label{spotSAR}
The projection slice theorem described in this section plays an important role in many synthetic aperture applications, including spotlight SAR, computer-aided tomography (CAT), and geophysical imaging.  With CAT scans a collimated beam of x-rays illuminates the human body from an angle.  The measured result is essentially the shadow of 3-D anatomical structures.  In this case, a 3-D object is projected onto a 2-D image.  We continue this section with a description of CAT and then show the connection to spotlight SAR.

\begin{figure}
%\centering
\includegraphics[width=1.0\columnwidth]{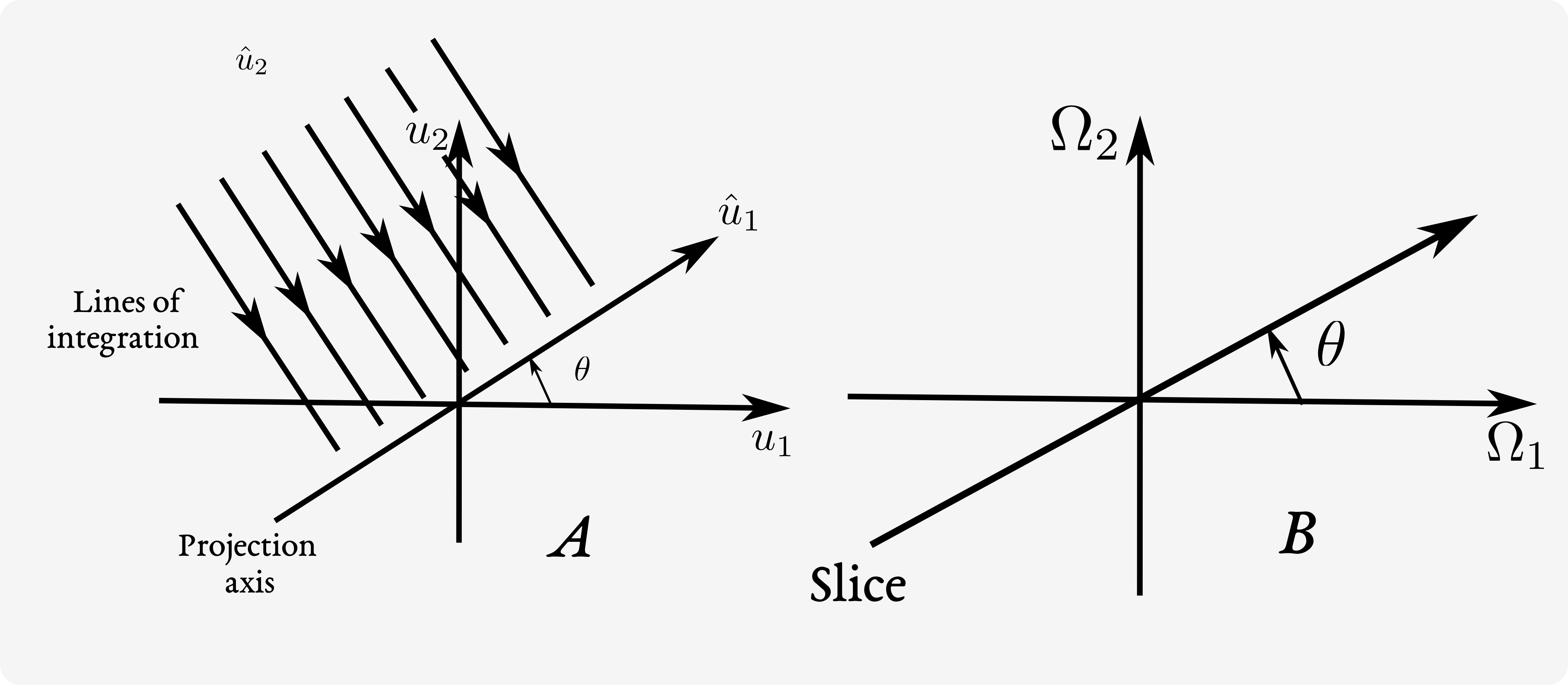}
\caption[example]{\label{fig:project} 
A) Projection of a 2-D function B) Slice of its Fourier transform \cite{Dudgeon84}}
\end{figure}

Assume the object to be imaged is located at the origin of the ${(u_1,u_2)}$ axes shown in Fig. \ref{fig:project}A.  Denote the unknown density function of the object being radiated as ${g(u_1,u_2)}$ and assume that the x-rays propagate in a direction that is normal to the ${\widehat{u}_1}$ axis which is rotated by an angle ${\theta}$ with respect to the ${u_1}$ axis.  The rotated ${(\widehat{u}_1,\widehat{u}_2)}$ coordinate system is described by
\begin{align}
    \widehat{u}_1 &= u_1 \cos{\theta} + u_2 \sin{\theta} \\
    \widehat{u}_2 &= -u_1 \sin{\theta} + u_2 \cos{\theta}.
\end{align}
The propagating radiation is parallel to the ${\widehat{u}_2}$ axis and normal to the ${\widehat{u}_1}$ axis.

The Fourier transform of ${g}$ is defined by 
\begin{equation}
    G(\Omega_1,\Omega_2) = \displaystyle\int_{-\infty}^{\infty}\displaystyle\int_{-\infty}^{\infty}g(u_1,u_2)e^{-\textrm{j}(u_{1}\Omega_1 + u_{2}\Omega_2)}du_1{du_2}
\end{equation}
and the inverse FT is 
\begin{equation}
\label{eqn:IFT}
    g(u_1,u_2) = \frac{1}{4{\pi}^2}\displaystyle\int_{-\infty}^{\infty}\displaystyle\int_{-\infty}^{\infty}G(\Omega_1,\Omega_2)e^{\textrm{j}(u_{1}\Omega_1 + u_{2}\Omega_2)}d{\Omega}_1{d{\Omega}_2}.
\end{equation}
The projection of ${g}$ at an angle ${\theta}$ is given by
\begin{equation}
    p_{\theta}(\widehat{u}_{1}) = \displaystyle\int_{-\infty}^{\infty}g(\widehat{u}_1{\cos}\theta - \widehat{u}_2{\sin}\theta, \widehat{u}_{1}{\sin}{\theta} + \widehat{u}_2{\cos}{\theta})d{\widehat{u}_2}.
\end{equation}
The projection ${p_{\theta}(\widehat{u}_{1})}$ evaluated at ${\widehat{u}_{1}=\widehat{u}_{1k}}$ is a line integral through the point ${\widehat{u}_{1k}}$ in the direction of the ${\widehat{u}_{2}}$ axis.  The function ${p_{\theta}(\widehat{u}_{1})}$ therefore represents a series of such line integrals for each value of ${\theta}$ \cite{Munson83}.

The projection-slice theorem states that
\begin{align}
    P_{\theta}(\widehat{\Omega}_{1}) &= \displaystyle\int_{-\infty}^{\infty}p_{\theta}(\widehat{u}_{1})e^{-{\textrm{j}}\widehat{u}_{1}{\widehat{\Omega}}_{1}}d{\widehat{u}}_{1} \\
    &=G(\widehat{\Omega}_{1}{\cos}{\theta},\widehat{\Omega}_{1}{\sin{\theta}}).
\end{align}
In other words, the 1-D FT of ${p_{\theta}(\widehat{u}_{1})}$ is a slice of the 2-D FT ${G(\Omega_{1},\Omega_{2})}$ taken at an angle ${\theta}$ with respect to the ${\Omega_{1}}$ axis.  The reconstruction problem is to invert a finite number of projections ${p_{\theta{j}}(\widehat{u}_{1})}$ for different values of ${\theta_{j}}$ to yield an estimate of ${g(u_{1},u_{2})}$.  In the context of CAT, projections of ${g}$ are defined as
\begin{equation}
    p_{\theta}(\widehat{u}_{1}) = -\log{\frac{I_{\theta}(\widehat{u}_{1})}{I_{0}}}
\end{equation}
where ${I_{0}}$ is the intensity of the x-ray source and ${I_{\theta}(\widehat{u}_{1})}$ is the measured intensity at the detector.  Projections ${p_{\theta{j}}(\widehat{u}_{1})}$ are obtained at equally spaced discrete angles ${\theta_{j}}$ by rotating the object or the array of x-ray sources and detectors.

A spotlight SAR illuminates a small patch of terrain  with a narrow beam as shown in Fig. \ref{fig:SAR_scans}.  The radar beam is continuously pointed at the ground patch as the aircraft flies.  At intervals corresponding to equal increments of the aspect angle ${\theta}$, high-bandwidth pulses, such as LFM chirps, are transmitted and echoes from the ground patch are recorded.  The radar return essentially yields a band-pass filtered projection of the ground-patch reflectivity function ${g(u_1,u_2)}$, assuming that the phasefront of the propagating signal does not exhibit significant curvature over the extent of the patch \cite{Munson83}. 

\begin{figure}
%\centering
\includegraphics[width=1.0\columnwidth]{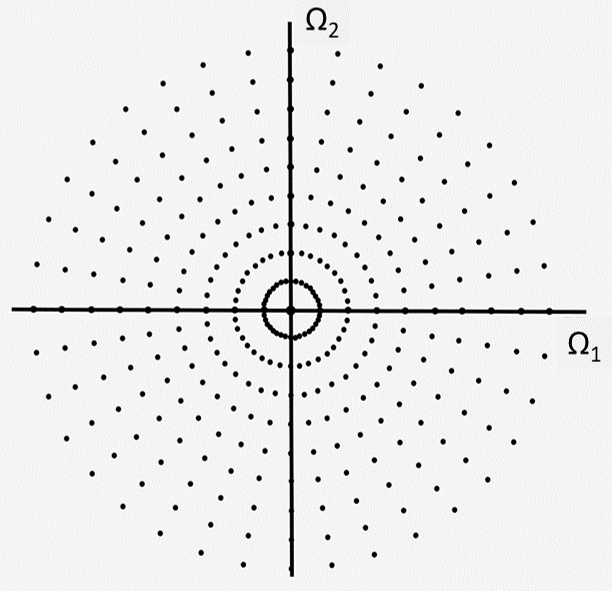}
\caption[example]{\label{fig:polar_grid} 
Spotlight SAR measures samples of the target reflectivity function ${G(\Omega_{1},\Omega_{2})}$ on a polar grid in the frequency domain  \cite{Munson83}}
\end{figure}

Consider the case where samples of the projections ${p_{\theta{j}}(\widehat{u}_{1k})}$ of the unknown target reflectivity function ${g(u_1,u_2)}$ are measured for discrete aspect angles in the interval ${0 \leq \theta_{j} < \pi}$.  For each angle ${\theta_{j}}$ the FFT is used to compute uniformly spaced samples of ${P_{\theta{j}}(\widehat{\Omega}_{1k})}$ from ${p_{\theta{j}}(\widehat{u}_{1k})}$ with ${k = 0,\ldots,K}$.  By the projection-slice theorem, the samples of ${P_{\theta{j}}(\widehat{\Omega}_{1k})}$ are samples of ${G(\Omega_{1},\Omega_{2})}$ along a line at an angle ${\theta}$ with respect to the ${\Omega_{1}}$ axis.  The collection of 1-D FFTs for various ${\theta_{j}}$ provides samples of ${G(\Omega_{1},\Omega_{2})}$ along the polar grid shown in Fig. \ref{fig:polar_grid}.  Interpolating the polar samples onto a Cartesian grid allows a 2-D IFFT to be used to efficiently compute ${g(u_1,u_2)}$.

\begin{figure}
%\centering
\includegraphics[width=1.0\columnwidth]{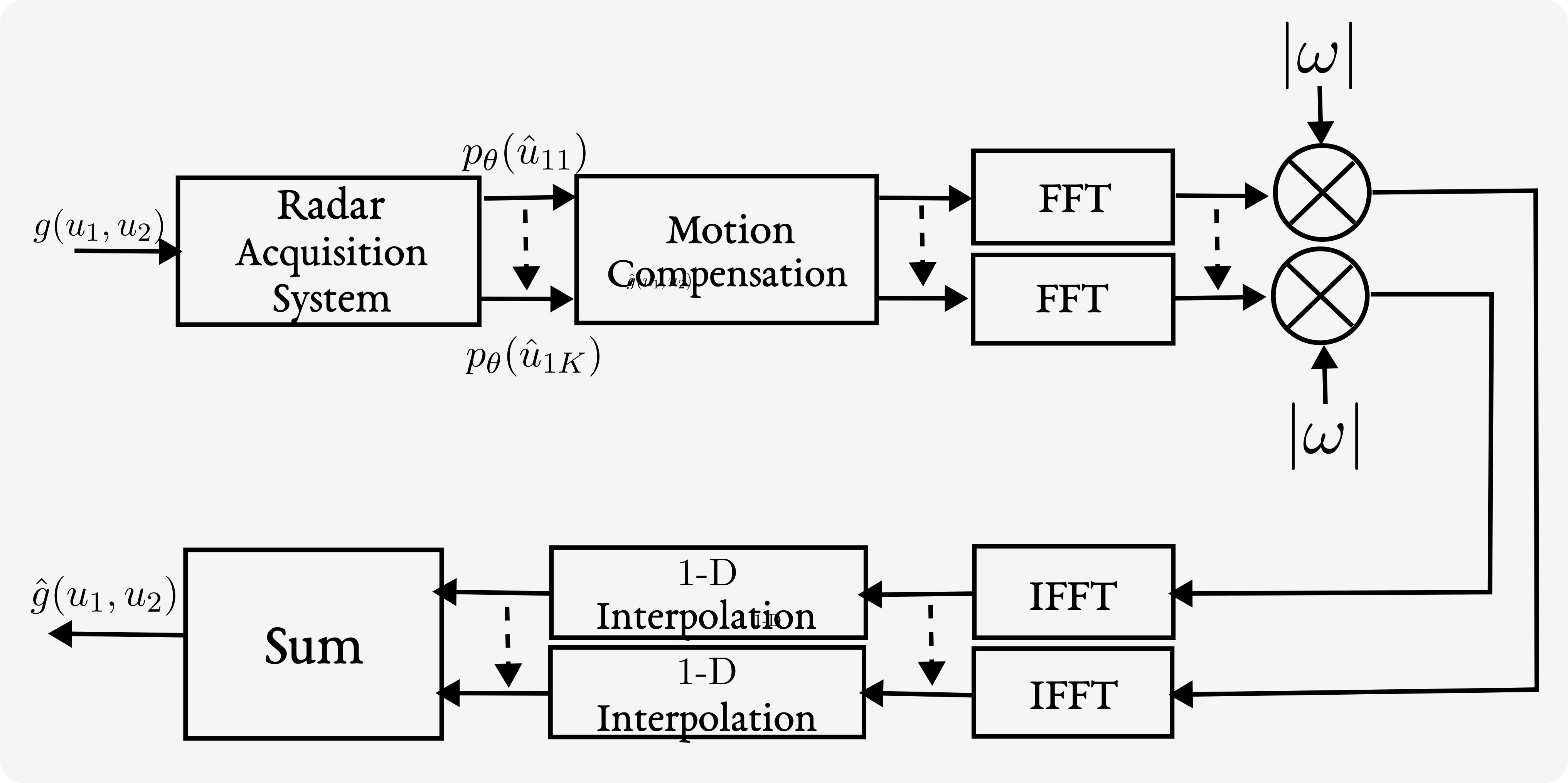}
\caption[example]{\label{fig:back_proj} 
Frequency-domain implementation of back-projection algorithm \cite{Martorella2010}.}
\end{figure}

In the continuous case, the 2-D Fourier plane can be rewritten using the polar coordinates ${(\omega,\theta)}$. Then the 2-D inverse FT given in (\ref{eqn:IFT}) becomes
\begin{align}
\label{eqn:polarIFT}
    &g(u_1,u_2) = \\  \nonumber
    &\frac{1}{4{\pi}^2}\displaystyle\int_{0}^{\pi}\displaystyle\int_{-\infty}^{\infty} G(\omega{\cos\theta},\omega{\sin \theta}) e^{\textrm{j}\omega(u_{1}{\cos\theta} + u_{2}{\sin\theta})}\vert \omega \vert d{\omega}d{\theta} \\  \nonumber
    &=\frac{1}{4{\pi}^2}\displaystyle\int_{0}^{\pi}\displaystyle\int_{-\infty}^{\infty} P_{\theta}(\omega) e^{\textrm{j}\omega(u_{1}{\cos\theta} + u_{2}{\sin\theta})}\vert \omega \vert d{\omega}d{\theta}.
\end{align}
The inner integral represents the 1-D inverse FT of the product of ${P_{\theta}(\omega)}$ and ${\vert{\omega}\vert}$.  Multiplication by ${\vert{\omega}\vert}$ corresponds to taking the derivative of the Hilbert transform of ${p_{\theta}(u_{1}\cos\theta + u_{2}\sin\theta)}$ \cite{Dudgeon84}.  Recall the Hilbert transform ${H[s(t)]}$ of a signal ${s(t)}$ is defined as
\begin{equation}
    H[s(t)] = s(t) \ast \frac{1}{\pi{t}} = \frac{1}{\pi}\displaystyle\int_{-\infty}^{\infty}\frac{g(\tau)}{t-\tau}d{\tau}
\end{equation}
which represents the convolution of ${s(t)}$ with a linear time-invariant filter that has impulse response ${1/{\pi}t}$.  Thus (\ref{eqn:polarIFT}) can be rewritten as
\begin{equation}
    g(u_1,u_2) = \frac{1}{2\pi}\displaystyle\int_{0}^{\pi}q_{\theta}(u_{1}\cos\theta + u_{2}\sin\theta)d\theta
\end{equation}
where
\begin{align}
    q_{\theta}(t) &\triangleq \frac{d}{dt}\displaystyle\int_{-\infty}^{\infty} \frac{p_{\theta}(\tau)}{t-\tau}d{\tau} \\ \nonumber
    &\triangleq p_{\theta}(t) \ast k(t).
\end{align}
The generalized function ${k(t)}$ is known as the Radon kernel and is the inverse FT of ${\vert{\omega}\vert}$.  The steps outlined above describe the frequency domain implementation of the back-projection algorithm.  A block diagram summarizing the functional steps is provided in Fig. \ref{fig:back_proj}.

%\subsection{SAR Focusing using Frequency-Wavenumber Techniques from Seismic Inverse Problems}
\subsection{Focusing via Seismic-Imaging-Based Frequency-Wavenumber Techniques}
As shown in Fig. \ref{fig:SAR_history}, when the SAR antenna flies past a scatterer on the ground, the signal peaks after matched filtering fall along a hyperbola due to the varying path length from the radar to the object.  A similar phenomenon occurs for seismic imaging, ground penetrating radar, and synthetic aperture sonar.  To create the desired high-resolution image, the synthetic aperture system must migrate the dispersed signal peaks to a single pixel that corresponds to an object's true location and reflectivity.  This task is also known as focusing.

In the case of SAR, a radiated pulse propagates to the ground and is reflected back towards the antenna.  The radar must then determine the location of targets and scatterers.  A similar inverse problem in seismic applications is to determine the location of signal sources after measuring an acoustic wavefield along different locations on the earth's surface.  It is possible to treat the SAR focusing problem using the same framework as the seismic inverse problem if one considers the received electromagnetic signal to originate from exploding scatterers on the ground before it travels towards the radar~\cite{Cafforio91}.  In this case, there is only one-way propagation from the signal source to the antenna so to maintain the same round-trip delay the speed of the traveling electromagnetic wave should be considered to be ${c/2}$, where ${c}$ is the speed of light.  The exploding sources assumption avoids additional complexity due to the radar radiating a pulse from one location along its trajectory and then receiving the pulse a short time and distance later.

Once SAR data has been downconverted to baseband and digitized, it can be arranged into a two-dimensional array.  The columns contain the discrete-time samples for successive pulse repetition intervals (PRIs) so that the column indices correspond to the along-track dimension and the row indices represent slant range.  Consider a signal source located at a slant range ${z_{0}}$ from the antenna.  Assume the source radiates at the time ${t=0}$.  One approach to focus the signal received by the radar from a slant range ${z_{0}}$ is to downward continue the wavefield along the ${z}$ axis to determine its source amplitude at time ${t=0}$ at the source location.  This operation is implemented in the frequency-wavenumber ${(\omega,k)}$ domain by decomposing the received signal into a sum of plane waves, downward continuing each plane wave separately, and then recombining the plane waves at the proper time~\cite{Cafforio91,Ozdemir2014}.  The concept is briefly summarized in the next paragraphs.   
\begin{figure}
%\centering
\includegraphics[width=1.0\columnwidth]{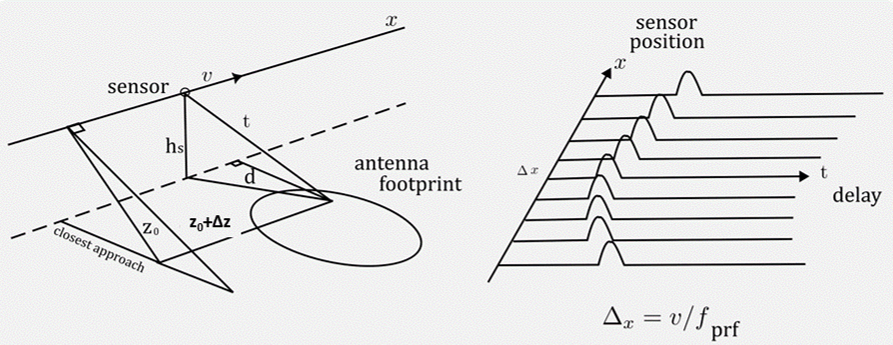}
\caption[example]{\label{fig:stolt_geometry} 
Geometry for downward continuing received SAR pulses shown on left.  The distance traveled by the radar between pulses is ${\Delta{x} = v/f_{PRF}}$ where ${f_{PRF}}$ is the pulse repetition frequency (PRF) and ${v}$ is the speed of the aircraft or satellite in the ${x}$ direction~\cite{Cafforio91}}. 
\end{figure}

Let ${a(x,z=0,t)}$ denote the received data after range compression as shown on the right in Fig. \ref{fig:stolt_geometry}.  The radar is flying along the ${x}$ direction and and ${z}$ is the slant range at the point of closest approach from the aircraft to a target on the ground.  The ${y}$-axis is orthogonal to both the ${x}$ and ${z}$ directions.  The case ${z=0}$ corresponds to the radar flying at ground level directly over the target.  The measured electromagnetic field values at the radar antenna correspond to the superposition of plane waves with different wavelengths and wavenumbers as in
\begin{equation}
    a(x,z=0,t) = \frac{1}{2{\pi}^2}\int{\int}A(k_{x},z=0,{\omega})e^{{\textrm{j}}({\omega}{t} + k_{x}x)}d{\omega}\text{ }dk_{x}
\end{equation}
where the plane wave amplitudes at ${z=0}$ are given by
\begin{equation}
    A(k_{x},z=0,\omega) = \int{\int}a(x,z=0,t)e^{-{\textrm{j}}({\omega}{t} + k_{x}x)}d{x}\text{ }dt.
\end{equation}
Further, if ${\theta}$ is the angle of incidence to the antenna, then the following relations hold
\begin{align}
    k^2 &= k_{x}^2 + k_{y}^2 + k_{z}^2, \\
    k_x &= \frac{2\pi}{\lambda}\sin{\theta}, \\
    k_z &= \frac{2\pi}{\lambda}\cos{\theta},  \\
    \omega &= \frac{c}{2}\sqrt{k_{x}^2 + k_{z}^2}.
\end{align}
For the geometry shown in Fig. \ref{fig:stolt_geometry}, ${k_{y}=0}$.

The phase operator required to back propagate a plane wave from ${z=0}$ to ${z=z_{0}}$ is
\begin{equation}
    K_p = e^{\textrm{j}k_{z}z_{0}} = e^{\textrm{j}z_{0}\sqrt{k^2 - k_{x}^2}} = e^{\textrm{j}kz_{0}\sqrt{1 - (k_{x}/k)^2}}.
\end{equation}
Recall that for an exploding signal source on the ground, the velocity of propagation is taken to be ${c/2}$.  Thus, ${K_p}$ becomes
\begin{equation}
    K_p = e^{\textrm{j}(2{\omega}/c)z_{0}\sqrt{1 - (c^2k_{x}^2/4{\omega}^2)}}.
\end{equation}
The wavefield at a depth ${z}$ can now be expressed as
\begin{equation}
A(k_{x},z,\omega) = A(k_{x},z=0,\omega)e^{\textrm{j}k_{z}z}.
\end{equation}
A map of the sources on the ground can be obtained by propagating the wavefield at ${z}$ backward in time to ${t_{0}}$ by computing
\begin{align}
\label{eqn:focus}
    a(x,z,t=-t_{0}) &= \frac{1}{(2\pi)^2}\int_{-\omega_{c}-\frac{\omega_b}{2}}^{\omega_{c}+\frac{\omega_b}{2}}d{\omega} \int_{k_{x}^{min}}^{k_{x}^{max}} A(k_{x},z=0,\omega) \\ \nonumber 
    &\cdot e^{\textrm{j}(k_{z}z + k_{x}x - {\omega}t_{0})} dk_{x}
\end{align}
where ${\omega_c}$ is the carrier frequency and ${\omega_b}$ is the signal bandwidth.

Equation (\ref{eqn:focus}) describes the focused output of the SAR but must be modified to use a two-dimensional inverse FFT for efficient computation.  This step entails a change of variables in the integral from ${\omega}$ to ${k_{z}}$ using the relations
\begin{align}
\label{eqn:omega-k}
    \omega &= \frac{c}{2}\sqrt{k_{x}^2 + k_{z}^2},  \\ \nonumber
    d{\omega} &= \frac{c^2k_{x}}{4\omega} dk_{x}.
\end{align}
The transformed version ${A(k_{x},z=0,(c/2)\sqrt{k_{x}^2 + k_{z}^2})}$ of the original frequency domain data ${A(k_{x},z=0,\omega)}$ will not lie on a uniform grid however due to the nonlinearity of the ${\omega}$ to ${k_{z}}$ relation in (\ref{eqn:omega-k}).  Therefore, the data must be interpolated before using the two-dimensional inverse FFT.  This procedure is known as Stolt interpolation.

\subsection{Chirp Scaling}% for SAR}
Wideband SAR systems operate over a large absolute or fractional bandwidth and have a wide antenna beamwidth.  Wideband SAR can yield high-resolution images for applications like mine-field detection and at low frequencies (20-90 MHz) it can detect changes in dense forested or camouflaged areas~\cite{Ulander05,Carin99}.  SAR processing may correct for the range cell migration (RCM) of scatterers via a time-domain interpolation procedure before azimuth compression is performed.  Since the RCM correction terms are range dependent, they must be updated for every range bin~\cite{Moreira96}.

The two-dimensional signal received by a SAR is the sum of back-scattered reflections from individual scatterers in the scene.  The azimuth coordinate of an object in the scene is along the direction of the radar's velocity vector, and an object's range is the orthogonal distance from the velocity vector.  Signal coordinates are similar except that range is defined as the instantaneous distance from the antenna to the object.  In general, range is not perpendicular to azimuth in signal space.  The SAR processor must filter the data to focus it in both the range and azimuth dimensions.

The baseband signal received from an isolated point scatterer at range ${r}$ and azimuth time ${t = 0}$ is described in signal space as~\cite{Raney94}
\begin{equation}
    \label{E:pp}
    y(\tau, t; r) = a(t,r) s \left( \tau  - \frac{2R(t;r)}{c} \right) e^{-\textrm{j}(4\pi/{\lambda})R(t;r)}.
\end{equation}
Here ${s(\cdot)}$ is the envelope of the transmit signal, ${a(t,r)}$ is the azimuth antenna gain (volts), ${t}$ denotes time along the radar's trajectory, ${\tau}$ is delay in the slant range direction, ${c}$ is the speed of light, and ${\lambda}$ is the transmitted wavelength.  The straight-line distance from the radar to the scatterer measured as a function of time with respect to the point of closest approach is
\begin{equation}
\label{E:eqnR}
    R(t;r) = \sqrt{r^2 + V^2(r)t^2}.
\end{equation}
For SAR deployed on an aircraft, the relative velocity term ${V(r)}$ is not dependent on range and approximately equal to the platform velocity.  For spaceborne applications, ${V(r)}$ will depend on range.

As the aircraft approaches and then flies past a fixed scatterer, the signal peak will migrate across different range resolution cells.  Range cell migration can be corrected by straightening the curved signal space or by equalizing the range term in (\ref{E:eqnR}).  If there are several scatterers at different azimuth locations in the scene, then it is impossible to correct all range cell migrations via a simple scaling procedure in the signal domain.  Once range cell migration correction (RCMC) has been applied, then compression along the azimuth axis is a simple one-dimensional operation.

In principle, a two-dimensional linear filter can be applied in the time domain to simultaneously correct range cell migration and perform compression, although new filter coefficients would be required for every range cell.  A more efficient implementation of RCMC is possible in the frequency domain.  The two-dimensional Fourier transform of the signal impulse response has the form,
\begin{equation}
    Y(f_{\tau},f;r) \propto A(\cdot) S(\cdot) e^{\textrm{j}\phi(f_{\tau},f;r)}
\end{equation}
where ${f_{\tau}}$ and ${f}$ denote the frequency variables corresponding to range delay ${\tau}$ and azimuth time ${t}$ respectively.  The functions ${A(\cdot)}$ and ${S(\cdot)}$ are the transforms of the antenna weighting and the pulse envelope.

An analytic expression for the baseband signal impulse response in the case of LFM is given by
\begin{equation}
    y(\tau,t;r) = a(t)s\left(\tau - \frac{2}{c}R(t;r)\right) e^{-\textrm{j}\pi{K}\left(\tau - \frac{2}{c}R(t;r)\right)^2} e^{-\textrm{j}\frac{4\pi}{\lambda}R(t;r)}
\end{equation}
where ${K}$ is the chirp rate.  To within a constant ${C}$ the corresponding transform along azimuth time yields the range-Doppler response as
\begin{align}
\label{E:eqnRD}
    y(\tau,f;r) &= C \cdot a\left(-\frac{rf\lambda}{2V^2} \right) s \left(\tau - \frac{2}{c}R_{f}(f;r) \right) \\ \nonumber
    & \cdot e^{-\textrm{j}\pi{K_{s}}(f;r)\left(\tau - \frac{2}{c}R_{f}(f;r)\right)^2} \\ \nonumber
    & \cdot e^{-\textrm{j} \frac{4\pi{r}}{\lambda} \left[ 1 - \left( \frac{\lambda{f}}{2V(r)} \right)^2 \right]^{0.5} }.
\end{align}
The function ${R_{f}(f;r)}$ in (\ref{E:eqnRD}) yields a family of curves that describe range migration in the range-Doppler domain through a position shift in the signal envelope ${s(\cdot)}$,
\begin{align}
    & R_{f}(f;r) = r \left[ 1 + C_{s}(f) \right],  \\  \nonumber
    & C_{s}(f) = \frac{1}{\sqrt{1 - \left(\frac{\lambda{f}}{2V(r)}\right)^2}} - 1.
\end{align}
The term ${C_{s}(f)}$ is known as the curvature factor and describes the Doppler-dependent portion of the signal trajectory.

The function ${K_{s}(f;r)}$ in (\ref{E:eqnRD}) is the effective chirp rate in range and has the form,
\begin{equation}
\frac{1}{K_{s}(f;r)} = \frac{1}{K} + r{\alpha}(f;r)
\end{equation}
where
\begin{equation}
    \alpha(f;r) = \frac{2\lambda}{c^2}\frac{\left(\frac{\lambda{f}}{2V(r)}\right)^2}{\left[1 - \left(\frac{\lambda{f}}{2V(r)}\right)^2 \right]^{3/2} }
\end{equation}
is the range distortion factor.  This factor is a function of geometry and leads to range defocusing if not compensated.

\begin{figure}
%\centering
\includegraphics[width=1.0\columnwidth]{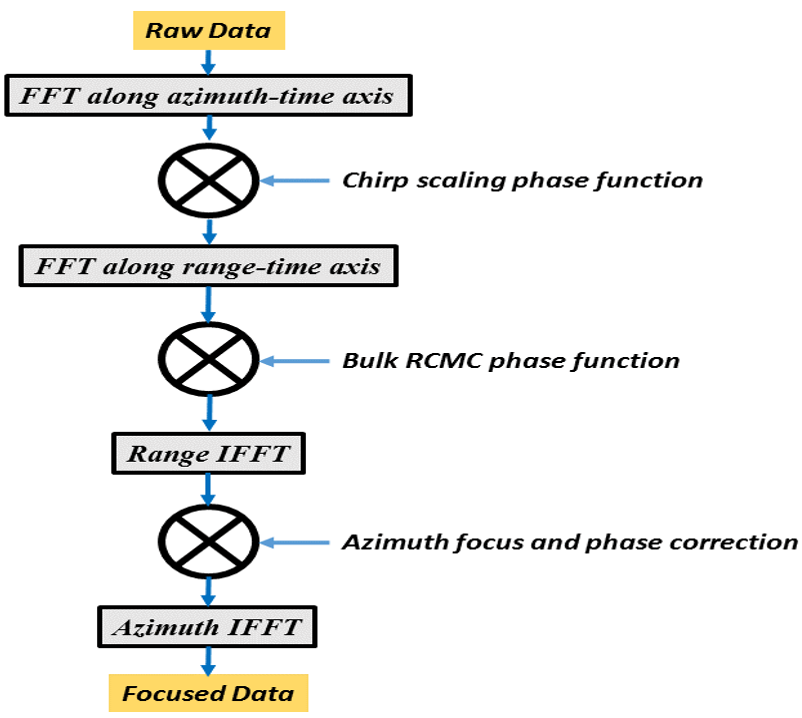}
\caption[example]{\label{fig:chirp_scaling} 
Processing flow of chirp scaling algorithm for focusing SAR images~\cite{Raney94}.}
\end{figure}

The processing flow of the chirp scaling algorithm is summarized in Fig. \ref{fig:chirp_scaling}.  The first step in the algorithm is a Fourier transform of the acquired data along the azimuth-time dimension which yields the result in (\ref{E:eqnRD}).  The second step is multiplication by the chirp scaling phase function, ${\Phi_{1}}$, that equalizes the range migration phase term of every scatterer to that of the reference range,
\begin{align}
    &\Phi_{1}(\tau,f;r_{ref}) = e^{-\textrm{j}{\pi}K_{s}(f;r_{ref})C_{s}(f)[\tau - \tau_{ref}(f)]^2},  \\  \nonumber
    &\tau_{ref}(f) = \frac{2}{c}r_{ref}[1 + C_{s}(f;r)].
\end{align}
Next a forward Fourier transform is computed along the range dimension to yield a two-dimensional frequency domain image.  Multiplication by the two-dimensional phase function, ${\Phi_{2}}$, implements RCMC and range focusing,
\begin{equation}
    \Phi_{2}(f_{\tau},f;r_{ref}) = e^{-\textrm{j}{\pi}\frac{f_{\tau}^{2}}{K_{s}(f;r_{ref})[1 + C_{s}(f)]}} e^{\textrm{j}\frac{4\pi}{c}f_{\tau}r_{ref}C_{s}(f)}.
\end{equation}

After an inverse Fourier transform along the range axis, the focused envelope of the signal appears at the correct delay ${2r/c}$.  In the last step of the algorithm, the remaining phase is matched after multiplying by ${\Phi_{3}}$ to focus the signal in azimuth~\cite{Raney94}, 
\begin{align}
&\Phi_{3}(\tau,f) = e^{-\textrm{j}\frac{2\pi}{\lambda}c{\tau}\left[1 - \left[1 - \left(\frac{\lambda{f}}{2V(r=c{\tau}/2)}\right)^2 \right]^{0.5} \right] + \textrm{j}\Theta_{\Delta}(f;r)}, \\ \nonumber
&\Theta_{\Delta}(f;r) = \frac{4\pi}{c^2}K_{s}(f;r_{ref})\left[1 + C_{s}(f)\right]C_{s}(f)(r-r_{ref})^2.
\end{align}
The final step is an inverse Fourier transform along the azimuth axis which yields the compressed image impulse response ${y_C(\tau,t)}$, for zero azimuth offset as
\begin{equation}
    y_{C}(\tau,t) = C \cdot S_{0}(\tau - 2r/c)A(t)
\end{equation}
where ${S_{0}}$ is the focused signal envelope at the correct delay ${2r/c}$ and ${A(t)}$ is the transformed envelope of the antenna weighting.

\subsection{Emerging Applications}%Extensions to Wideband SAR}
As discussed in \cite{Vu2010}, chirp scaling or the Stolt interpolation procedure for wideband SAR data requires significant computational effort.  Furthermore, chirp scaling performance degrades at high PRFs and large fractional bandwidths.  Several algorithms have been developed to improve RCM correction and autofocusing performance in the wideband case, including \cite{Sheen94,Ulander99}.  Techniques to reduce the computational time are discussed in \cite{Vu2008,Li2014-2}.  Moving targets illuminated by a wideband SAR can be both displaced and defocused in the final image due to the long integration time.  A novel approach described in \cite{Vu2010-2} can be used to detect moving targets in wideband SAR by focusing.

In the following sub-sections, we briefly describe additional popular and recent wideband SAR configurations.
\subsubsection{Wideband Tomographic SAR}
\label{subsubsec:tomsar}
The tomographic SAR follows the principle of computed tomography (CT) in medical imaging through the use of diversity in the scan geometry. The transmitters and receivers are deployed in multiple locations to provide additional angular information about the targets leading to spatial diversity. This is useful for both SAR-based ground-penetrating radar (GPR) as well as space-based SAR or inverse SA radar (ISAR) \cite{gassot2021ultra}. For wideband tomographic SAR \cite{almutiry2019wideband}, the transmitter emits a step-frequency waveform leading to a different wavelength and resolution at each step. This method trades-off the image resolution at each stepped-frequency for the computational cost. Coherent processing of all low-resolution images obtained at each stepped frequency is then used to construct a high-resolution tomographic image of the scene.  A novel approach for tomographic SAR inversion via compressive sensing is described in \cite{Zhu2010}.  The use of deep learning to achieve super-resolution in tomographic SAR is discussed in \cite{Qian2010}.

\subsubsection{Ultrawideband SAR}
Low-frequency ultra-wideband (UWB) SAR has become very popular recently largely because it offers a unique capability of detecting complex hidden objects such as landmines and other explosive hazards.  However, the sizes of the targets-of-interest are relatively small compared to the wavelengths of the radar signals within the operating frequency band. As a result, in the reconstructed SAR image, these targets (even when detected) only show up in a few pixels as point-like targets without any specific structure. Moreover, other manmade and clutter objects of a similar size as the targets-of-interest also result in point-like responses in SAR images. Thus, discriminating these targets from confusers or clutter objects in SAR imagery is a highly challenging task in the emerging low-frequency UWB SAR technology used for this application \cite{nguyen1998mine}. In general, techniques ranging from dictionary learning to neural networks (NNs) are employed for object classification in the wideband SAR mode \cite{vu2018classifying,giovanneschi2019dictionary}.  Through-the-wall imaging is another important application of ultrawideband and polarimetric SAR \cite{Wang2018, Ahmad2016}.

\subsubsection{Millimeter-Wave SAR}
%-----------------------------------------------------------------------------
\begin{figure}
%\centering
\includegraphics[width=1.0\columnwidth]{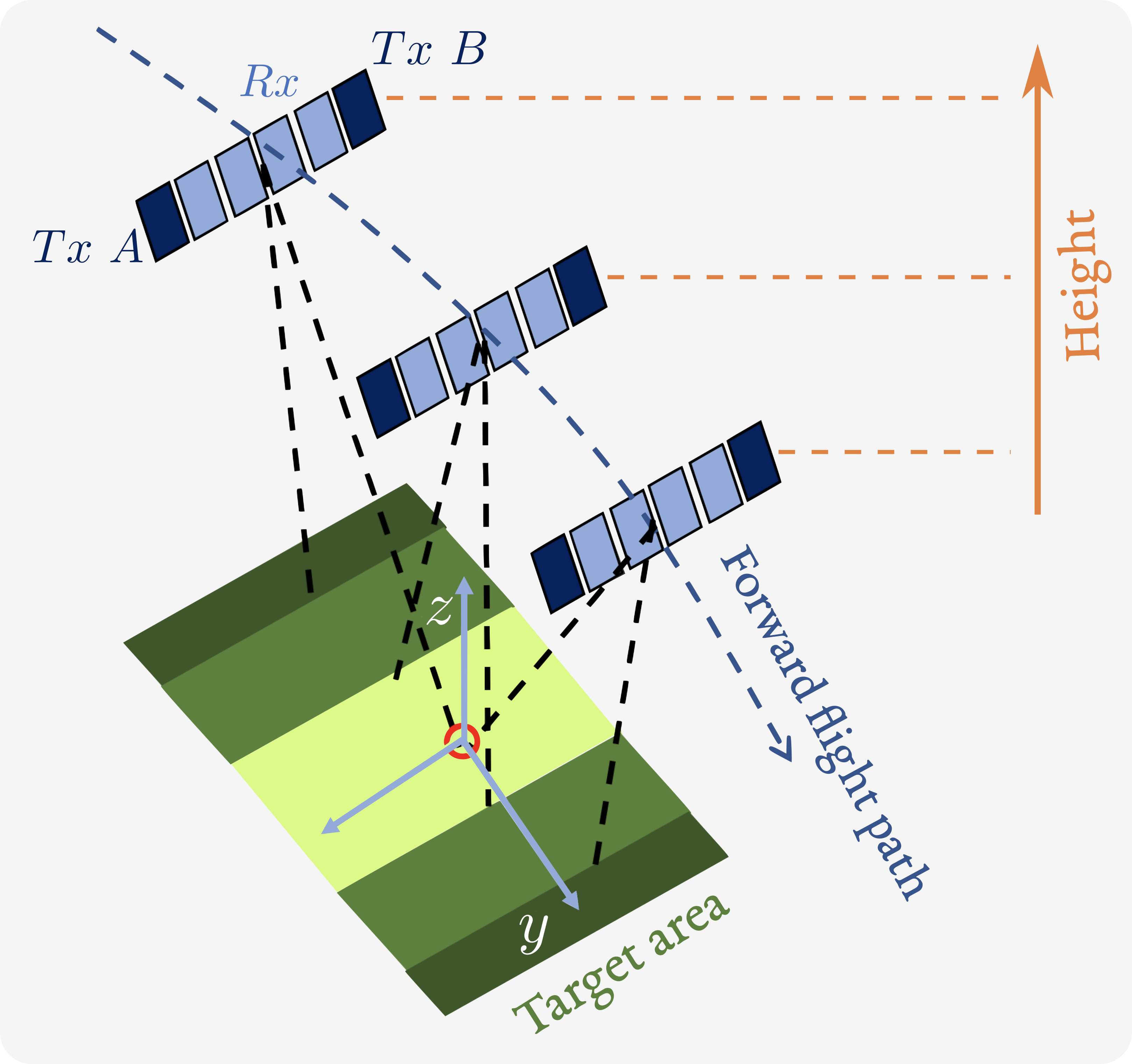}
\caption[example]{\label{fig:flosar} 
Imaging geometry of the FLoSAR. The receive antenna array is flanked on both sides by a transmit antenna. As the radar traverses a curvilinear downward motion, it exploits the virtual array along the height dimension \cite{mishra2019autofocus}.}
\end{figure} 
%-----------------------------------------------------------------------------
Toward higher frequencies, there is growing interest in millimeter wave (mm-Wave) forward-looking SA radar (FLoSAR) technology because the very wide, unlicensed bandwidth available at mm-Wave band has potential for very high-resolution applications. In addition, the mm-Wave components have reduced dimensions and the signal experiences little attenuation at close-ranges. Yet substantial challenges remain in deploying such a system on airborne platforms whose motion is not stable within subwavelength levels because the coherent SAR processing requires subwavelength knowledge of platform position from pulse to pulse relative to the target scene. In general, coherent SAR processing relies on motion sensors such as an inertial measurement unit (IMU) or the global positioning system (GPS) for this information \cite{lim2008sar}. However, at mm-Wave, GPS accuracy is insufficient thereby leading to inaccurate or \textit{defocused} image reconstructions. Therefore, it becomes imperative to resort to signal or data-driven motion compensation algorithms to \textit{autofocus} SAR images \cite{ash2012autofocus,wang2006sar}.

\subsubsection{THz SAR}
\label{subsubsec:thz}
The free space path loss and atmospheric attenuation are severe at THz spectrum. Hence, THz band is currently explored for short-range applications such as automotive, non-destructive testing, food processing, body scanners, and indoor room profiling. The THz band offers contiguous wide bandwidths up to 15 GHz. In automotive SAR, the forward looking mode is not very useful because of relatively slight change in aperture motion. The side-mounted SAR is rendered ineffective for guiding the driver in the incoming traffic. Therefore, squint-mode with side-mounted SAR has been the preferred mode for THz automotive SAR \cite{gishkori2019imaging}. Apart from high-resolution, THz electromagnetic waves exhibit good penetration depth and are, therefore, employed for applications such as through-material scans. The near-optical performance of the resulting images makes these devices very useful. The spatial resolution is further enhanced through the use of MIMO-SAR at these frequencies \cite{phippenheight}.  The use of photonic devices in a THz SAR capable of three-dimensional imaging is described in \cite{Yi2022}.

\subsection{Wideband Autofocusing}
\label{subsec:autofocus}
There is a large body of literature on SAR autofocus algorithms (see e.g. \cite{koo2005comparison} and references therein). The principle of autofocusing algorithms is as follows. The range measurement introduces two artifacts: defocusing in the azimuthal domain arising from azimuth phase errors and 2-D defocusing due to range cell migration. At mm-Wave wavelength $\lambda$, wherein $4\pi/\lambda \gg 1$, the azimuth defocusing is a more serious effect and, as long as the range measurement error is less than the range resolution itself, range cell migration is negligible. Most autofocusing techniques estimate an equivalent phase error in the measured signal by modeling the effect of the position error as a linear time-invariant filter \cite{mao2013autofocus}.

There are several approaches toward data-driven SAR image autofocus processing. The most common phase gradient autofocus (PGA) \cite{wahl1994phase, kuzniak2007implementation} does not assume any specific model of the phase error function and estimates phase errors from echoes reflected from multiple strong scatterers. The method has several variations such as the eigenvector method \cite{mao2013autofocus} and its fast computation counterparts \cite{liu2018fast}. Alternatively, a few approaches consider minimizing the image entropy to obtain a sharp image. These algorithms exploit the fact that a focused image will yield lower entropy than its blurry counterparts \cite{koo2005comparison}. More recently, autofocusing techniques based on compressed sensing (CS) \cite{jian2011forward}, blind deconvolution \cite{mansour2018sparse}, and deep learning \cite{mason2017deep} have been proposed. In the context of autofocusing in FLoSAR, very few works exist \cite{feng2018extended,wu2014focusing}; further, there have not been in-depth investigations into autofocus algorithms for mm-Wave FLoSAR.

\subsection{Multiband Processing}
\label{subsec:multi}
		\begin{figure}[t]
		\centering
		\includegraphics[width=0.5\textwidth]{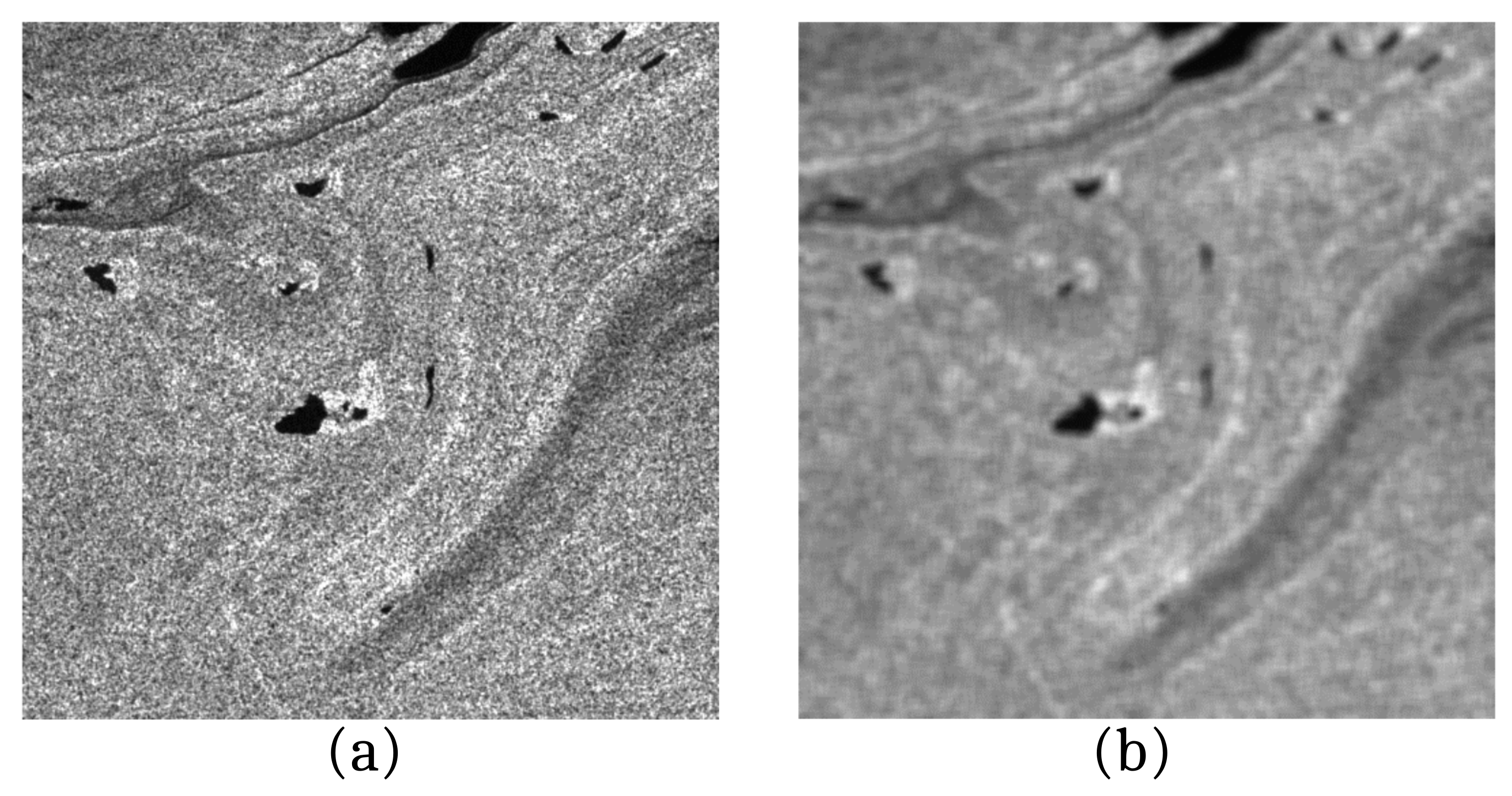}
		\caption{Example of (a) noisy and (b) denoised SAR images using Lee filters for a window of size $5\times 5$. In this particular example, component images from channels 5 and 11 are used of the Sentinel-2 multiband imager \cite{rubel2021selection}. }
		\label{fig:speckle_noise}
	\end{figure}
	
%Radar remote sensing (RS) has found numerous applications in ecological monitoring, agriculture, forestry, hydrology, etc. \cite{schowengerdt2007remote,kussul2016parcel,joshi2016review}. This can be explained by the following reasons \cite{kussul2016parcel}. First, radar sensors can be used in all-weather conditions during day and night. Second, modern radars (mostly SARs) provide high spatial resolution and data (image) acquisition for large territories, often with high periodicity since many existing systems perform the frequent observations (monitoring) of terrains under interest. Third, 
Modern SARs produce valuable information content especially if they operate in multichannel \cite{guerra2016computationally}, multi-polarization \cite{schowengerdt2006remote,kussul2016parcel}, or multi-temporal mode \cite{ferrentino2020use}. The multiband SAR typically provides images with higher spatial resolution but poorer spectral resolution. On the other hand, hyperspectral SAR acquires images in the form of a  set of reflectance spectra in many contiguous and very narrow bands thereby yielding a high spectral resolution but trading off the spatial resolution. Sometimes, the SAR may be fitted with both sensors and employ techniques for hyperspectral and multiband image fusion \cite{guerra2016computationally}. In wideband/multi-frequency InSAR, split-bandwidth interferomtery may be employed as a fast method for absolute phase determination in interferograms  \cite{bamler2005accuracy}. %Fourth, SAR images are often provided after their pre-processing that includes co-registration, geometric and radiometric correction, and calibration. This is convenient for their further processing and interpretation.
%Radar imaging has become a useful tool to collect data from scenes of large territories through the use of polarimetric SAR . 

The multi-band applications often suffer from the presence of a multiplicative, non-Gaussian and spatially correlated \cite{joshi2016review} \textit{speckle} noise \cite{schowengerdt2006remote} in SAR images. Mathematically, a noisy image $\boldsymbol{Z}\in \mathbb{R}^{m\times n}$ is modeled as 
\begin{align}
 \boldsymbol{Z} = \boldsymbol{Y}\odot \boldsymbol{\zeta},
 \label{eq:noisyImage}
\end{align}
where $\boldsymbol{Y}\in \mathbb{R}^{m\times n}$ is the target clean image, $\boldsymbol{\zeta}\in \mathbb{R}^{m\times n}$ stands for multiplicative noise with assumed variance equal to $\sigma_{\mu}^{2}$, and $\odot$ represents point-wise multiplication. The image/noise model relies on general information about SAR image/speckle properties \cite{rubel2021selection}. 

In practice, it is desired to remove this specific noise $\boldsymbol{\zeta}$ in \eqref{eq:noisyImage} (see Fig.~\ref{fig:speckle_noise}(a)) through \textit{denoising} or \textit{despeckling} filters \cite{ferrentino2020use}. However, it is not always possible, at least not without degrading useful information \cite{ferrentino2020use,nascimento2018detecting}. In other words, a positive effect of speckle suppression takes place simultaneously with a negative effect of smearing of image edges and details. Depending on the properties of an image, the filter used and characteristics of speckle, there can be a different proportion of positive and negative effects. When this proportion is about equal, despeckling becomes an unreasonable procedure \cite{nascimento2018detecting}.
	
Thus, it is important to predict the filtering performance before applying image filtering. A recent successful example is the Lee filter, which its output is expressed as
\begin{align}
 \boldsymbol{Z}^{\text{Lee}}[i,j] = \hat{\boldsymbol{Z}}[i,j] + \frac{\sigma^{2}_{i,j}}{\hat{\boldsymbol{Z}}^{2}[i,j]\sigma_{\mu}^{2} + \sigma^{2}_{i,j}}(\boldsymbol{Z}[i,j] - \hat{\boldsymbol{Z}}[i,j]),
\end{align}
where $\boldsymbol{Z}^{\text{Lee}}[i,j]$ is the filtered image, $\hat{\boldsymbol{Z}}[i,j]$ denotes the local mean in the scanning window centered on the $i,j$-th pixel, $\boldsymbol{Z}[i,j]$ denotes the central element in the window, and $\sigma^{2}_{i,j}$ is the variance of the pixel values in the current window. In Fig. \ref{fig:speckle_noise}(b) we present an example of Lee filter outputs.

In \cite{touzi2002review}, it was demonstrated that such a prediction is possible for filters based on the discrete cosine transform (DCT) with application to SAR images acquired by the Sentinel-1 sensor. Here, data provided by the Sentinel-1 sensor have been already used for several important applications \cite{oliver2004understanding}. Then, there are numerous papers dealing with estimation of image quality \cite{frost1982model} including visual quality and prediction of filtering efficiency \cite{touzi2002review}. For the corresponding methods, there is a clear tendency to apply neural networks (NNs) \cite{touzi2002review}. Then, it is increasingly popular to employ visual quality metrics in analysis of image original quality and filter performance \cite{touzi2002review}. Finally, it has been shown that filtering efficiency can be predicted for different types of noise (additive, pure multiplicative, and, in general, signal dependent; white and spatially correlated) and for different types of filters \cite{cozzolino2013fast}.

Filtering based on the DCT \cite{touzi2002review} is one type of filtering used to remove speckle. Meanwhile, there are many other methods to deal with SAR image denoising and the prediction of filter efficiency. A predictor based on a trained NN for the well-known Lee filter \cite{gupta2013despeckling} is included in many existing tools for SAR image despeckling. %As an example, Sentinel-1 SAR images are considered as a potential application with the reasons given above.

\subsection{Quantum Systems for SAR}
\label{ssec:quantumSAR}

While, at present, %the time of writing this overview paper, 
a fully quantum SAR system is yet to be demonstrated, steps toward practical quantum SAR have been made in recent years that bear mentioning. A quantum SAR (QSAR) system is any SAR system that exploits the effects of quantum mechanics. Generally speaking, QSAR systems employing entanglement are the primary schemes being considered by the community. The benefit of entanglement QSAR is the enhanced ability to distinguish signals from noise especially in low SNR scenarios. This allows for the use of very weak transmitter powers with such systems showing excellent potential for covert applications~(see, e.g., \cite{Tahmasbi2021}) or cases where radiation dose must be limited, e.g., imaging of human tissue and other biomedical applications.

Entanglement QSAR is based on the principle that two entangled signals have a higher degree of correlation than their classical counterparts. This enhanced correlation means that any matching that takes place following the injection of noise through a measurement activity is more robust and likely to return a correct positive match. Even though the process of launching one of the two entangled signals into free space destroys the entanglement, from the combination of noise and loss mechanisms, successful detection is still enhanced by the degree of entanglement~\cite{Lloyd2008}. For example, consider the engtangled state $\Psi_{SI}=(1/\sqrt{d})\Sigma_k\ket{k}_S\ket{k}_I$ for a signal photon, sent in the direction where an object is expected to be, and idler photon, where $d$ is the number of signal and detector modes. Assuming noise is injected into the system, where $b$ is the number of noise photons, an object with reflectivity $\eta$ is likely to be detected when $\eta/b>1$ whether entanglement is used in the measurement or not. However, when $\eta/b<1$, the SNR is low and a simple analysis~\cite{Lloyd2008} shows that on average $8b/n^2$ photons must be collected to distinguish the signal from noise in a classical measurement, whereas only $8b/n^2d$ photons on average are needed when entanglement is used. In other words, the degree of entanglement defined by the number of modes $d$ enhances the SNR and reduces the number of trials needed to distinguish a signal photon from noise.

Following a theoretical study using SNR and error detection probability calculations, Lanzagorta et al. predicted the benefit of entanglement-based QSAR over coherently integrated classical SAR~\cite{Lanzagorta2017}. They define SNR as 
\begin{equation}
    SNR = \frac{PG^2\lambda^3\sigma_0\delta_r}{2(4\pi)^3 R^3 k_b T_0 F_n l_a v \cos(\theta)},
\end{equation}
where $P$ is the average transmitted power in the classical regime or $P=M\hbar \omega$ in the quantum regime defining $M$ signal photons of frequency $\omega$ and $\hbar=h/2\pi$, $h$ is Planck's constant. In this expression, $G$ is the antenna gain, $\lambda$ is the radar wavelength, $\sigma_0$ is the target radar cross-section, $\delta_r$ is the range resolution, $R$ is the range to the target, $k_b$ is Boltzmann constant, $T_0$ is the normal scene noise temperature, $F_n$ is the dimensionless noise figure, $l_a$ is the loss due to atmospheric attenuation, $v$ is the speed of the radar platform, and $\theta$ is the grazing angle. They also define the detection error probability of the classical and quantum systems, respectively, in the low brightness, high noise, low reflectivity regime as 
\begin{equation}
    \epsilon_c = \frac{1}{2}e^{-SNR/4}, \hspace{1cm}
    \epsilon_q = \frac{1}{2}e^{-SNR}.
\end{equation}
Comparing the classical and quantum detection error probabilities, advantages of QSAR are found in range, speed, target size, and grazing angle. For example, by defining a clear image as one having an SNR of at least 5~dB, QSAR returns clear images over a range of $125~\mathrm{km}$ while classical SAR does not in their analysis. This and other theoretical studies of microwave entanglement applied to radar, ranging, and SAR motivate the investigation of practical realizations of these quantum systems.

In quantum illumination (QI)~\cite{Lloyd2008}, a signal and idler pair of entangled photons are generated by some parametric converter. The signal is sent into free space while the idler is held in memory at the point of the receiver. At the time when the signal is expected to return, the received signal is compared with the saved idler to determine if what was received is noise or the returned signal. Holding the idler in quantum memory is non-trivial and generally limits the detection range due to losses in that memory. Therefore, two groups have demonstrated schemes where a quadrature measurement is made on the idler to digitize its information for more convenient storage. 

Luong et al. presented experimental measurements in 2019 demonstrating their so-called quantum two-mode squeezing (QTMS) technique where the in-phase (I) and quadrature (Q) voltage signals of the retained idler and returned signal are mixed to enhance sensitivity~\cite{Luong2019}. They used a Josephson parametric amplifier to generate an entangled pair at 6.1445~GHz (idler) and 7.5376~GHz (signal). Both signals were first passed through a chain of amplifiers before being split into two paths, detected, and compared. The experimental demonstration did not include a target as the transmitter and receiver horns of the radar system were pointing directly at each other; nevertheless, they demonstrated the process of a detection by performing matched filtering between the stored 6~GHz and launched 7~GHz signals. The technique showed some quantum benefit when the team exchanged the quantum signal generator (the Josephson parametric amplifier) with a classical signal generator. The correlated classical signals underwent the exact same amplification and propagation chains followed by the same matched filtering. Based on receiver operating characteristic curves, Luong et al. found that, at low SNR, the classical measurement required longer integration time to reach the signal performance of the QTMS measurement~\cite{Luong2019}. The authors note there are similarities between their QTMS radar technique and noise radar in \cite{Chang2019} and discuss spaces for future development of this and related quantum radar systems in \cite{Luong2020}. 

In 2020, Barzanjeh et al. published a thorough investigation of a QI setup with a digital receiver~\cite{Barzanjeh2020}. Their use of the digital detection scheme, where again I and Q voltages are obtained of the signal and idler, circumvents the memory requirements of the traditional QI schemes~\cite{Barzanjeh2015}. In this setup, a Josephson parametric converter (JPC) was used to generate the entangled signals through three-wave mixing producing the signal photons at $\omega_S/2\pi=10.09~\mathrm{GHz}$ and idler photons at $\omega_I/2\pi=6.8~\mathrm{GHz}$. Following amplification, the signal and idler are down converted to an intermediate frequency of $20~\mathrm{MHz}$ and digitized with a sample rate of $100~\mathrm{MHz}$. They applied a fast Fourier Transform (FFT) and postprocessing to obtain I and Q voltages for the signal and idler paths, respectively. These quadrature voltages are related to the complex amplitudes $a_j$ and their complex conjugate $a_j^*$ of the signal ($j=S$) and idler ($j=I$) modes at the output of the JPC by
\begin{equation}
    a_j = \frac{I_j+iQ_j}{\sqrt{2\hbar\omega_j B\Omega G_j}}, \hspace{1cm}
    a_j^* = \frac{I_j-iQ_j}{\sqrt{2\hbar\omega_j B\Omega G_j}},
\end{equation}
where $\Omega = 50~\mathrm{ohms}$ is the resistance, $B=200~\mathrm{kHz}$ is the measurement bandwidth, and $(G_S,G_I)=(93.98(01),94.25(02))~\mathrm{dB}$ is the measured system gain for each channel. They also measured the added system noise to be $(n_S,n_I)=(9.61(04),14.91(1))$ referenced to the JPC output. The degree of entanglement is measured using the nonseparability criterion $\Delta\coloneqq\expval{\hat{X}_{-}^2}+\expval{P_{+}^2}<1$, where $\hat{X}_{-}=(\hat{a}_S+\hat{a}_S^\dagger-\hat{a}_I-\hat{a}_I^\dagger)/2$, $\hat{P}_{+}=(\hat{a}_S-\hat{a}_S^\dagger+\hat{a}_I-\hat{a}_I^\dagger)/(2i)$, $\expval{\hat{O}}$ defines the mean of the operator $O$, and $O^\dagger$ is the transpose conjugate of the operator $O$. They measure $\Delta$ as a function of the signal photon number $N_S=\expval{\hat{a}_S^\dagger \hat{a}_S}$, and find that at low photon number, $\Delta$ is below one meaning the outputs of the JPC are entangled, while at larger photon number obtained with large pump powers, entanglement gradually degrades and vanishes at $N_S=4.5~\mathrm{photons/sHz}$. 

With this confirmation of entanglement, Barzanjeh et al. then analyzed the SNR of the QI detection (Eq.~\ref{eq:SNRqici})  with comparisons to classical illumination (also Eq.~\ref{eq:SNRqici}), subjected to the same noise and loss conditions as the QI measurement, and to a coherent-state illumination scheme (the classical benchmark) with digital homodyne (Eq.~\ref{eq:SNRhomo}) and digital heterodyne (Eq.~\ref{eq:SNRhet}) detection  also following the same measurement chain, signal bandwidth, and signal power. Their analysis showed marginal quantum enhancement of the SNR over the classical benchmark with perfect microwave photon counting of the idler, which they simulate by calibrating the idler path. 
\begin{equation}\label{eq:SNRqici}
    SNR_{QI/CI} = \frac{\left(\expval{\hat{N}_1}-\expval{\hat{N}_0}\right)^2}{2\left(\sqrt{\sigma_{N_1}^2}+\sqrt{\sigma_{N_0}^2}\right)^2}
\end{equation}
\begin{equation}\label{eq:SNRhomo}
    SNR_{CS}^{homo} = \frac{\left(\expval{\hat{X}_{S,1}^{det}}-\expval{\hat{X}_{S,0}^{det}}\right)^2}{2\left(\sqrt{\sigma_{X_{S,1}^{det}}^2}+\sqrt{\sigma_{X_{S,0}^{det}}^2}\right)^2}
\end{equation}
\begin{equation}\label{eq:SNRhet}
    SNR_{CS}^{het} = \frac{\left(\expval{\hat{X}_{S,1}^{det}}-\expval{\hat{X}_{S,0}^{det}}\right)^2+\left(\expval{\hat{P}_{S,1}^{det}}-\expval{\hat{P}_{S,0}^{det}}\right)^2}{2\left(\sqrt{\sigma_{X_{S,1}^{det}}^2+\sigma_{P_{S,1}^{det}}^2}+\sqrt{\sigma_{X_{S,0}^{det}}^2+\sigma_{P_{S,0}^{det}}^2}\right)^2},
\end{equation}
where $\hat{N}_j=\hat{a}_{j,+}^\dagger\hat{a}_{j,+}-\hat{a}_{j,-}^\dagger\hat{a}_{j,-}$ is the annihilation operator of the mixed signal and idler modes in the absence ($j=0$) or presence ($j=1$) of a target, where $\hat{a}_{j,\pm}=(\hat{a}_{S,j}^{det\dagger}+\sqrt{2}\hat{a}_v\pm\hat{a}_I^{det})/\sqrt{2}$, $\hat{a}_v$ is the vacuum noise operator, $\hat{a}_{S,j}^{det}$ is the detected radiation, and $\sigma_O^2$ is the variance of the operator $O$. Also, $\hat{X}_{S,j}^{det}=(\hat{a}_{S,j}^{det}+\hat{a}_{S,j}^{det\dagger})/\sqrt{2}$ and $\hat{P}_{S,j}^{det}=(\hat{a}_{S,j}^{det}-\hat{a}_{S,j}^{det\dagger})/\sqrt{i2}$ are the field quadrature operators. To calibrate the number counting of the idler, Barzanjeh et al. reduce the variance in the denominator of Eq.~\ref{eq:SNRqici} by the calibrated idler vacuum and amplifier noise as $\expval{\hat{a}_I^\dagger \hat{a}_I}=\expval{\hat{a}_I^{det\dagger} \hat{a}_I^{det}}/(G_I-(n_I+1))$. 

There are significant challenges still to overcome before QSAR becomes a reality. The generation of entangled microwave signals and the quantum detection of those signals, both likely requiring cryogenic temperatures and, for the moment struggling with heavy amplification noise, are the main technological barriers to practical implementations of QSAR or any quantum illumination application in the microwave regime~\cite{Sanz2018,Brandsema2018,Balaji2018,Karsa2021}. Plus, the synchronization of the signal and idler places some constraints on the measurement acquisition process that are noteworthy~\cite{Karsa2021_conference}. That said, new innovations are consistently put forward, meaning there is reason to continue to track developments in this field and look forward to new advancements.

\section{Single Aperture Joint Communications-Radar}
\label{sec:jrc}
Historically, there has been strong interest in combining the functions of radar and data communications using a single aperture \cite{Mishra2019}. While it is true that conceptually any radar is capable of performing a communications function by using modulated waveforms for target detection, in reality the attainable detection performance will be far from optimal.  The unique feature of a SAR system that makes it a candidate for a dual-use communications platform is that it transmits with high average power from an airborne vehicle.  This section will describe how the signal processing chain for a communications system is very different than for a radar and how ultimately a communications waveform, such as OFDM, would have to be processed using sophisticated techniques, like adaptive pulse compression, to make it suitable for target detection.  Nevertheless, the joint execution of radar and communication functions through a single wideband aperture is an important emerging technology focus area for wideband synthetic apertures. 

If one considers only the output of a single beamformer channel and the waveforms that are generated by modulating one carrier, then it seems possible to combine the detection functions of a radar and a communications systems into a single aperture. This capability is especially practical for software defined systems where the transmitted and received signals are digitized as close to the antenna as possible, allowing for much of the necessary functionality to be built into software. Ideally, digitization would occur behind each element of a phased array, as in digital beamforming (DBF) architectures, such that software selects the necessary processing functions for executing a desired task. This section provides a brief description of the processing similarities and differences between a notional radar as compared to a digital communication system assuming both systems rely on a binary frequency shift keying (BFSK) waveform. The situation is more complex for multi-carrier waveforms or for systems that can generate multiple beamformer output channels, including simultaneous beams. Since wideband SAR systems also operate at high power levels they are well suited for high data-rate communications. References are provided to highlight some of the recent advances in joint radar-communications processing.

\begin{figure}[t]
	%\centering
	\includegraphics[width=1.0\columnwidth]{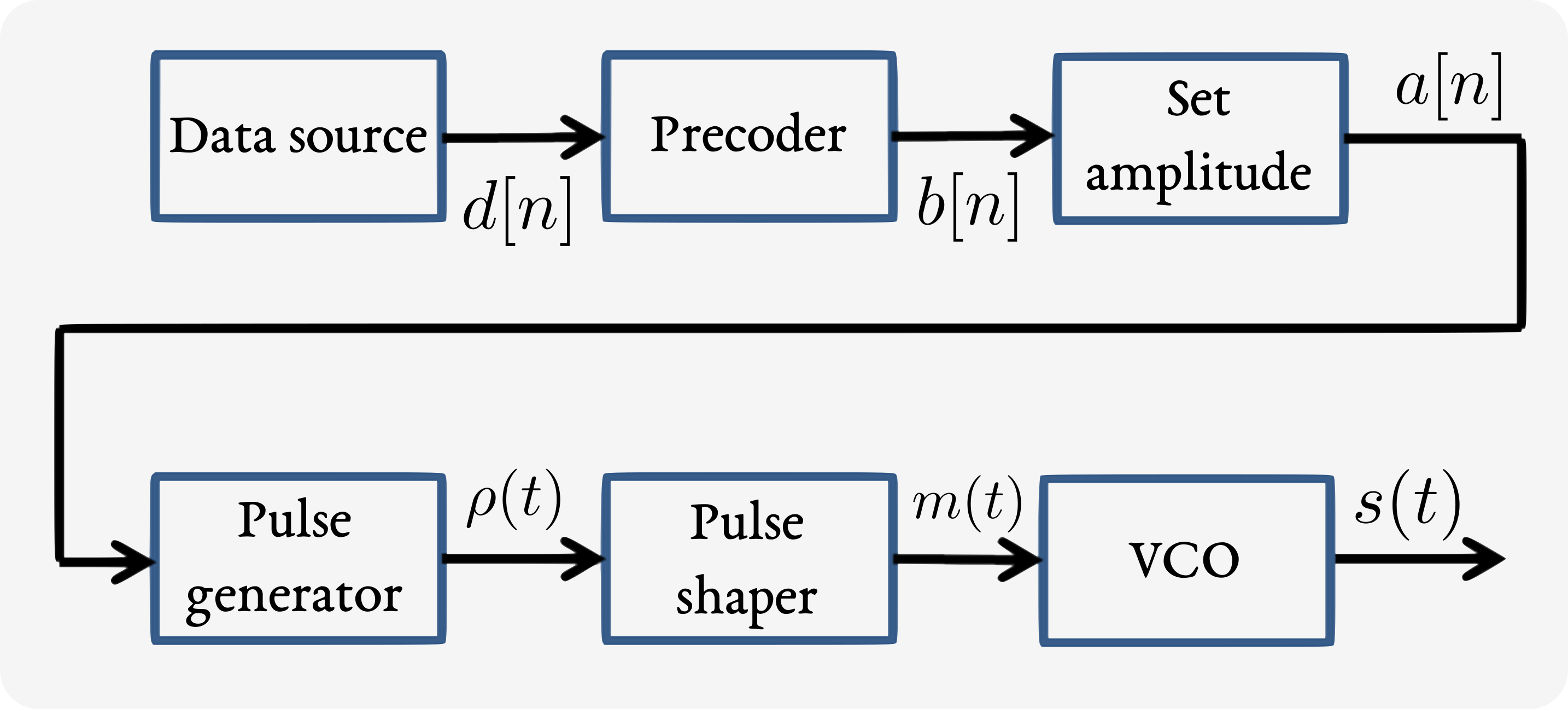}
	\caption{Simplified illustration of BFSK transmit chain. Wideband SAR systems are suitable for combining such a communications system in the existing hardware.}
	\label{fig:BFSK}
\end{figure}

For the case of BFSK modulation, a notional block diagram describing the transmit chain is shown in Fig. \ref{fig:BFSK}. With BFSK two tones at ${f_{0}}$ and ${f_{1}}$ are used to transmit two symbols ${A}$ and ${B}$. The symbol rate, ${f_{SY}}$, is known as the Baud rate and ${|f_{1} - f_{0}|}$ is the frequency excursion. The modulation index ${h}$ is,
\begin{equation}
 h = \frac{|f_{1} - f_{0}|}{f_{SY}}.
\end{equation}
If the tone spacing equals one-half the symbol rate, then ${h = 0.5}$ and the modulation is known as minimum shift keying (MSK). The minimum value that ${h}$ can take is ${0.5}$ because any smaller values will violate the orthogonality of the tones ${f_{0}}$ and ${f_{1}}$. Notice that since ${|f_{1} - f_{0}| = 2f_{SY}}$ for the case of MSK, transmitting at higher data rates will require a higher signal bandwidth.

Referring to Fig. \ref{fig:BFSK}, the precoder converts the input data ${d[n]}$ into a discrete sequence of bits ${b[n]}$ with values of ${0}$ or ${1}$. Absolute encoding is used if the bit sequence ${b[n]}$ corresponds directly to points in a symbol constellation. Differential encoding is used if the values of ${b[n]}$ correspond to the changes in the data sequence ${d[n]}$. For a radar, typical values for ${b[n]}$ might be a pseudo-noise (PN) sequence with low autocorrelation sidelobes. The PN codes allow peak-power constrained radars to illuminate targets with high average-power, long-duration waveforms that also provide high delay resolution after matched filtering. The Amplitude block in Fig. \ref{fig:BFSK} converts the precoded bits ${b[n]}$ into amplitude levels ${-1}$ or ${1}$ according to ${a[n] = (-1)^{b[n]}}$. The pulse generator creates a continuous-time waveform of pulses ${p(t)}$ which are then filtered by the pulse shaper to control the waveform's bandwidth. The filter's output ${m(t)}$ drives a voltage controlled oscillator (VCO) whose frequency varies between ${f_{c}-f_{0}}$ and ${f_{c}+f_{1}}$. The final transmitted signal ${s(t)}$ can be represented as,
\begin{align}
\label{E:IQ}
 s(t) & = m_{\textrm{I}}(t)\cos(2{\pi}f_{c}t + \theta_{c}) - m_{\textrm{Q}}(t)\sin(2{\pi}f_{c}t + \theta_{c}) \\ \nonumber
 & = \text{Re}\left[(m_{\textrm{I}}(t) + \mathrm{j}m_{\textrm{Q}}(t))(\cos(2{\pi}f_{c}t + \theta_{c}) + \mathrm{j}\sin(2{\pi}f_{c}t + \theta_{c}))\right] \\ \nonumber
 & = \text{Re}\left[(m_{\textrm{I}}(t) + \mathrm{j}m_{\textrm{Q}}(t))e^{\mathrm{j}(2{\pi}f_{c}t + \theta_{c})}\right].
\end{align}
The term ${(m_{\textrm{I}}(t) + jm_{\textrm{Q}}(t)}$ is known as the complex envelope of the signal with in-phase component ${m_{\textrm{I}}(t)}$ and quadrature component ${m_{\textrm{Q}}(t)}$. The complex envelope contains all the information of the signal. For the case of BFSK,
\begin{align}
 & \text{Symbol A} \rightarrow a[n] = -1 \rightarrow m_{\textrm{I}}(t) = \cos(2{\pi}(f_{c} - f_{0})t + \theta(t)), \\ \nonumber
 & \text{Symbol B} \rightarrow a[n] = +1 \rightarrow m_{\textrm{Q}}(t) = \cos(2{\pi}(f_{c} + f_{1})t + \theta(t)),
\end{align}
where the phase ${\theta(t)}$ may change randomly or deterministically with each symbol depending on the VCO.

\subsection{Intersymbol Interference}
The role of the pulse-shaping filter in the transmit chain highlights the major differences between the mission requirements of a radar and a digital communication system. An optimal waveform for a radar produces a thumb-tack ambiguity diagram and measures the Doppler and range of a target unambiguously. Typically, a matched filter is used on receive because it achieves the maximum output SNR for a signal in additive white noise.

In digital communications, especially in environments with congested spectrum, such as with cellular telephones or wireless local area networks (LANs), transmitting a sequence ${p(t)}$ of ideal brick-wall pulses that never overlap in time would require infinite bandwidth due to the ${\sin(f)/f}$ spectrum for each pulse. Since there is an inverse relationship between bandwidth and the temporal extent of a signal, limiting the signal bandwidth will increase the duration of each symbol and cause it to interfere with neighboring symbols, creating ISI. Choosing an appropriate pulse-shaping filter limits the ISI created when a finite bandwidth pulse spreads into the time bin of an adjacent pulse.

In general, the maximum possible symbol rate without ISI for a baseband receiver with a frequency bandwidth of ${f_{\textrm{SY}}}$ Hz is ${f_{\textrm{SY}} = 1/T}$ symbols per second, where ${T}$ is the Baud interval or duration of a single symbol. Note that ${1/T}$ is the symbol or Baud rate, not the bit rate, since there may be multiple bits per symbol. This result is known as the Nyquist bandwidth constraint and should not be confused with the Nyquist sampling criterion which states that a signal can be reconstructed from its samples provided that the sampling frequency ${f_{\textrm{SAM}} \geq 2f_{\textrm{MAX}}}$, where ${f_{\textrm{MAX}}}$ is the highest frequency component of the signal.

Three spectral-shaping filters are typically used to control the spectral splatter of symbols and to limit ISI. The zero-ISI ${\sin(t)/t}$ or ${\textrm{sinc}(t)}$ filter for a symbol rate ${f_{SY}}$ is
\begin{equation}
 h_{\textrm{SINC}}(t) = \frac{\sin({\pi}f_{SY}t)}{({\pi}f_{SY}t)}.
\end{equation}
This filter produces a ${\sin(t)/t}$-shaped pulse which is equal to unity at time ${t=0}$ and is also zero at the sampling instants corresponding to
\begin{equation}
t = \frac{n}{f_{\textrm{sy}}}, \quad n = \ldots -2, -1, 0, 1, 2, \ldots
\end{equation}

A more common spectral control filter is generated by multiplying the ${\sin(t)/t}$ function by a raised cosine. This filter provides less passband ripple and lower sidelobes in the pulse spectrum. The impulse response is given by,
\begin{equation}
  h_{\textrm{COS}}(t) = \frac{\sin({\pi}f_{\textrm{SY}}t)}{({\pi}f_{\textrm{SY}}t)} \space \cdot \space \frac{\cos({\beta}{\pi}f_{\textrm{SY}}t)}{1-4{\beta}^2t^{2}f_{\textrm{SY}}^{2}},
\end{equation}
where ${\beta}$ is a rolloff factor that describes how steeply the filter's passband transitions. If ${f_{\textrm{BW}}}$ denotes the null-to-null bandwidth of the filter's frequency response, then ${\beta = (f_{\textrm{BW}} - f_{\textrm{SY}})/f_{\textrm{SY}}}$. Values of ${\beta > 0}$ allow excess bandwidth that enable the receiver to recover the symbol or Baud timing from the transmitted waveform. Typical values of ${\beta}$ are in the range ${0.3}$ to ${0.5}$.

The raised cosine waveform is ideal but it is not possible to match filter such a signal and still maintain zero ISI. Therefore, the square root of the raised cosine frequency response ${H_{\textrm{COS}}^{0.5}(f)}$ is applied at the transmitter and also at the receiver to yield the desired raised cosine response ${H_{\textrm{COS}}(f)}$.
\begin{figure}[t]
	\centering
	\includegraphics[width=1.0\linewidth]{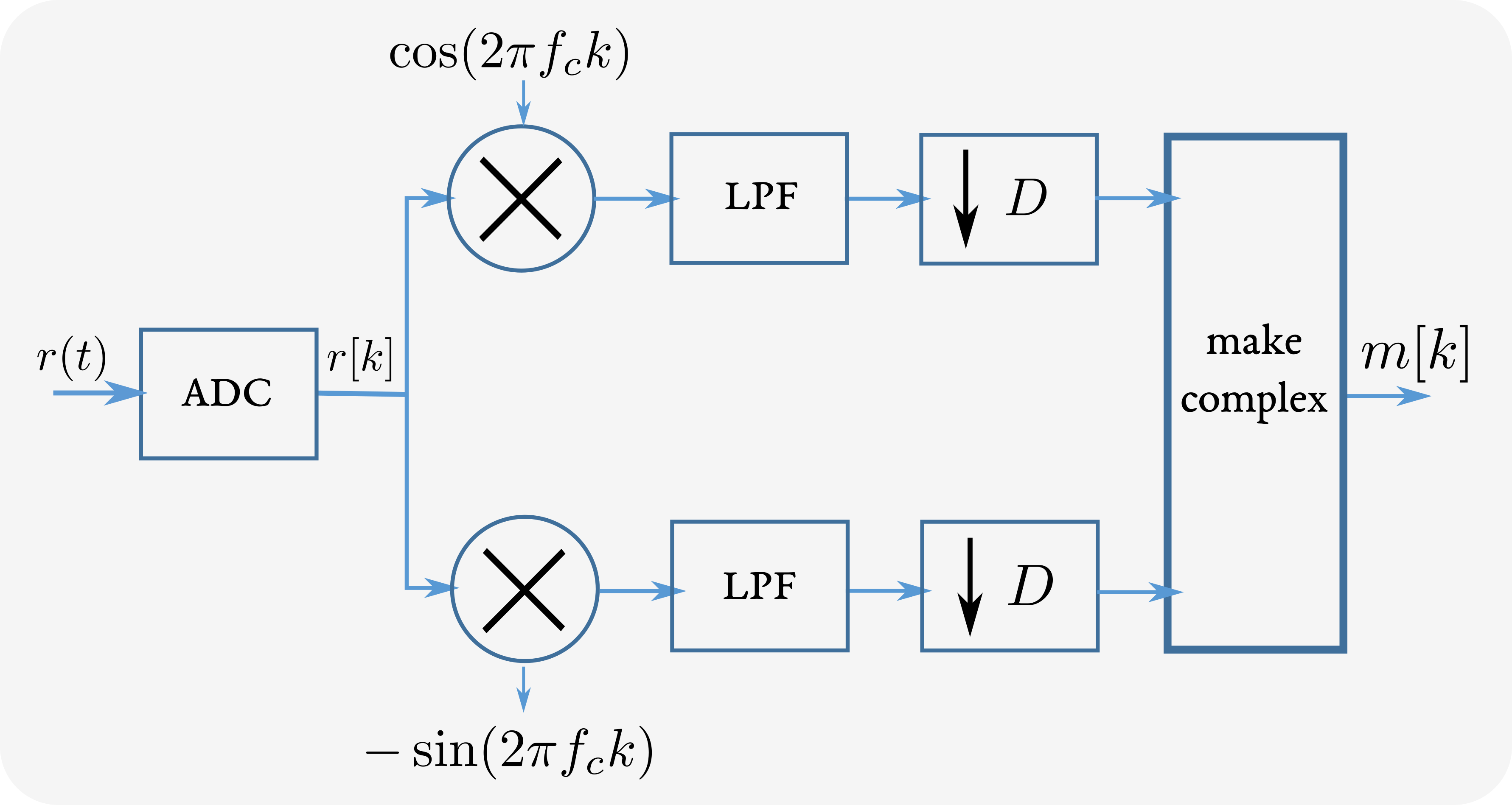}
	\caption{The I/Q demodulator for wideband signals replaces the standard decimator after filtering by a polyphase filter bank.}
	\label{fig:downconvert}
\end{figure}
\subsection{Wideband Signal Basebanding}
\label{ssec:wideband_basebanding}
Demodulating a received communications signal is not the same process as traversing the transmit chain backwards. The demodulator must recover the carrier frequency and the baud rate from the received signal. In the absence of any multipath or interference, the received signal ${r(t)}$ is equal to the transmitted signal ${s(t)}$ given in (\ref{E:IQ}). The transmitted data is contained in the complex envelope of the modulation ${m(t)}$ which must be recovered from ${r(t)}$. If the carrier frequency ${f_{c}}$ and carrier phase ${\phi_{c}}$ are perfectly known, then multiplying ${r(t)}$ by ${\cos(2{\pi}f_{c}t + \phi_{c})}$ and ${\sin(2{\pi}f_{c}t + \phi_{c})}$ yields,
\begin{align}
 &r(t)\cos(2{\pi}f_{c}t + \phi_{c})  \\ \nonumber
 &=0.5\left[m_{\textrm{I}}(t)\left[1 + \cos(2{\pi}2f_{c}t + 2\phi_{c})\right] - m_{\textrm{Q}}(t)\left[\sin(2{\pi}2f_{c}t + 2\phi_{c})\right]\right],
\end{align}
and
\begin{align}
 &-r(t)\sin(2{\pi}f_{c}t + \phi_{c})  \\ \nonumber
 &=-0.5\left[m_{\textrm{I}}(t)\sin(2{\pi}2f_{c}t + 2\phi_{c}) + m_{\textrm{Q}}(t)\left[1 - \cos(2{\pi}2f_{c}t + 2\phi_{c})\right]\right].
\end{align}
Thus, low pass filtering these products yields ${0.5m_{\textrm{I}}(t)}$ and ${0.5m_{\textrm{Q}}(t)}$. Typically, the filter output is then also decimated to a lower sample rate. This process is summarized in Fig.~\ref{fig:downconvert}. For wideband signals sampled at a high rate, decimation is not performed after filtering since that wastes computations. Instead, a polyphase filter bank is used to split the input signal ${r[k]}$ into ${D}$ sub-bands operating at a sample rate reduced by a factor of ${D}$. Rather than convolving all signal samples with a filter and then retaining only the ${D}$th sample, the polyphase filter bank only calculates the convolution samples that are retained \cite{Skolnik2008}. Note that if the system designer has the flexibility to choose the sampling frequency in relation to the intermediate frequency (IF) according to
\begin{equation}
 f_s = \frac{4f_{\textrm{IF}}}{2M- 1},
\end{equation}
where ${M}$ is an integer, then digital downconversion schemes can generate I and Q samples directly from the output ${r[k]}$ of the ADC \cite{Waters1985, Waters1982}.

If the estimated carrier phase ${\tilde{\phi_{c}}}$ does not equal the true value ${\phi_{c}}$, then multiplying ${r(t)}$ by ${\cos(2{\pi}f_{c}t + \tilde{\phi_{c}})}$ and ${\sin(2{\pi}f_{c}t + \tilde{\phi_{c}})}$ and low-pass filtering as before yields,
\begin{align}
\tilde{m_{\textrm{I}}}(t) &= 0.5\left[ m_{\textrm{I}}(t)\cos(\phi_{c} - \tilde{\phi_{c}}) - m_{\textrm{Q}}(t)\sin(\phi_{c} - \tilde{\phi_{c}}) \right], \\ \nonumber
\tilde{m_{\textrm{Q}}}(t) &= 0.5\left[ m_{\textrm{I}}(t)\sin(\phi_{c} - \tilde{\phi_{c}}) + m_{\textrm{Q}}(t)\cos(\phi_{c} - \tilde{\phi_{c}}) \right].
\end{align}
In the complex plane, ${(\tilde{m_{\textrm{I}}}(t),\tilde{m_{\textrm{Q}}}(t))}$ is a rotated version of ${(m_{\textrm{I}}(t),m_{\textrm{Q}}(t))}$. Thus, the effect of not correctly compensating for carrier phase is that after demodulation is complete, the symbol constellation will appear rotated. 

Alternatively, if the estimated carrier frequency ${\tilde{f_{c}}}$ does not equal the true value ${f_{c}}$, perhaps due to oscillator drift, then multiplying ${r(t)}$ by ${\cos(2{\pi}\tilde{f_{c}}t + \phi_{c})}$ and ${\sin(2{\pi}\tilde{f_{c}}t + \phi_{c})}$ and low-pass filtering as before yields,
\begin{align}
\tilde{m_{\textrm{I}}}(t) &= 0.5\left[ m_{\textrm{I}}(t)\cos(2{\pi}(f_{c} - \tilde{f_{c}})t) - m_{\textrm{Q}}(t)\sin(2{\pi}(f_{c} - \tilde{f_{c}})t) \right], \\ \nonumber
\tilde{m_{\textrm{Q}}}(t) &= 0.5\left[ m_{\textrm{I}}(t)\sin(2{\pi}(f_{c} - \tilde{f_{c}})t) + m_{\textrm{Q}}(t)\cos(2{\pi}(f_{c} - \tilde{f_{c}})t) \right].
\end{align}
Geometrically, the point ${(\tilde{m_{\textrm{I}}}(t),\tilde{m_{\textrm{Q}}}(t))}$ will rotate continuously in the complex plane at a rate equal to the frequency error, ${f_{c} - \tilde{f_{c}}}$. In short, the impact of not knowing the carrier frequency correctly is that the symbol constellation will appear spinning after demodulation.

\begin{figure}[t]
	\centering
	\includegraphics[width=1.0\linewidth]{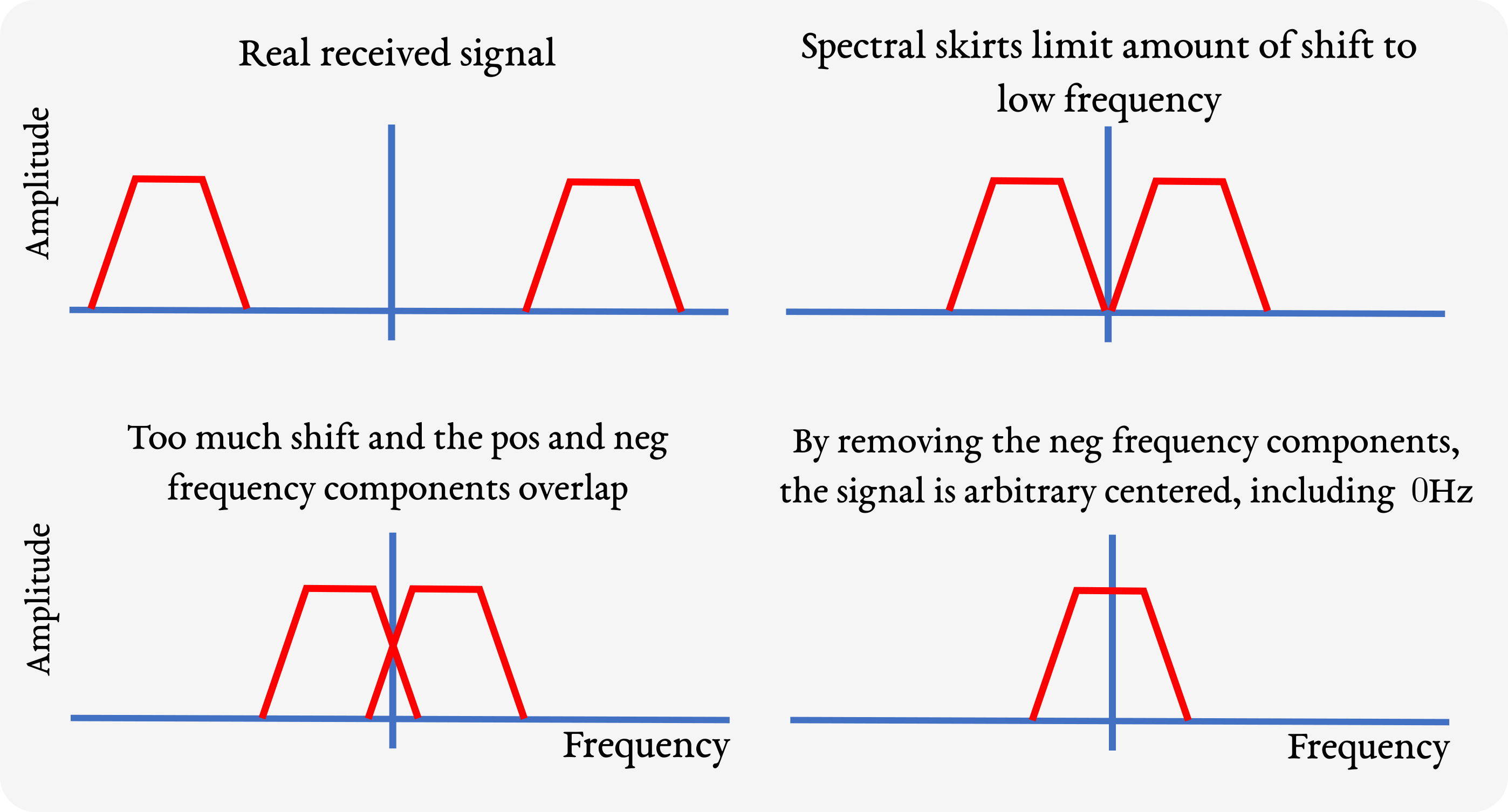}
	\caption{Complex basebanding of wideband signals. The ADC employs subsampling techniques \cite{mishra2012frequency} at the receiver. }
	\label{fig:basebanding}
\end{figure}

The process of estimating the complex envelope ${m(t) = m_{\textrm{I}}(t) + \mathrm{j}m_{\textrm{Q}}(t)}$ can also be performed after digitizing the real-valued receive signal mixed down to the IF. When sampling a wideband signal, the analog-digital-converter (ADC) ideally operates at the lowest practical sampling rate without aliasing the signal. The top-left corner of Fig.~\ref{fig:basebanding} illustrates the spectrum of the real-valued received signal after mixing down to the IF. Because the signal is real, the spectrum is double-sided. If the Nyquist sampling criterion is obeyed then the analog-to-digital sampling frequency must be at least twice the highest frequency component of the signal. However, such a high sampling rate is unnecessary given that the signal only occupies a band-pass region and the information in the positive and negative frequency sidebands is redundant. To reduce the required sampling frequency, the two sidebands could be moved closer together until they touch at 0 Hz as shown in the top right. If the sidebands move any closer, they will overlap which creates aliasing as shown in the bottom left. The optimal solution is to eliminate one of the sidebands (since retaining both is redundant) and then shift the remaining sideband to 0 Hz. This signal could then be represented at the lowest sampling rate and is equivalent to the desired complex envelope of the transmitted signal. Equivalent implementations that use band-pass instead of low-pass filtering can be constructed using the Hilbert transform \cite{Skolnik2008}.

\subsection{Low-Bit Sampling of Wideband Signals}
\label{ssec:wideband_sampling}
In narrowband digital receivers, the primary consideration driving the choice of ADC hardware is dynamic range. Typically, the system design chooses an ADC with the maximum available dynamic range that also satisfies constraints on sampling rate, power dissipation, and cost. In wideband applications however, it may be desirable to sample with fewer bits so as to increase the sample rate \cite{Rodenbeck2019,Akos1996,Rodenbeck2014,Rodenbeck2009,Fornaro1997,Li2012,Mo2016}.

\begin{figure}[t]
	\centering
	\includegraphics[width=1.0\linewidth]{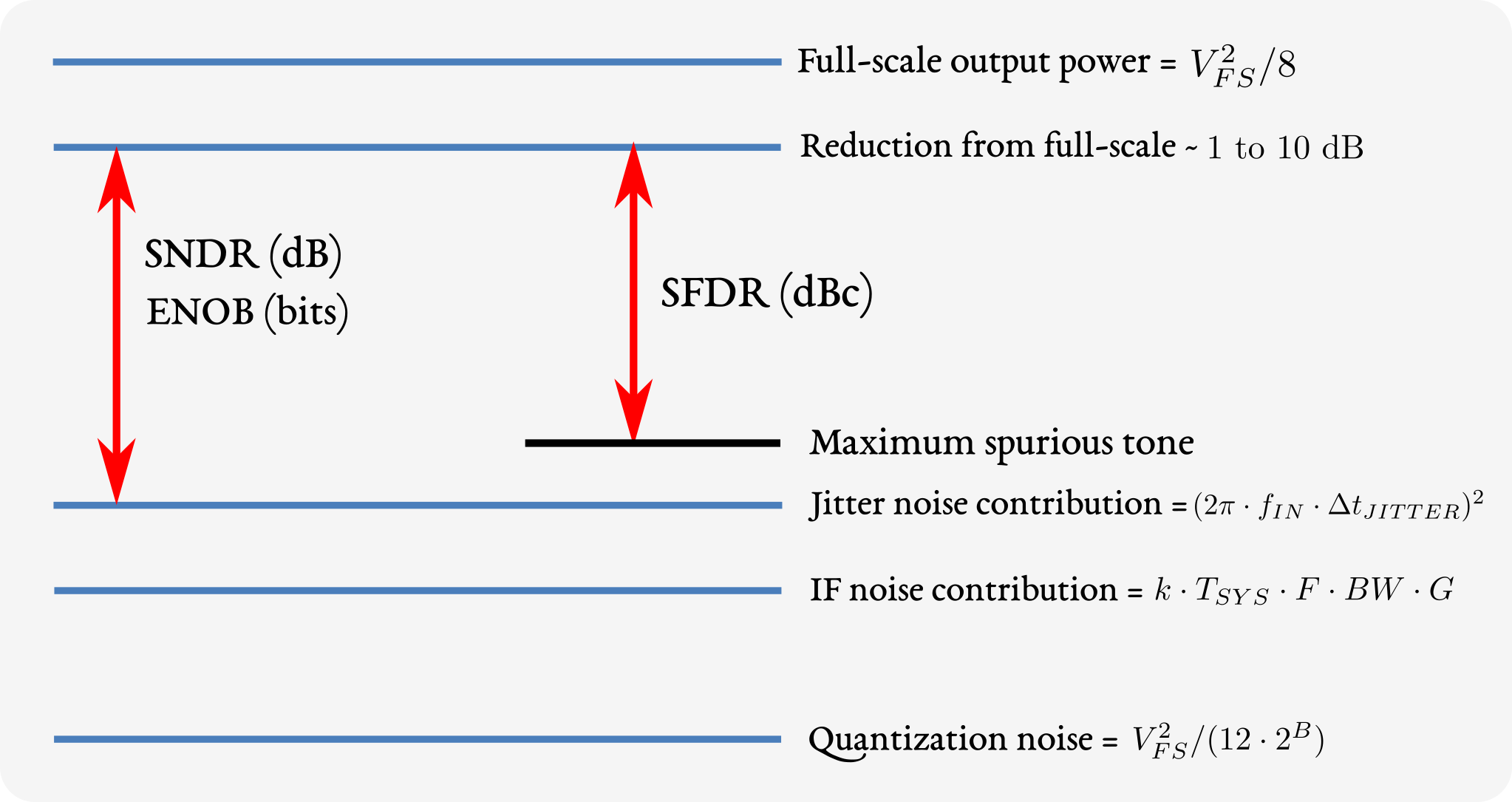}
	\caption{Receiver dynamic range contributions. A receiver with a wide dynamic range is able to handle high in-band power levels. Although there are several techniques to increase the dynamic range of a digital receiver \cite{tsui2004digital,tsui2010special,mishra2012frequency}, it is primarily decided by the choice of the ADC.}
	\label{fig:ADC_noise}
\end{figure}

Fig. \ref{fig:ADC_noise} illustrates the contributions to the dynamic range of an ADC that samples input signals at a rate of ${1/T_{s}}$, where ${T_{s}}$ is the sample period and is chosen proportionally to signal bandwidth. The output of the ADC is quantized to one of ${2^B}$ levels, where ${B}$ is the number of bits. The amplitude quantization intrinsic to an ADC creates a random error known as quantization noise. The quantization noise power for a sinusoidal input signal is approximately equal to \cite{Oppenheim2009}
\begin{equation}
 N_{\textrm{QU}} = \frac{\textrm{LSB}^{2}}{12} = \frac{1}{12}\left( \frac{V_{\textrm{FS}}}{2^B} \right)^{2},
\end{equation}
where ${LSB}$ denotes the least significant bit voltage value and ${V_{\textrm{FX}}}$ is the peak-to-peak full scale range at the ADC input.
If the input signal is a full scale sinusoid, the ideal output signal power is ${V_{\textrm{FS}}^2/8}$ which implies the SNR at the output of the quantizer is 
\begin{equation}
\label{eqn:snr}
 \textrm{SNR}_{\textrm{OUT}} = \frac{V_{\textrm{FS}}^2/8}{N_{\textrm{QU}}} = 1.5(2^{2B}) = 6.02B + 1.76 \quad \text{dB}.
\end{equation}
A more common figure of merit is the signal-to-noise-and-distortion ratio (SNDR or SINAD), which includes the power contribution from nonlinear distortions at the ADC output. Substituting SNDR in (\ref{eqn:snr}) yields an expression for the effective number of ADC bits (ENoB),
\begin{equation}
 \textrm{ENoB} = \frac{\textrm{SNDR}_{\textrm{OUT}} - 1.76}{6.02}.
\end{equation}

The ADC's full-scale output power and quantization noise floor determine the upper bound of receiver dynamic range. The total gain from the RF and IF components should be chosen such that the maximum
input signal to the ADC is less than full-scale by about 1 to 10 dB to avoid signal distortion. The RF gain should be sufficiently large so that the receiver's thermal noise is dominated by the noise figure and noise bandwidth of the RF front end rather than the thermal noise contributions from gain or loss in the IF section or ADC input stages. The thermal noise power adds linearly to the
quantization noise floor as shown in Fig. \ref{fig:ADC_noise}. It is often desirable to design the thermal noise to be approximately 10 dB higher than the quantization noise power in order to guarantee dithering the LSB and to preclude any hysteretic effects. Jitter noise from the sample clock and RF waveform can further degrade the noise floor at the output of the ADC \cite{Shinagawa1990}. The resulting SNDR, or ENOB, establishes the useful dynamic range of the receiver prior to any digital signal processing gain. The maximum power of any spurious tone within the full Nyquist bandwidth
then defines the spurious free dynamic range (SFDR).

Many of these narrowband principles are not accurate for few-bit or monobit ADCs \cite{Rodenbeck2019}. For example, the conventional rule-of-thumb that ADC output SNDR increases by about 6.02 dB/bit for sinusoidal inputs, is only approximate for less than 4 bits and mostly invalid for the quantization of low-SNR signals. Also, two-tone intermodulation distortion is a significant issue for ADCs with
less than 4 bits operating in the presence of strong in-band interfering signals. Lastly, if the ADC sample rate is not sufficiently high, the noise floor at the ADC output increases since wideband noise will fold into the bandwidth of the ADC.

\subsection{SAR Processing of Communications Waveforms}
A substantial amount of recent research has been devoted to joint communication and radar sensing (JCRS) system design \cite{dokhanchi2019mmwave,bhattacharjee2022evaluation,vargas2022joint}. Applications at the higher mmWave band frequencies between 70 to 100 GHz include environmental sensing for autonomous vehicles \cite{Mishra2019,duggal2020doppler}. In some designs the radar and communication functions proceed along separate paths through the receive chain. Other designs strive to implement both functions on a common platform to reduce system cost, size, weight and power (SWAP)~\cite{Mishra2019,wu2022resource}. In systems with the available hardware design flexibility, a powerful approach is to jointly optimize the transmit waveform and the receive filter such that constraints on power amplifier outputs and radar detection performance are satisfied \cite{elbir2021terahertz}. Other schemes for JCRS leverage a full-duplex capability that allows for simultaneously transmitting and receiving signals \cite{liu2020co}. Novel approaches for cancelling the self-interference between transmit and receive antennas are described in \cite{Li2014, Rajagopal2014, Xiao2017, Singh2020, Lin2006}. The IEEE 802.11ad wireless standard for 60 GHz has also been explored for JCRS because of its available 2 GHz bandwidth \cite{Ieee802, Grossi2018, Kumari2018, Muns2019, duggal2020doppler}. Some recent wideband JCRS designs employ intelligent reflecting surfaces for non-line-of-sight (NLoS) sensing and communications \cite{elbir2022rise,wei2022,esmaeilberg2022irs}.

As described previously, the matched filter is the optimal detector for a single target with known impulse response in additive white Gaussian noise since it maximizes SNR at the output. However, matched filtering waveforms that use typical modulations for wideband digital communications, such as Phase Shift Keying (PSK) or OFDM, can result in high range sidelobes that mask nearby weak targets. In these scenarios matched filtering is suboptimal and a mis-matched filter is preferred. 

To eliminate the effects of masking, the receive filter must be adaptively estimated from the received signal independently for every delay bin. The re-iterative minimum mean-square error (RMMSE) algorithm described in \cite{Blunt2006, Rossler2013} alternates between estimating the true range profile impulse response and the respective receive filters. For every delay bin ${k}$ the RMMSE algorithm minimizes the standard minimum mean-square error (MMSE) cost function,
\begin{equation}
 J(k) = \mathbb{E}\{|x(k) - \mathbf{w}^{H}(k)\mathbf{y}(k)|^{2}\},
\end{equation}
where ${x(k)}$ is the sample of the range profile to be estimated, ${\mathbf{y}(k)}$ is a blocked vector of received signal samples, ${\mathbf{w}(k)}$ is the adaptive filter for the ${k}$th bin, and $\mathbb{E}\{\cdot\}$ is the statistical expectation operator. The optimal MMSE filter that minimizes ${J(k)}$ is the standard Wiener solution,
\begin{equation}
 \mathbf{w}(k) = (\mathbb{E}\{\mathbf{y}(k)\mathbf{y}(k)^{H}\})^{-1}\mathbb{E}\{\mathbf{y}(k)x^{*}(k)\}.
\end{equation}
More approaches to adaptive filtering in the range domain include for ISAR, interferometric SAR (InSAR), and interferometric ISAR (InISAR) applications \cite{Gabriel1990, Liao1998}. Additional methods for range processing adaptively subtract off the effects of large targets via the CLEAN algorithm \cite{Tsao1998, Bose2002}.

\section{SA Channel Sounding}
\label{sec:channel_sound}
Communication at mmWave frequencies with high bandwidths and high data transfer rates is enabling a new era of wireless applications. To effectively utilize the wider channel capacities available at mmWave frequencies the signal propagation environment must be comprehensively analyzed. Multipath signals at the receiver created by numerous propagation paths can significantly increase bit error rate and degrade data transfer performance. Alternatively, multipath can be leveraged to improve spatial diversity and to enable multiple-input multiple-output (MIMO) communications. The most important method for characterizing the signal propagation environment is channel sounding. Channel sounding refers to the process of estimating the impulse response of a communication channel and yields information on the source of signal echoes caused by reflections, the extent of diffuse scattering and diffraction, and the amount of shadow effects or signal blocking created by stationary objects or moving people and vehicles in the scene.

\begin{figure}[t]
	\centering
	\includegraphics[width=0.5\textwidth]{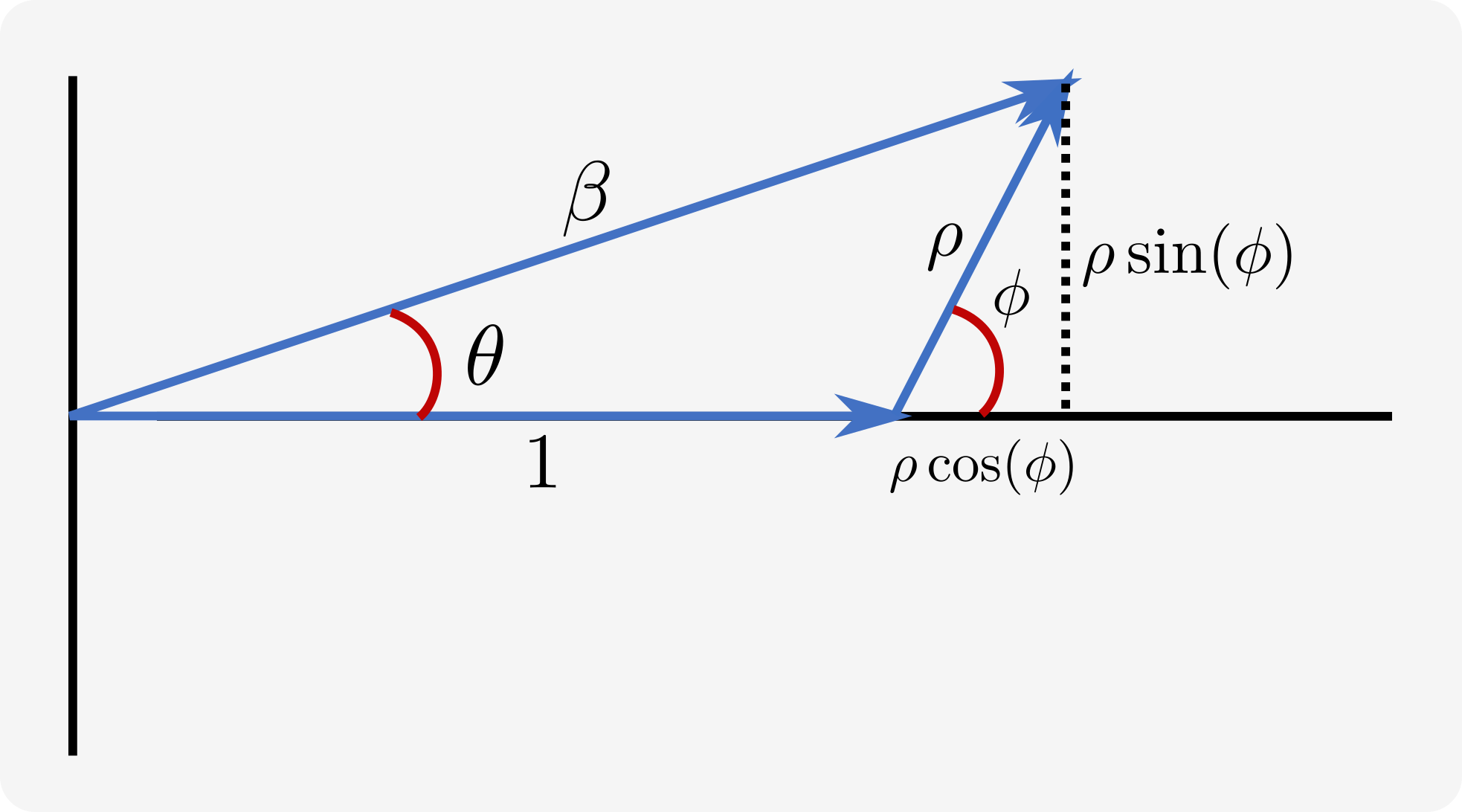}
	\caption{Two-ray multipath geometry.  The complex LoS signal and a multipath replica combine as vectors in the complex plane. }
	\label{fig:two_ray}
\end{figure}

To illustrate the impact of multipath consider the case where two signals arrive at the receiver. The direct path signal is ${V_{\textrm{D}}(t) = \cos(\omega_{0}t)}$. The scattered multipath signal is ${V_{\textrm{R}}(t) = {\rho}\cos(\omega_{0}(t - \tau)) = {\rho}\cos(\omega_{0}t + \phi)}$, where ${\phi = -{\omega_{0}}{\tau}}$ and ${\rho}$ and ${\tau}$ are random variables. The complete signal ${V_{\textrm{RX}}(t)}$ at the receiver is given by the phasor sum shown in Fig.~\ref{fig:two_ray},

\begin{align}
 V_{\textrm{RX}}(t) &= \cos(\omega_{0}t) + {\rho}\cos(\omega_{0}t + \phi) \\ \nonumber
 &= {\beta}\cos(\omega_{0}t + \theta).
\end{align}

\begin{figure}[t]
	\centering
	\includegraphics[trim={0 0 0 25},clip,width=0.5\textwidth]{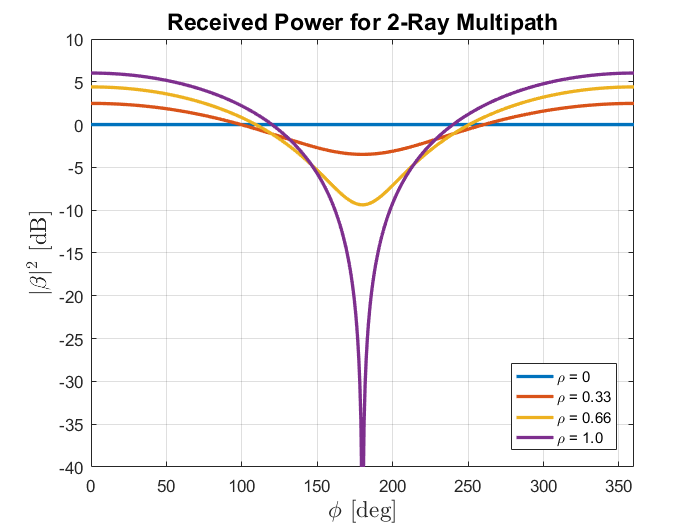}
	\caption{Receiver impact of multipath.  When the direct LoS signal and a delayed maultipath replica are out-of-phase the result can be a deep null in received power due to destructive interference.}
	\label{fig:path_loss}
\end{figure}

The path loss ${{\beta}^{2}}$ and phase of the received signal are given by,
\begin{align}
 {\beta}^{2} &= \frac{P_{\textrm{RX}}}{P_{\textrm{TX}}} = 1 +2{\rho}\cos{\phi} + {\rho}^2, \\
 \theta &= \text{tan}^{-1}\left[\frac{{\rho}\sin{\phi}}{1 +{\rho}\cos{\phi}}\right],
\end{align}
where, ${P_\textrm{TX}}$ and ${P_\textrm{RX}}$ are transmit and received powers, respectively. For signals that are highly correlated ${\rho \approx 1}$ and the destructive sum of the two incident signals at the receiver results in high path loss if the signals are close to ${180^{\circ}}$ out of phase as shown in Fig.~\ref{fig:path_loss} (data in this section is available on GitHub~\cite{SAMURAIData}). Characterizing the severity of multipath scattering in a wireless channel is the primary motivation behind channel sounding which is described next.

\subsection{Frequency Domain SA Sounders}
\begin{figure}[t]
% \begin{minipage}[b]{1.0\linewidth}
 \centering
 \includegraphics[width=1\linewidth]{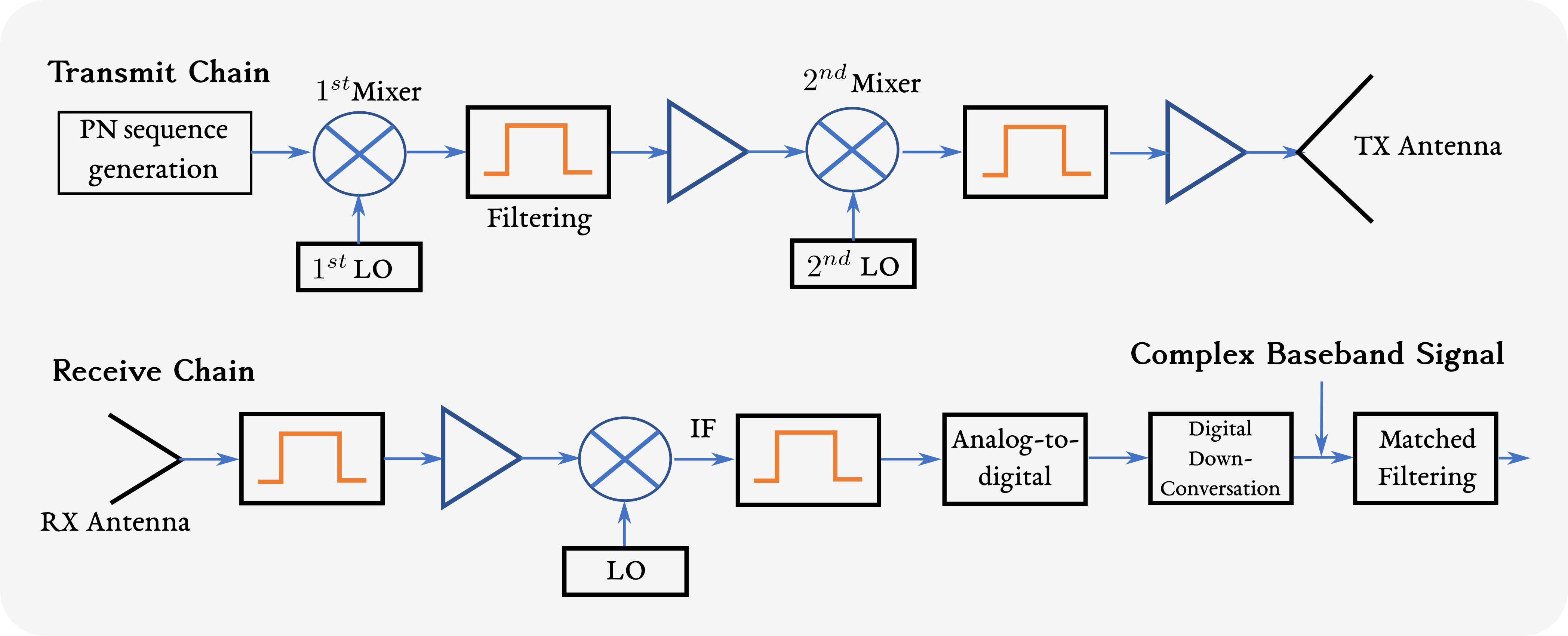}
% \vspace{2.0cm}
 \caption{Channel sounder architecture.  The baseline channel sounder configuration is a transmit and receive antenna in a bistatic configuration.}%\medskip
 \label{fig:channelSounderArchitecture}
% \end{minipage}
\end{figure}

The simplest channel sounding systems rely on directional antennas placed in a bistatic geometry to transmit and receive a probe signal that is matched filtered to produce an estimate of the channel impulse response. The concept of synthesizing an aperture larger than the physical size of an antenna can be leveraged to yield improved angular resolution performance for channel sounders ~\cite{Ranvier2009,Nguyen2016,Nguyen2016_2,Hayashi2015,Mannesson2014,Dias2005,Qiao2020,Chen2015,Zhou2018}. Furthermore, a higher measurement bandwidth than the instantaneous bandwidth of the signal can be synthesized to improve delay resolution. Fig.~\ref{fig:channelSounderArchitecture} illustrates the architecture of a typical channel sounding system.

\begin{figure}%[htb]
% \begin{minipage}[b]{1.0\linewidth}
 \centering
 \includegraphics[width=8.5cm]{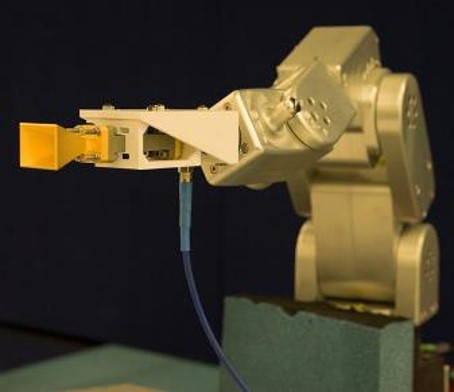}
% \vspace{2.0cm}
 \caption{Receive antenna mounted on robot.  The robot moves the antenna to precise locations in space to measure the spatial distribution of signal phase.  These coherent measurements form the basis for a synthetic aperture.}%\medskip
 \label{fig:receiveAntennaMountedOnRobot}
% \end{minipage}
\end{figure}

The nominal angular resolution of the sounder in Fig.~\ref{fig:channelSounderArchitecture} is equal to the receive antenna beamwidth, or ${\Delta{\theta} = \lambda/D}$, where ${\lambda}$ represents wavelength and ${D}$ is the dimension of the antenna in the principal plane. The delay resolution ${\Delta{\tau}}$ is inversely proportional to the signal bandwidth ${B}$, or ${\Delta{\tau} = 1/B}$. A SA channel sounder with greater resolution in both the angular and delay domains can be constructed by attaching the receive antenna to a precise mechanical positioner such as the robot arm shown in Fig.~\ref{fig:receiveAntennaMountedOnRobot}.

The premise behind a SA channel sounder is that the mechanical positioner moves the receive antenna (also called a probe) to points along a spatial sampling lattice. In the most general sense, the lattice can be arbitrary but in typical cases it is chosen to be planar or cylindrical. At each spatial sample point, the receiver digitizes the antenna output and writes the data to memory. The availability of the digitized receive signal at every spatial sample location allows the SA to emulate the functionality of an element-level digital beamforming (DBF) array.

A conventional heterodyne receiver can be used behind the antenna to detect the signal or a vector network analyzer (VNA). If a VNA is used, then a discrete frequency grid is specified of carrier frequencies. VNA receivers have been investigated in a number of channel sounder configurations. For example, wideband channel measurements using a VNA are discussed in ~\cite{Santella2014,Howard1990}. A cubic SA with a VNA is utilized in ~\cite{Mbugua2018} to estimate LoS and NLoS propagation paths in an indoor environment. A long-range wideband sounder using a VNA was described in ~\cite{Kyro2010} for outdoor measurements.

The VNA-based SA channel sounder (known as SAMURAI) developed at the National Institute of Standards and Technology (NIST) and described in ~\cite{Weiss2020,Loh2021} radiates 1351 sinusoidal tones spaced 10 MHz apart in the range from 26.5 to 40 GHz and measures ${S_{21}}$ parameters at each spatial sample. Thus, the total synthesized measurement bandwidth ${B}$ is 13.5 GHz even though each radiated tone is very narrowband. An advantage of the VNA sounding approach is that the channel is illuminated with a uniform power spectral density since each radiated sinusoidal tone is of equal amplitude. With some channel sounding waveform modulations, such as pseudo-random noise sequences, the shape of the signal spectrum allocates more power to some frequencies than to others.

\begin{figure}%[htb]
% \begin{minipage}[b]{1.0\linewidth}
 \centering
 \includegraphics[trim={0 0 0 25},clip,width=90mm]{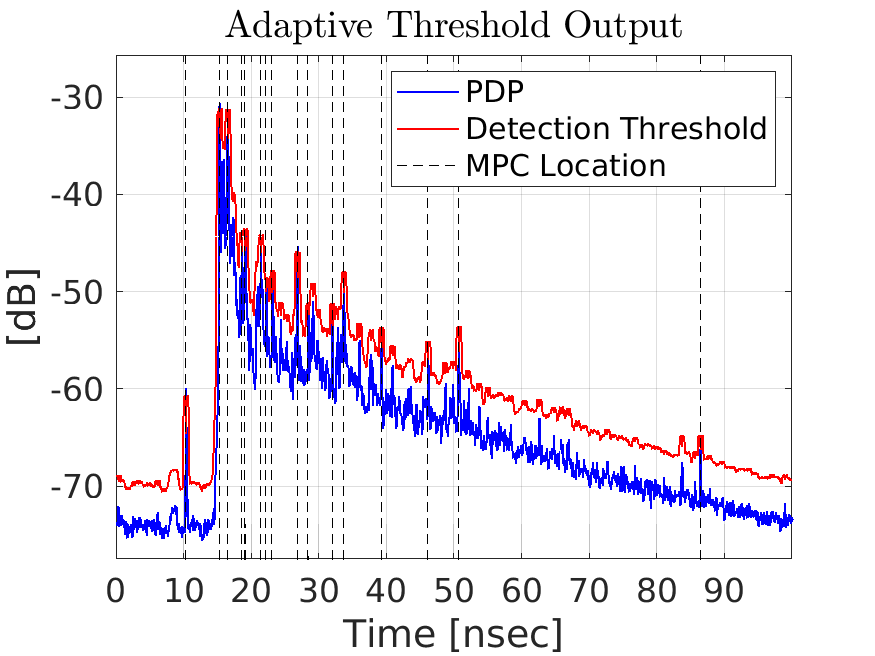}
% \vspace{2.0cm}
 \caption{Directional PDP showing normalized power received (path loss) as a function of delay for a specified look direction.  The long fading time constant of diffuse multipath is clearly evident for this wireless scenario.}%\medskip
 \label{fig:directionalPDP}
% \end{minipage}
\end{figure}

After beamforming the wideband data towards a specified direction, a power delay profile (PDP) is generated that represents the beam's temporal output. The resulting delay resolution of the system is approximately ${1/B}$ or 2.2 cm and this value is much less than the delay resolution available using a typical narrowband channel sounder that radiates an instantaneous signal bandwidth in the range of 1-3\% of the carrier. The total unambiguous delay ${T_{\textrm{dur}}}$ that can be measured by the SA is determined by the frequency step size ${\Delta{f_{\textrm{sa}}}}$ as in ${T_{\textrm{dur}} = 1/\Delta{f_{\textrm{sa}}}}$. An example of a directional PDP is shown in Fig.~\ref{fig:directionalPDP} for sounding data measured in an industrial environment with dense multipath. A frequency varying phase taper was applied across the aperture to steer the beamformed output to the direction ${-0.8^{\circ}}$ azimuth and ${-0.8^{\circ}}$ elevation and an inverse Discrete Fourier Transform (DFT) was used to transform the frequency domain data to the delay domain. The blue curve represents signal path loss (or normalized receive power) as a function of delay and the red curve is an adaptive threshold that rides over the data. Data values that exceed the threshold are multipath detections and are marked using a dashed line.

To avoid any spatial aliasing, the maximum distance between points on the spatial sampling grid of the SA is chosen to be ${\lambda/2}$ at the highest frequency of the measurement bandwidth. If using conventional beamforming or Fourier techniques to coherently combine the received data across the aperture, the nominal angular resolution of the system is equal to the beamwidth, or ${\theta_{B} = \lambda/D}$ for a planar aperture. An important observation is that the size, ${D}$, of a SA can be made almost arbitrarily large. For example, the baseline NIST SA channel sounding configuration is a square 35-by-35 grid of spatial samples that yields a half power beamwidth of ${2.9^{\circ}}$ at 40 GHz. This aperture contains 1225 spatial sample points which would be a challenge to build in hardware. Aside from the benefits of finer angle and delay resolution, SA channel sounders have other desirable features that will be described next.

\subsubsection{Detection Range}
\label{ssec:detect_range}
The Friis equation defined in \cite{Vouras2020} can be used to compute the maximum detection range possible given the transmit and receive antennas of a channel sounder. In a LoS geometry, the Friis equation for detection range is
\begin{equation}
 \label{E:eqn08}
 R^2 = \frac{P_{t}GA_{e}}{4{\pi}S_{\textrm{min}}},
\end{equation}
where ${P_{t}}$ refers to the transmit power, ${R}$ is the distance between antennas, ${G}$ is the power gain of the transmit antenna, ${A_{e}}$ is the effective aperture area on receive, and ${S_{\textrm{min}}}$ is the minimum detectable signal level in the receiver. In a multipath or non-LoS scenario, the Friis equation yields
\begin{equation}
 \label{E:eqn9}
 R_{1}^2R_{2}^2 = \frac{P_{t}GA_{e}\sigma}{(4{\pi})^{2}S_{\textrm{min}}},
\end{equation}
where ${R_{1}}$ denotes the distance from the transmit antenna to an object with backscatter area ${\sigma}$ (in units of ${m^2}$), and ${R_{2}}$ is the distance from the scatterer to the receive aperture. Both equations show that to maximize detection range the effective area of the receive aperture, ${A_{e}}$, should be as big as possible, which is most practical using a SA. With a SA, the number of spatial sample points can be made large and, by coherently combining the signals received across all the samples, ${A_{e}}$ can be maximized subject only to constraints such as the available measurement time or the range of motion of the mechanical positioner.

\subsubsection{Array Coordinate Systems}
\label{ssec:coordinates}
\begin{figure}%[htb]
% \begin{minipage}[b]{1.0\linewidth}
 \centering
 \includegraphics[width=8.5cm]{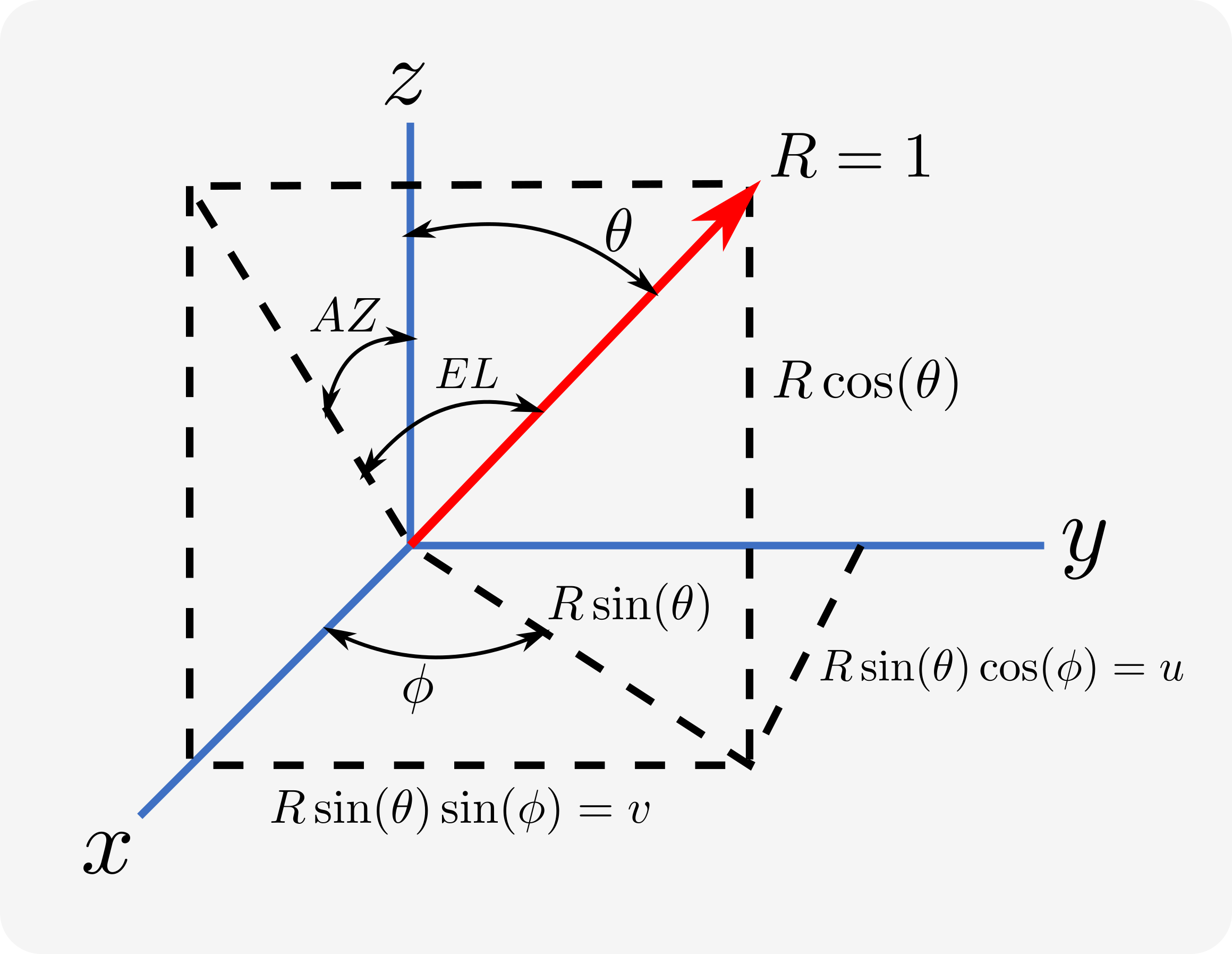}
% \vspace{2.0cm}
 \caption{Array coordinate systems.  The beam pattern is scan-angle  invariant when computed in sine space or ${uv}$-coordinates.}%\medskip
 \label{fig:array_coord}
% \end{minipage}
\end{figure}

Fig.~\ref{fig:array_coord} shows the angles used to describe beam scanning directions for three common array coordinate systems. In this illustration, the array lies in the x-y plane and the z-axis points along the normal to the plane of the array, also known as the boresight direction. In a spherical coordinate system, the angles ${\theta}$ and ${\phi}$ define points on the surface of the forward unit hemisphere. The angle ${\theta}$ is measured from boresight and ${\phi}$ extends to the plane of scan from the x-axis. The projection of points from the forward hemisphere onto the ${xy}$ plane yields the coordinate axes labeled as ${u}$ and ${v}$. The coordinates ${u}$ and ${v}$ can also be used to describe beam directions and are known as sine-space coordinates. The relations used to transform angles between the spherical, azimuth (AZ)/elevation (EL), and sine-space coordinate systems are defined below. 
\begin{align}
 \label{E:eqn10}
 &u = \sin\theta \cos\phi, \quad v = \sin\theta \sin\phi \\
 &u = \cos{EL}\sin{AZ}, \quad v = \sin{EL} \\
 &{\sin}{^2}\theta = u^2 + v^2, \quad \tan\phi = v/u \\
 &\cos\theta = \cos{EL}\cos{AZ}, \quad \tan\phi = \tan{EL}/{\sin{AZ}} \\
 &\tan{AZ} = u/\sqrt{1 - u^2 - v^2}, \quad \sin{EL} = v \\
 &\tan{AZ} = \tan\theta\cos\phi, \quad \sin{EL} = \sin\theta\sin\phi
\end{align}
where ${0 \leq \theta \leq \pi}$ and ${-\pi \leq \phi \leq \pi}$. Note that azimuth angle is defined with respect to the boresight axis and elevation angle is defined with respect to the projection onto the ${xz}$ plane.

\subsubsection{Dynamic Range and Mutual Coupling}
\label{ssec:dyn_range}
In typical hardware phased arrays, an analog beamforming network coherently combines the RF signals received at each array element before downconversion to baseband and digitization. One consequence of this architecture is that the coherent integration gain due to beamforming (equal to ${10\log_{10}N}$, where ${N}$ is the number of array elements) can limit the dynamic range of the system. For example, if a strong signal is incident on an array of 1000 elements then the 30 dB of additional gain at the output of the beamformer may cause the ADC to saturate. With a SA however, the signal received at each spatial sample position is digitized separately and the only additional gain before the ADC is the gain of the receive probe. This architecture maximizes system dynamic range, especially if a low gain probe is used, because there is no coherent integration gain before the digitizer. Low gain antennas are desirable as probes in SAs because the wider antenna beam allows for higher array factor scan angles.

An additional benefit due to the fact that SAs digitize each spatial sample position sequentially is that there is no mutual coupling created between array elements that can affect the overall array beampattern. With hardware arrays, the pattern of each array element is perturbed once the element is embedded in the array structure. For example, the elements in the interior of the array will exhibit a different pattern compared to the elements at the edge of the array, even if all the element patterns are identical outside the array. This phenomenon is caused by currents that re-radiate between the elements. Mutual coupling can create angle estimation and beam pointing errors in hardware arrays but does not exist in SAs.

\subsection{Frequency and Scanning Behavior of the Array Factor}
\label{ssec:freq_behavior}
The far-field response in spherical coordinates ${(\theta,\phi)}$ for an array of ${M \times N}$ homogeneous elements located in the ${xy}$ plane is given by,
\begin{equation}
\label{E:eqn01}
{B}(\theta,\phi) = E(\theta,\phi) \sum\limits_{m=0}^{M-1} \sum\limits_{n=0}^{N-1} w_{mn} e^{\mathrm{j}k(x_{m}\sin\theta\cos\phi + y_{n}\sin\theta\sin\phi)},
\end{equation}
where ${E(\theta,\phi)}$ is the array element pattern, the wavenumber ${k = 2{\pi}/{\lambda}}$, ${\lambda}$ is the operating wavelength, and ${w_{mn}}$ is the array element weighting. If the array elements are uniformly spaced on a rectangular grid then the element locations are given by ${x_{m} = md_{x}}$ and ${y_{n} = nd_{y}}$ where ${d_{x}}$ and ${d_{y}}$ denote the distance between elements in the ${x}$ and ${y}$ directions. This equation can be rewritten as a 2-D spatial Fourier Transform by using the sine space coordinates ${u = \sin\theta\cos\phi}$ and ${v = \sin\theta\sin\phi}$,
\begin{equation}
\label{E:eqn02}
{B}(u, v)= E(u, v) \sum_{m=0}^{M-1}\sum_{n=0}^{N-1}w_{mn}e^{\mathrm{j}k(md_{x}u + nd_{y}v)}.
\end{equation}

The summation term is known as the array factor. The array factor is periodic in the ${u}$ dimension with period ${\lambda/d_{x}}$ and repeats in the ${v}$ dimension with period ${\lambda/d_{y}}$. A single period in ${uv}$ space of the array factor is equal to the rectangular region ${-0.5\lambda/d_{x} \leq u < 0.5\lambda/d_{x}}$ and ${-0.5\lambda/d_{y} \leq v < 0.5\lambda/d_{y}}$. The visible region of the array factor that exists in physical space corresponds to the interior of the unit circle ${u^2 + v^2 \leq 1}$. Replicas of the mainlobe outside the unit circle are known as grating lobes. The Nyquist spatial sampling rate that avoids spatial aliasing or grating lobes is given by ${d_{x} = d_{y} = \lambda/2}$. If the element spacing is greater than ${\lambda/2}$ then the array is undersampled and the grating lobes move closer to the unit circle and may even enter the visible region. If the element spacing is less than ${\lambda/2}$ then the array is oversampled and the grating lobes move farther away from the unit circle. The peak of the mainbeam depends only on the number of array elements and is equal to ${10 \log_{10}(MN)}$.

All of the properties of Fourier Transforms apply to the array factor and in particular the Fourier shift and the Fourier scaling properties. The Fourier shift property is exercised when the main beam is steered to a direction (${u_0,v_0}$). Beam steering is accomplished by applying the linear phase taper ${e^{-\mathrm{j}k(md_{x}u_{0} + nd_{y}v_{0})}}$ across the aperture. The main beam shifts by an angular distance equal to the slope of the linear phase taper in the ${u}$ and ${v}$ directions. Steering the main beam for a single frequency is a linear transformation that does not affect the amplitude of the array factor or the shape of the main beam. When the beam scans, sidelobes that were outside the unit circle in the invisible region of ${uv}$ space will enter the visible region and therefore the sidelobe structure of the beampattern changes.

In wideband regimes where the linear phase taper is computed for a single center frequency but applied to the aperture over a wider bandwidth, then the main beam will squint, or point to different directions, as the operating frequency changes. Beam squint is an undesired effect that can be mitigated by using true time delay beam steering or a frequency invariant beamformer \cite{Frigyes1995}. True time delay beam steering applies a frequency independent time delay or a frequency dependent linear phase shift between array elements to steer the beam and can be easily implemented on SAs.

\begin{figure}[t]
% \begin{minipage}[b]{1.0\linewidth}
 \centering
 \includegraphics[trim={0 0 0 25},clip,width=8.5cm]{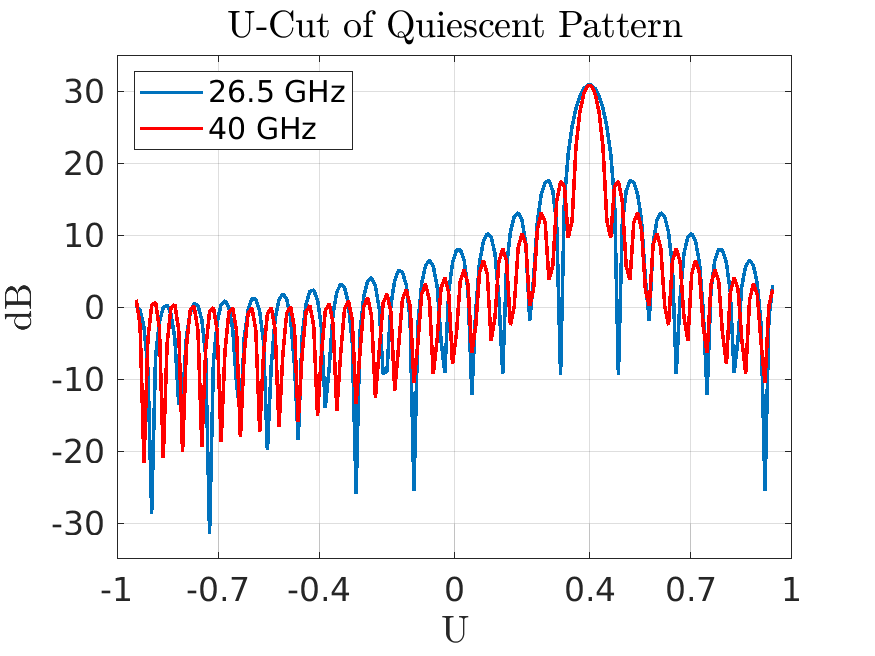}
% \vspace{2.0cm}
 \caption{U-Cut of beampattern illustrating Fourier scaling property.  Since array element spacing is fixed, the width of angular lobes decreases as the frequency increases because array becomes electrically larger.}%\medskip
 \label{fig:U_cut}
% \end{minipage}
\end{figure}

The Fourier scaling property states that changing the sample spacing of a 2D discrete sequence ${h(m,n)}$ expands or contracts the output of the Fourier transform ${H(u,v)}$ according to
\begin{equation}
%\label{E:eqn03}
H\left(\frac{u}{a},\frac{v}{b}\right) \iff \frac{1}{|ab|}h(am,bn).
\end{equation}
This property implies that if the physical spacing between array elements is held fixed while the operating frequency decreases, then the width of every angular lobe in the beampattern (main beam and sidelobes alike) will increase. Conversely, if the element spacing is held fixed while the frequency increases, then the width of every angular lobe decreases. Consequently, a SA of fixed dimensions attains higher angular resolution at 40 GHz than at 26.5 GHz due to the narrower width of the main beam (approximately equal to ${\lambda/D}$ radians, where ${D}$ is the largest dimension of the aperture in the principal planes). Fig.~\ref{fig:U_cut} illustrates the Fourier scaling property by comparing a ${u}$-dimension cut of the array beampattern for 26.5 and 40 GHz with the main beam steered to ${(u=0.4,v=0.3)}$. In channel sounding applications it is important to maintain a frequency invariant array response such that it does not obfuscate the estimated channel frequency response. 

\subsubsection{Beam-Squint in Wideband Arrays}
\label{ssec:beam_squint}
As described by (\ref{E:eqn01}) or (\ref{E:eqn02}) forming the coherent sum of the signals collected across the SA forms a directional beam in space. This beam may be steered to different directions by applying the appropriate phase shift between successive array elements. To steer the beam in the direction ${(\theta,\phi)}$ the phase shift applied at the ${mn}$th array element is given by 
\begin{equation}
\label{E:eqn3}
\psi_{mn}(\theta,\phi) = \frac{2\pi}{\lambda}(x_{m}\sin\theta\cos\phi + y_{n}\sin\theta\sin\phi),
\end{equation}
where ${x_{m}}$ denotes the x-coordinate of the element's location and ${y_{m}}$ denotes the y-coordinate. For the case of a SA with a VNA acting as the signal receiver the exact phase shift required can be applied in the post-processing since the received signal is a single monochromatic sinusoidal tone for each measurement frequency. If frequency is substituted into (\ref{E:eqn3}) rather than wavelength then it is clear that the required steering phase at any array element varies linearly with frequency \cite{Frigyes1995}.
\begin{align}
\label{E:eqn11}
\psi_{mn}(\theta,\phi) &= \frac{2\pi{f}}{c}(x_{m}\sin\theta\cos\phi + y_{n}\sin\theta\sin\phi) \\
&= \frac{\omega}{c}(x_{m}\sin\theta\cos\phi + y_{n}\sin\theta\sin\phi).
\end{align}

\begin{figure}[t]
 \centering{\includegraphics[trim={0 0 0 20},clip,width=85mm]{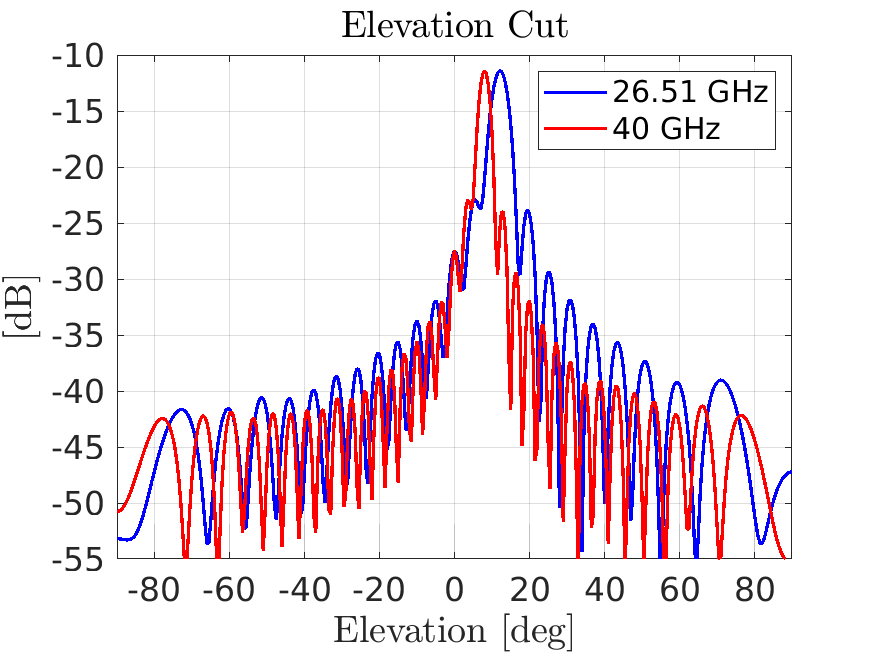}}
 \caption{Beampattern elevation cut showing beam squint.  The beam-steering phase taper computed for 26.51 GHz is also applied at 40 GHz and as a result the beam pointing direction changes.}
 \label{fig:el_cut}
\end{figure}

If the steering phase is computed for a single frequency only and the frequency of the received signal is allowed to vary without adjusting the steering phase accordingly, then the beam squints, or points in a slightly different direction for each frequency. Fig.~\ref{fig:el_cut} illustrates the effect of beam squint using measured NIST data collected in a conference room setting. In this case, a steering vector is computed for 26.51 GHz and applied to the data measured at 40 GHz. The elevation cut shown in the plot illustrates how the beam squints, or changes pointing direction because the computed steering phase is not matched to the frequency of the received signal.  Beam squint can have an impact in millimeter-wave systems as described in \cite{Wang2019TSP}.

To implement true time-delay beamforming, if the phase shift ${\psi_{mn}(\theta_{0},\phi_{0})}$ is computed exactly for an initial frequency ${\omega_{0}}$ then at all other frequencies a differential phase shift proportional to the frequency difference should be applied. The steering phase versus frequency is then given by
\begin{align}
\label{E:eqn12}
\psi_{mn}(\omega;\theta_{0},\phi_{0}) &= \psi_{mn}(\omega_{0};\theta_{0},\phi_{0})[1 + (\omega - \omega_{0})/\omega_{0}] \\
&= \psi_{mn}(\omega_{0};\theta_{0},\phi_{0})\omega/\omega_{0}.
\end{align}
The slope of the linear phase ramp in (\ref{E:eqn12}) corresponds to a time delay of
\begin{align}
\label{E:eqn13}
\tau &= \frac{d{\psi}}{d{\omega}} = \frac{\psi_{mn}(\omega_{0};\theta_{0},\phi_{0})}{\omega_{0}} \\
&= \frac{1}{c}(x_{m}\sin\theta_{0}\cos\phi_{0} + y_{n}\sin\theta_{0}\sin\phi_{0}).
\end{align}
The wideband processing algorithm described in the next section eliminates beam squint by applying a phase shift proportional to frequency, or equivalently a time delay, at each array element to steer the beam.

\subsubsection{Wideband Power Angle Delay Profile (PADP)}
\label{ssec:wideband_padp}
True time delay beam steering refers to the practice of inserting a pure time delay instead of a phase shift behind each array element to steer the beam in wideband arrays. True time delay beam steering can be implemented on wideband SAs to avoid beam squint by applying a frequency dependent phase taper to the array output vector as described in (\ref{E:eqn12}). After computing the dot product of the beam steering phase taper and the array output vector for every frequency, an Inverse Fourier Transform is computed to yield the beam output power received from the direction ${(\theta_{0},\phi_{0})}$ as a function of delay. This beam output is also known as the PDP for the direction ${(\theta_{0},\phi_{0})}$. The process is summarized in Algorithm \ref{alg:alg_1} below. Note that a frequency invariant beamformer may replace the phase steering vectors ${\mathbf{w}(\omega_{k};u_{0},v_{0})}$ with optimized weight vectors computed at every frequency for the desired beam-steering direction.
%\noindent\begin{minipage}{1.0\columnwidth}
\begin{algorithm}%[H]
\label{alg:alg_1}
\caption{PADP and delay slice creation}\label{PADP}
\begin{algorithmic}[1]
\Statex \textbf{Input:} Array output vector ${\mathbf{y}(\omega_{k})}$ at each frequency ${\omega_{k}}$ for ${k = 0,\ldots,S-1}$ and desired beam pointing direction ${(u_{0},v_{0})}$
\Statex \textbf{Outputs:} PDP ${x(\tau_k;u_{0},v_{0})}$ in the fixed direction ${(u_0,v_0)}$ and delay slice ${x(\tau_{0};u,v)}$ at the fixed delay ${\tau_0}$. %For a fixed delay ${\tau = \tau_{0}}$, ${x(\tau_{0};u,v)}$ is the spatial frequency spectrum of all signal sources impinging on the array (also called a delay slice) and can be used to estimate angles of arrival
\State Compute the phase steering vector ${\mathbf{w}(\omega_{k};u_{0},v_{0})}$ at each frequency
\State \textit{(Beamforming)} %the array output vector ${\mathbf{y}(\omega_{k})}$ at each frequency by forming the dot product 
${b(\omega_{k};u_{0},v_{0}) \gets \mathbf{w}(\omega_{k};u_{0},v_{0})^{H}\mathbf{y}(\omega_{k})}$ 
\State Compute a window function ${c_k}$ of length ${S}$ with low sidelobes $\triangleright$ This reduces high-frequency time-domain ripple in wide bandwidth measurements and increases sampling resolution; Hamming window is commonly used
\State Zero-pad the sequence ${c_{k}b(\omega_{k};u_{0},v_{0})}$ to ${L}$ times its original length %before computing the IDFT
\State %Compute the Inverse Fourier Transform (temporal) to obtain the beam output 
\textit{(Directional PDP)} ${x(\tau_{k};u_{0},v_{0}) \gets IDFT[b(\omega_{k};u_{0},v_{0})]}$
\State \Return  $x(\tau_{k};u_{0},v_{0})$
\end{algorithmic}
\end{algorithm}
%\end{minipage}
\vspace{4pt}

\subsubsection{Spatial Wideband Effect}
In large phased arrays with many elements, the propagation time of an impinging electromagnetic wave travelling across the aperture is non-negligible.  More precisely, if the distance between elements on opposite corners of the array is large compared to the carrier wavelength, then there will be noticeable offsets between the signal arrival times across the array elements.  When all the array element signals are coherently combined during the beamforming operation, the effect of the non-uniform delay offsets will be to limit the instantaneous bandwidth of the array.

For a uniform linear array of length ${L}$, the time ${\tau}$ required to fill the aperture with energy for radiation arriving from an angle ${\theta_0}$ is given by
\begin{equation}
    \tau = \frac{L}{c}\sin{\theta_{0}}.
\end{equation}
For a pulsed waveform, as the array mainbeam is scanned away from boresight, each spectral component is steered to a slightly different direction.  To determine the overall effect on antenna gain, it is necessary to add the far-field patterns of all the individual spectral components.  The result is that a loss of 0.8 dB in energy on target occurs due to frequency-scanned spectral components when the mainbeam is scanned to an angle of ${60^{\circ}}$ and the pulse width is equal to the array fill time; or equivalently the signal bandwidth is equal to ${1/{\tau}}$ \cite{Skolnik2008}.

Thus a large hardware array that achieves high angular resolution will be necessarily bandwidth-limited and not capable of supporting delay resolutions less than the array fill time. With VNA-based synthetic apertures however, high angle and delay resolutions are simultaneously compatible since the VNA inherently measures ${S}$-parameters (signal ratios) at each spatial location and the beamforming operation is carried out in post-processing.  A tutorial description of the spatial wideband effect is provided in \cite{Wang2019Mag}.

\subsection{Delay Resolution and Maximum Unambiguous Delay}
As previously mentioned, the SAMURAI synthetic aperture system uses a VNA as the signal transmitter and receiver.  In this configuration, the transmit antenna is connected through a coaxial cable to port 1 of the VNA and the receive probe located some distance away is connected to port 2.  The VNA signal source launches a sinusoidal tone towards port 1.  Some of the incident energy will reflect off port 1 and travel back to the signal source, although most of the energy will radiate through the transmit antenna and propagate into the environment. Some of the radiated energy will arrive at the receive probe antenna, where it is either absorbed in a load, or is reflected back into the environment. The transmission parameter, ${S_{21}}$, is used in the computations described in this paper to represent the complex transfer function between the transmit and receive antennas at the radiated frequency.

In the SAMURAI SA configuration, the VNA measures ${S_{21}}$ in ${\Delta{f} = 10}$ MHz increments between 26.5 and 40 GHz.  The bandwidth ${B}$ of 13.5 GHz determines the delay resolution ${\Delta\tau}$ as in~\cite{Porat1997},
\begin{align}
\label{E:BW}
    &\Delta\tau = \frac{1}{B} = 0.074 \text{ nsec (or 2.2 cm)}.
\end{align}

 Theoretically, the inverse Fourier transform ${y(t)}$ of the frequency domain channel measurements over all frequencies may be time aliased.  However, in the event that the channel impulse response ${h(t)}$ is time limited to a duration no greater than ${1/\Delta{f}}$, then the shifted replicas ${h(t - 2\pi{n}/\Delta\omega)}$ do not overlap and ${y(t)}$ is not time aliased.  Thus, the maximum unambiguous delay ${T}$ that can be measured using frequency domain sampling is equal to one over the frequency step size, or ${T = 1/\Delta{f} = 1/10 \text{ MHz} = 100}$ nsec (29.96 meters).

In the case of SAMURAI, the channel frequency response is sampled over a band-limited range, and not over all frequencies.  The default frequency step size of 10 MHz provides 1351 samples over the 13.5 GHz measurement bandwidth.  After an inverse Fourier Transform, there will be 1351 delay samples between 0 and ${T=100}$ nsec which effectively results in a sample rate of ${f_{sam} = 13.5}$ GHz, which is less than the highest frequency component at 40 GHz.  Thus, it seems that the power delay profiles (PDPs) computed using SAMURAI may still be time aliased.  For band-limited signals, however, sampling at a rate ${f_{sam}}$ lower than Nyquist does not create aliasing if,
\begin{equation}
\label{E:eqn03}
f_{\text{sam}} \geq qB \text{ where } 1 \leq q \leq \bigg\lfloor \frac{f_{\text{max}}}{B} \bigg\rfloor.
\end{equation}

Sampling the PDPs at a temporal rate equal to the measurement bandwidth, or ${f_{sam} = B = 13.5}$ GHz, satisfies the bandpass sampling constraints given in (\ref{E:eqn03}) since the ratio ${40/13.5 = 2.96 \approx 3}$ is nearly an integer.  Thus, any temporal aliasing in the measured PDPs will be negligible.

With VNA receivers, the dynamic range of a single measurement will depend on the intermediate frequency (IF) bandwidth setting and the amount of sweep-to-sweep averaging performed.  The SAMURAI configuration described here uses a 100 Hz IF bandwidth and no averaging applied to the data.  The resulting dynamic range calculated over the measurement bandwidth is close to 90 dB~\cite{Keysight}. The maximum specified VNA dynamic range of 120 dB corresponds to a 10 Hz IF bandwidth with no averaging.

\subsection{Delay Slices}
\begin{figure}%[htb]
% \begin{minipage}[b]{1.0\linewidth}
 \centering
 \includegraphics[width=8.5cm]{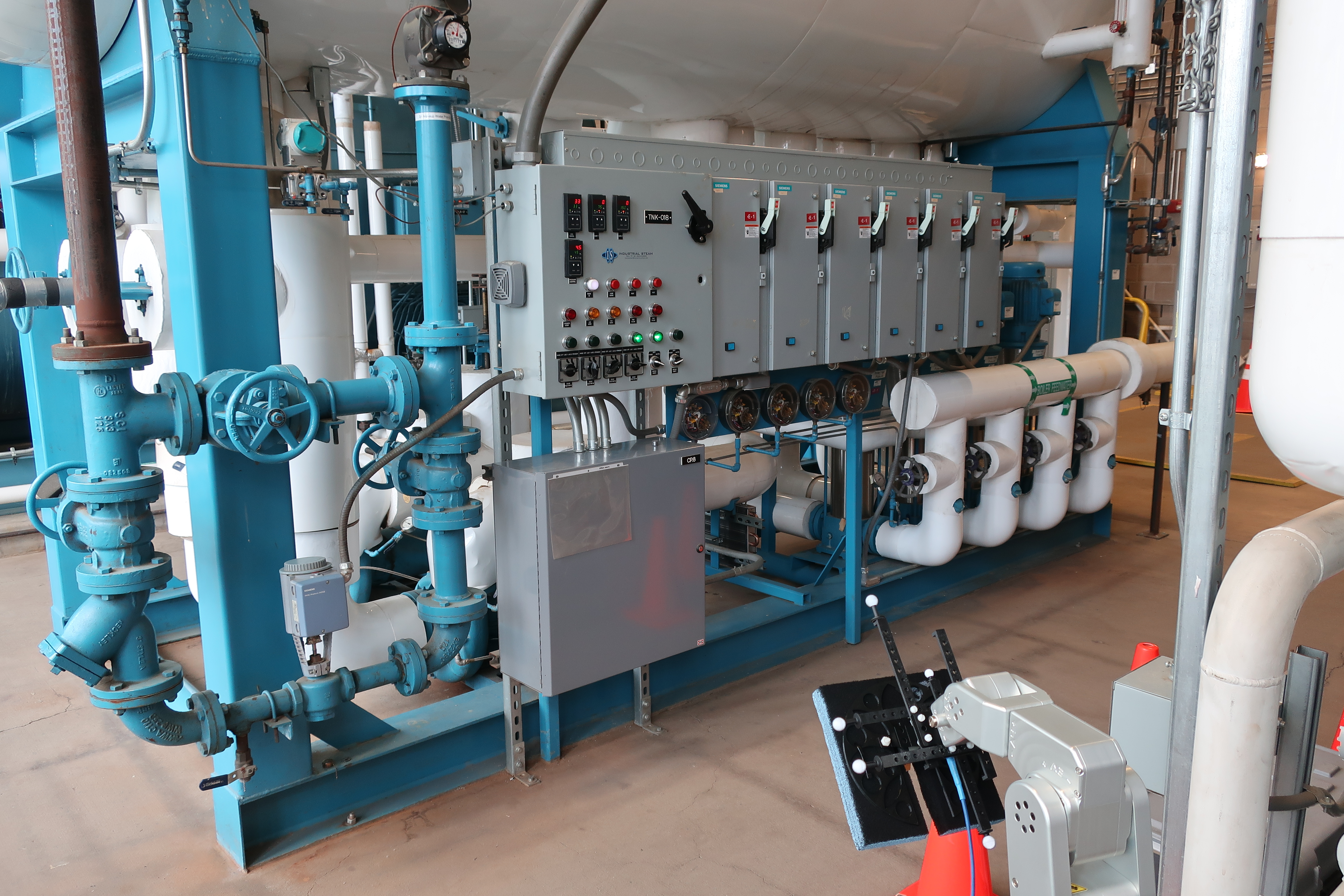}
% \vspace{2.0cm}
 \caption{NIST central utility plant (CUP).  Channel sounding experiments in this location were conducted to characterize dense multipath scattering environments.  Photo credit: NIST.}%\medskip
 \label{fig:CUP}
% \end{minipage}
\end{figure}

\begin{figure}[t]
% \begin{minipage}[b]{1.0\linewidth}
 \centering
 \includegraphics[trim={0 0 0 20},clip,width=9.5cm]{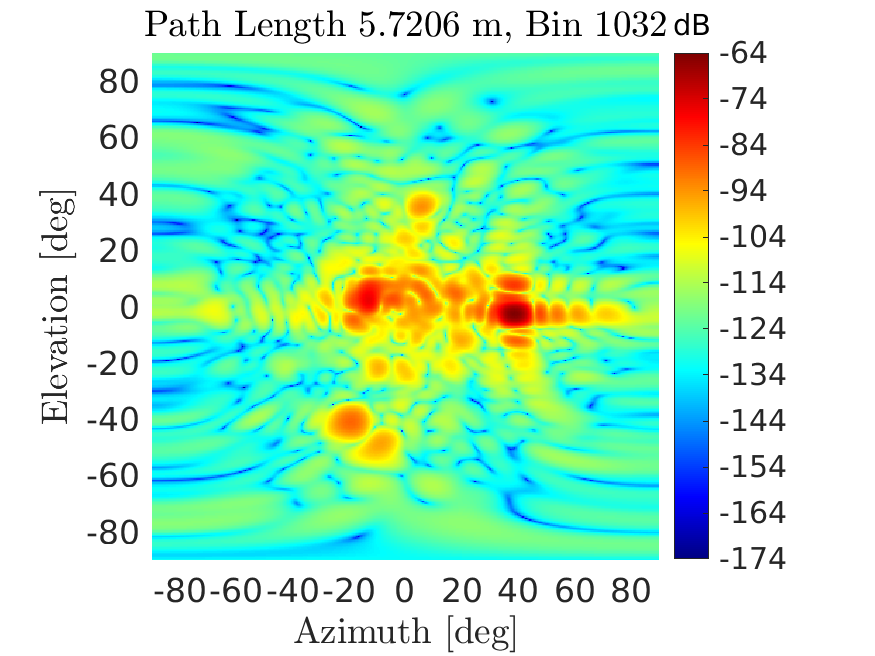}
% \vspace{2.0cm}
 \caption{Delay slice (dB) at $19.08$ ns.  A strong signal source is visible at ${40^{\circ}}$ azimuth as well as lots of diffuse scattering.}%\medskip
 \label{fig:azel_delay_slice_1032}
% \end{minipage}
\end{figure}

\begin{figure}[t]
% \begin{minipage}[b]{1.0\linewidth}
 \centering
 \includegraphics[trim={0 0 0 20},clip,width=9.5cm]{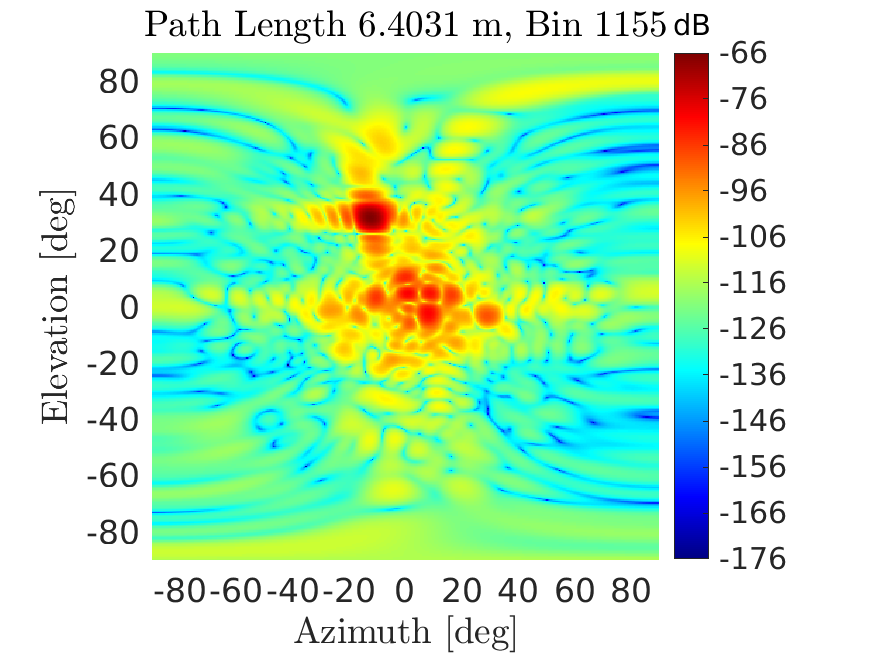}
% \vspace{2.0cm}
 \caption{Delay slice (dB) at $21.36$ ns.  A strong signal source is clearly visible at approximately ${30^{\circ}}$ elevation.  Many other specular returns are also visible near boresight.}%\medskip
 \label{fig:azel_delay_slice_1155}
% \end{minipage}
\end{figure}

The wideband true-time-delay algorithm can be leveraged to evaluate delay slices of the four-dimensional channel impulse response by computing directional PDPs at directions ${(\theta_{k}, \phi_{k})}$ or ${(u_{k}, v_{k})}$ on a discrete angular grid of ${k = 0, \ldots, K-1}$ angles that encompass the entire forward hemisphere. If these PDPs are evaluated over all the angles at the fixed delay ${\tau = \tau_{m}}$, then ${x(\tau_{m};u,v})$ is the spatial frequency spectrum of all signal sources impinging on the array and can be used to estimate strong angles of arrival for the delay bin ${\tau_{m}}$. The equation for evaluating the Inverse Discrete Fourier Transform of the beam output ${b(f_{s}; u_{k}, v_{k})}$ at only the ${m}$th delay bin ${\tau_{m}}$ is
\begin{equation}
x(\tau_{m}; u_{k}, v_{k}) = \frac{1}{S} \sum_{s=0}^{S-1}b(f_{s}, u_{k}, v_{k})e^{\mathrm{j}2{\pi}ms/S},
\end{equation}
where ${S}$ is the total number of frequency samples ${f_{s}}$ and ${0 \leq m \leq S-1}$. Fig.~\ref{fig:azel_delay_slice_1032} and Fig.~\ref{fig:azel_delay_slice_1155} illustrate delay slices for measurements taken in a utility plant at the NIST Boulder campus. The utility plant environment is shown in Fig.~\ref{fig:CUP}. In these figures, we see the signal received from the utility plant environment as a function of azimuth and elevation for two different time delay values. The relative received power level is given by the color bar in dB. These delay slices show a detailed view of the angular power spectrum created by multipath scattering as a function of time.

Practical experience suggests that the delay slice algorithm is most effective when a candidate set of delay bins to search for multipath components is determined in advance. A particularly useful method for determining candidate delay bins is to integrate (using summation) each available delay slice ${x(\tau_{m}; u,v)}$ over all angles in order to compute the total energy received versus delay. This procedure yields an aggregate power delay profile ${r(\tau_{m})}$ that describes the total energy impinging on the SA from the entire forward hemisphere as a function of delay,
\begin{equation}
r(\tau_{m}) = \sum_{k=0}^{K-1}|x(\tau_{m};u_{k},v_{k})|^2.
\end{equation}
If the summation is taken over a subset of ${(u_{k},v_{k})}$ samples smaller than the entire forward hemisphere, then the total power received versus delay is computed for an angular sector.

\subsection{Frequency-Invariant Array Response}
\label{ssec:wideband_beam}
When the complex field of a propagating wave is incident on a phased array or crosses the observation plane of a SA, the beamformed array response will be angle and frequency dependent. To avoid distorting the estimated wireless channel it is desirable to equalize the array frequency response such that it is constrained along specified directions. Consider the field of a propagating monochromatic wave as a function of position ${\mathbf{x}}$ and time ${t}$ given by, \cite{Scholnik}
\begin{equation}
 U(\mathbf{x}, t) = e^{\mathrm{j}2{\pi}(-\mathbf{k}^{T}\mathbf{x} + ft)},
\end{equation}
where ${\mathbf{k}}$ is the propagation direction (spatial frequency) vector and ${f}$ is temporal frequency. The Helmholtz equation relates the spatial and temporal frequencies by ${f = c{\lvert}{\lvert}\mathbf{k}{\rvert}{\rvert}}$, where ${c}$ is the speed of propagation. The output of an antenna element at position ${\mathbf{x}}$ is the time function
\begin{equation}
 y(t) = G(\mathbf{k},f)U(\mathbf{x},t),
\end{equation}
where ${G(\mathbf{k},f)}$is the complex gain of the element as a function of spatial and temporal frequency. The coherent sum of an array of ${N}$ elements yields
\begin{equation}
 s(t) = \sum_{k=0}^{N-1} G_{k}(\mathbf{k},f)U(\mathbf{x},t).
\end{equation}

Consider a linear array with ${K}$ elements where each element is a length-N FIR filter. Each element is located at ${md_{x}}$ where ${d_{x} = \lambda/2}$ at the highest frequency of interest to avoid grating lobes and ${0 \leq m \leq K-1}$. The filter taps are spaced at ${nT}$ intervals where ${0 \leq n \leq N-1}$. The array pattern is then
\begin{equation}
 H(\mathbf{k}, f) = \sum_{m}\sum_{n}c_{kn}e^{-\mathrm{j}2{\pi}(mk_{x}d_{x}+nfT)}.
\end{equation}
This equation describes the 3D response of a 2D FIR filter and does not depend on ${k_{y}}$ or ${k_z}$ since the array has no extent in the ${y}$, ${z}$ dimensions. The array pattern is periodic in ${k_{x}}$ and ${f}$, with period ${1/T}$ in ${f}$ and period ${l/d_{x}}$ in spatial frequency ${k_{x}}$. The tap spacing ${T}$ is chosen so that ${1/T}$ is larger than the desired instantaneous bandwidth of the array response.

Since (54) is linear in the filter coefficients ${c_{kn}}$ even with arbitrary element locations, many common constraints can be expressed as upper bounds of convex functions of the coefficients. Consequently, the array pattern can be designed using convex optimization tools \cite{Scholnik2007}.

Another frequency invariant array architecture that avoids the use of temporal filtering behind each element is the use of sensor delay lines (SDLs). For planar sampling, SDLs are constructed by creating consecutive sample planes spaced along regular spatial intervals. For cylindrical sampling, SDLs can be constructed using concentric circular lattices. SDLs are particularly straightforward to implement for SA channel sounders that use a robot positioner. For example, if an initial planar lattice lives in the ${xy}$ plane, then additional planar lattices would be created along the ${z}$-axis (boresight) dimension spaced a distance ${\lambda/2}$ or ${\lambda/4}$ apart. An optimized coefficient can be computed for each spatial sample in a 3-D SDL to create the frequency invariant array response \cite{WeiLiu2007, WeiLiu2010}.

It is also possible to generate frequency invariant beampatterns by placing a filter bank behind each array element and partitioning the wideband signal spectrum into sub-bands. Then each sub-band can be optimized independently using narrowband techniques to create the frequency invariant array response as descibed in \cite{LinRadar,Yeiss}. The filter banks used could be as simple as a Discrete Fourier Transform Filter Bank (DFTFB) or more complicated designs including Cosine Modulated Filter Banks (CMFBs). One advantage of this approach is that the sub-band processing can be implemented at a lower sample rate than the digitized signal at the filter bank input.

\begin{figure}%[htb]
% \begin{minipage}[b]{1.0\linewidth}
 \centering
 \includegraphics[trim={0 0 0 20},clip,width=8.5cm]{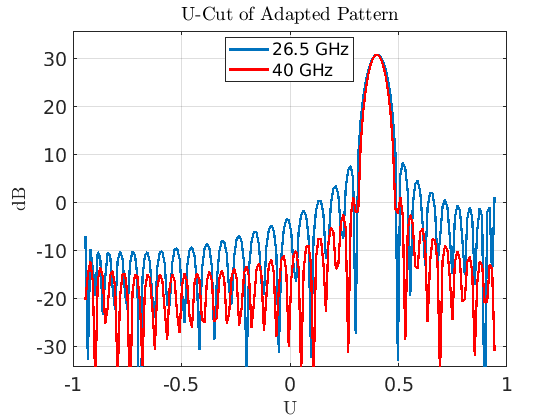}
% \vspace{2.0cm}
 \caption{Frequency invariant beampattern (dB) showing constant beamwidth and reduced sidelobes at $26.5$ and $40$ GHz for fixed array element spacing.}%\medskip
 \label{fig:FIB_U_cuts}
% \end{minipage}
\end{figure}

\begin{figure}%[htb]
% \begin{minipage}[b]{1.0\linewidth}
 \centering
 \includegraphics[trim={0 0 0 18},clip,width=8.5cm]{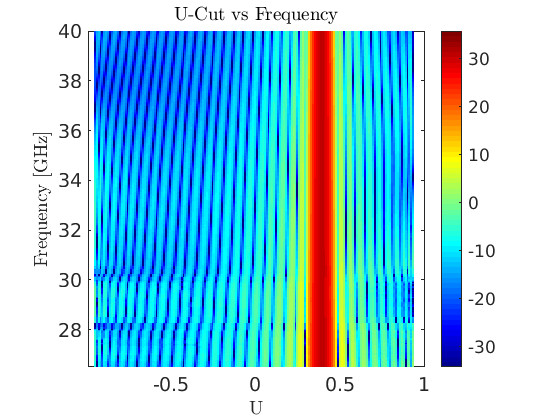}
% \vspace{2.0cm}
 \caption{Frequency invariant beampattern (dB) showing constant beamwidth across entire bandwidth from $26.5$ to $40$ GHz.}%\medskip
 \label{fig:FIB_vs_freq}
% \end{minipage}
\end{figure}

With a frequency domain channel sounder one can design an optimized beamformer for every beam-steering direction at each discrete measurement frequency as described in \cite{Vouras2020}. Fig.~\ref{fig:FIB_U_cuts} illustrates the case for a beamformer that has been designed to reduce ambient sidelobe levels while also maintaining a constant beamwidth over the frequency range from 26.5 to 40 GHz. Fig.~\ref{fig:FIB_vs_freq} shows constant beamwidth maintained over the entire bandwidth. In the absence of any optimized beamforming, the width of the mainbeam would decrease by ${33\%}$ from 26.5 to 40 GHz.

Note that in a conventional narrowband or wideband array design, a prototype array pattern is designed at boresight, and then used to generate beam patterns steered to other directions by applying direction-dependent phase shifts (narrowband) or time delays (wideband) to the signal at each array element. The result is a set of array patterns that at each temporal frequency are spatial-frequency-shifted copies of the prototype pattern. This approach is computationally efficient but not truly optimal in the sense that the beam pattern has not been jointly optimized for beam-steered and frequency response.

\subsection{Sparse Sampling Lattices}
An advantage of using a precision robot to position the receive antenna is that almost arbitrary spatial sampling lattices can be created. The simplest case is to create a planar lattice such as previously described. However, lattices with rotational symmetry, such as circular, cylindrical, or spherical, are also desirable because they offer omnidirectional signal reception. 

\begin{figure}
	\centering
	\includegraphics[trim={0 0 0 20},clip,width=0.5\textwidth]{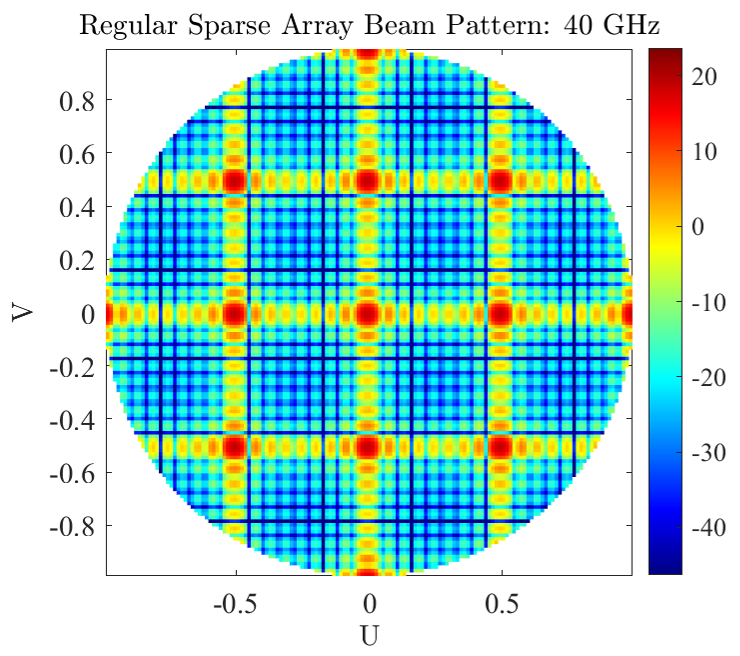}
	\caption{Grating lobes in array pattern (dB) due to a sparse sampling grid.  The grating lobes are essentially copies of the mainbeam and create spatial ambiguities when estimating angles of arrival.}
	\label{fig:grating_lobes}
\end{figure}

\begin{figure}
	\centering
	\includegraphics[trim={0 0 0 32},clip,width=0.5\textwidth]{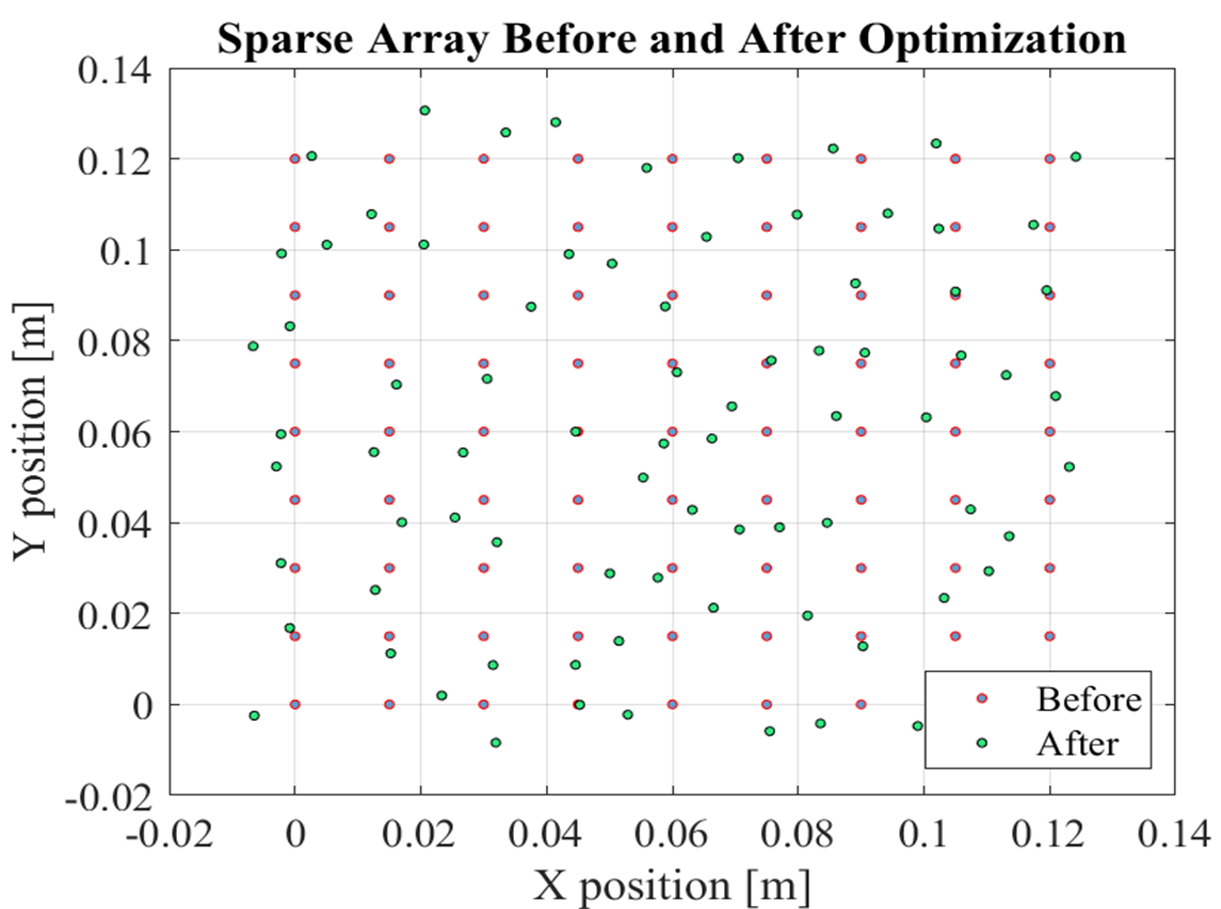}
	\caption{Sparse lattice sampled on a regular grid (red) versus optimized locations of spatial samples (green).}
	\label{fig:opt_array}
\end{figure}

\begin{figure}
	\centering
	\includegraphics[trim={0 0 0 30},clip,width=0.5\textwidth]{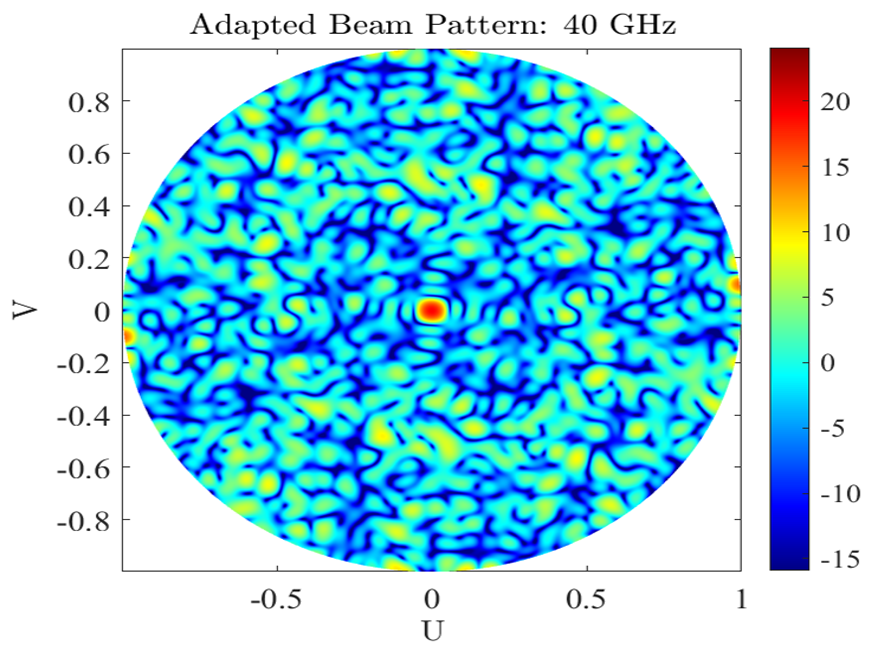}
	\caption{Beampattern corresponding to optimized sparse sampling lattice.}
	\label{fig:optimized_array}
\end{figure}

\begin{figure}
	\centering
	\includegraphics[trim={0 0 0 30},clip,width=0.5\textwidth]{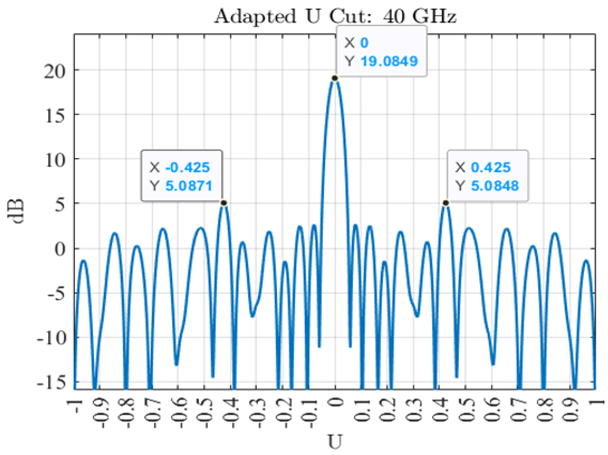}
	\caption{Beampattern U-cut showing grating lobes have been reduced 14 dB below the mainbeam peak.}
	\label{fig:opt_u}
\end{figure}

A drawback to measuring signals with a synthetic aperture over large spatial volumes is that the data acquisition time can be on the order of hours.  Sparse sampling lattices offer the potential to significantly reduce the time required to measure across large sampling lattices, provided an adequate SNR can be maintained.  In all sparse apertures care must be taken to mitigate the level of grating lobes introduced by spatial aliasing.  Grating lobes effectively act as replicas of the mainbeam and create spatial ambiguites as shown in Fig.~\ref{fig:grating_lobes} (model code and data publicly available~\cite{SparseData}).  Grating lobes are supported by any sampling periodicity across the aperture so the simplest approaches to reduce peak grating lobe randomize the locations of spatial samples.  An approach based on optimization is described in \cite{Vouras22} and Fig. \ref{fig:opt_array} illustrates the derived sparse lattice with the beampattern shown in Fig. \ref{fig:optimized_array}.  The beampattern U-cut shown in Fig. \ref{fig:opt_u} confirms that the peak sidelobe level has been reduced to at least 13 dB below the peak of the mainbeam when steered to boresight.

Other approaches investigated for sparse array design include simulated annealing and genetic algorithms. Simulated annealing is a stochastic optimization method analogous to the manner in which a metal cools and anneals \cite{Kirkpatrick1983}. The algorithm seeks to minimize an energy function which for sparse arrays is set proportional to the peak sidelobe level. At each algorithm iteration, the location of array elements is randomized by moving one element at a time. The peak sidelobe level of the perturbed array is found and compared to the best solution of the last iteration. The new solution is accepted if it lowers the peak sidelobe level, or it may also be accepted with some finite probability if it raises the sidelobe level. In this way, the algorithm is less likely to be trapped in a local minimum. As the cost function is progressively minimized, the probability of accepting an inferior solution is reduced and ultimately the algorithm converges to a solution that may be close to optimal, provided the optimization parameters are well chosen. Simulated annealing has been applied to the optimization of sparse lattices in \cite{Murino1996, Hopperstad1998, Trucco1996, Trucco1997}.

Genetic algorithms iteratively operate on the individuals in a population \cite{Holland1992, Goldberg1989}. Each member of the population represents a potential solution to the optimization problem. Initially, the population is randomly generated. The individuals are evaluated by means of a fitness function and then either retained or replaced. New individuals are created through either a cross-over operation or a mutation. Genetic optimization has been applied to the layout of sparse arrays in \cite{Haupt1994, Weber1994, ONeill1994}. 

\subsection{Sparse Fourier Transform Algorithms}
Sparse Fourier Transform (SFT) algorithms have been investigated for several
applications including fast Global Positioning System (GPS) receivers, wide-band spectrum sensing, and multidimensional radar signal processing \cite{Hassanieh2012,Hassanieh2014,Wang2017,Shi2014,Hassanieh2015}. In all SFT algorithms, the reduction of sample and computational
complexity is achieved by reducing the input data samples using a well-designed, randomized subsampling procedure \cite{Wang2019}. The significant frequencies contained in the original data are then localized and the corresponding Discrete Fourier Transform (DFT) coefficients are estimated with low-complexity operations. Iterative subsampling-localization-estimation SFT algorithms are described in \cite{Gilbert2008,Gilbert2014,Pawar2018,Ghazi2013}. Other approaches estimate sparse DFT coefficients in a single pass after obtaining sufficient copies of subsampled signals \cite{Hassanieh2012_2,Wang2017,Shi2014,Hassanieh2015}.

\subsection{Near-Field Beamforming and 3-D Imaging}
In some synthetic aperture applications, the far field assumption can yield suboptimal results after beamforming, especially for large arrays at THz frequencies.  One approach to account for near-field effects is to use spherical beamforming.  Spherical beamforming replaces the constant and delay-agnostic interelement phase shifts shown in (\ref{E:eqn01}) and (\ref{E:eqn02}) with interlement phase shifts that depend on the exact straight-line distance between a point signal source and the receive antenna.  Consider the wave propagation equation given in (\ref{E:wave_sph}) that describes a range-dependent phase shift.  By computing range-dependent phase shifts for a virtual point source along discrete angular locations, a modified steering vector can be defined for near-field beamforming \cite{Vouras22sph}.  The complex dot product of the modified steering vector and the array output vector yields the spherically beamformed output of the array as in,
\begin{equation}
\label{E:B_sph}
{B}_{xyz} = \sum\limits_{m=0}^{M-1} \sum\limits_{n=0}^{N-1} w_{mn} e^{jkD{mn}}.
\end{equation}
Recall that ${k=2{\pi}/{\lambda}}$ and the factor ${w_{mn}}$ refers to the product of the complex weight and the datum sample at the ${mn}$th array element.  For isotropic array elements with unity gain and in the absence of adaptive beamforming, ${w_{mn}}$ is strictly equal to the received signal sample.  Also, ${D_{mn}}$ represents the straight-line distance between a virtual signal source at coordinates ${(s_{x},s_{y},s_{z})}$ and the ${mn}$th array element at ${(x_{m},y_{n},z)}$,
\begin{equation}
\label{E:D_mn}
    D_{mn} = \sqrt{(s_x - x_{m})^2 + (s_y - y_{n})^2 + (s_z - z)^2}.
\end{equation}
The spherical steering vector given in Cartesian coordinates by ${\mathbf{w}(x,y,z)}$ or in spherical coordinates by ${\mathbf{w}(u,v,R)}$ effectively acts as a spatial filter matched to both the angle and distance of the signal source,
\begin{equation}
    \mathbf{w}(x,y,z) = \left[ \begin{array}{ccc} e^{jkD_{00}} \ldots e^{jkD_{MN}} \end{array} \right]^T.
\end{equation}

The pseudocode listed for Algorithm \ref{alg:alg_3}3 provides computational steps for near-field beamforming with a planar synthetic aperture that lies in the ${xy}$ plane with constant ${z}$ coordinates.  To compute spherical steering vectors corresponding to all delays and angles a virtual point source is moved on a circle of constant radius around the receive antenna.  Algorithm \ref{alg:alg_3}3 can be used to create both directional PDPs and delay slices of the channel impulse response.
\begin{algorithm}%[H]
\label{alg:alg_3}
\caption{Spherical Phasefront PADP and Delay Slice Creation}\label{alg:sphPADP}
\begin{algorithmic}[1]
\Statex \textbf{Input:} Array output vector ${\mathbf{y}(\omega_{k})}$ at each frequency ${\omega_{k}}$ for ${k = 0,\ldots,S-1}$ and desired beam pointing direction ${(Az_{0},El_{0})}$ corresponding to ${(u_{0},v_{0})}$.
\State Starting from an initial range ${R_{0}}$ and proceeding to a final range ${R_{1}}$ in increments of ${\Delta{R}}$, compute the Cartesian coordinates ${(\tilde{x}_{k},\tilde{y}_{k},\tilde{z}_{k})}$ corresponding to the spherical coordinates ${(R_{k},u_{0},v_{0})}$.
\State Compute the distance from ${(\tilde{x}_{k},\tilde{y}_{k},\tilde{z}_{k})}$ to each spatial sample in the synthetic aperture.
\State Compute the spherical steering vector, ${\mathbf{w}(\omega_{k};u_{0},v_{0},R_{k})}$, for each frequency.  Each component of the spherical steering vector corresponds to the propagation phase ${e^{jkD{mn}}}$ where ${k=2{\pi}/{\lambda}}$.  Here ${(m,n)}$ denotes the indices of each spatial sample in the synthetic aperture and ${D_{mn} = \sqrt{(\tilde{x}_{k} - x_{m})^2 + (\tilde{y}_{k} - y_{n})^2 + (\tilde{z}_{k} - z)^2}}$.
\State Stack all the frequency-dependent steering vectors into ${\widehat{\mathbf{w}}(\omega;u_{0},v_{0},R_{k})^{H}}$ and all the array output vectors into ${\widehat{\mathbf{y}}(\omega)}$.  Then beamform the wideband array output by forming the dot product ${b(u_{0},v_{0},R_{k}) = \widehat{\mathbf{w}}(\omega;u_{0},v_{0},R_{k})^{H}\widehat{\mathbf{y}}(\omega)}$
\State Repeat steps 1 through 4 for all angles on a discrete grid at a fixed range ${R_{k}}$ to create a delay slice ${x(u,v;R_{k})}$.
\end{algorithmic}
\end{algorithm}
%\end{minipage}
%\vspace{4pt}

\begin{figure}
 %\begin{minipage}[b]{1.0\linewidth}
  \centering
  \includegraphics[width=7cm]{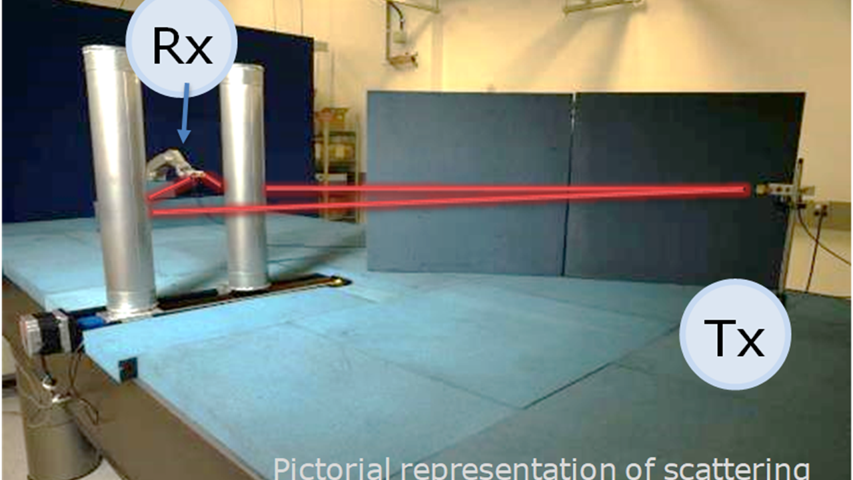}
%  \vspace{2.0cm}
  \caption{Scattering experiment using 2 cylinders.}%\medskip
\label{fig:2_cylinders}
\end{figure}

\begin{figure}
% \begin{minipage}[b]{1.0\linewidth}
  \centering
  \vspace{0.2cm}
  \includegraphics[width=7cm]{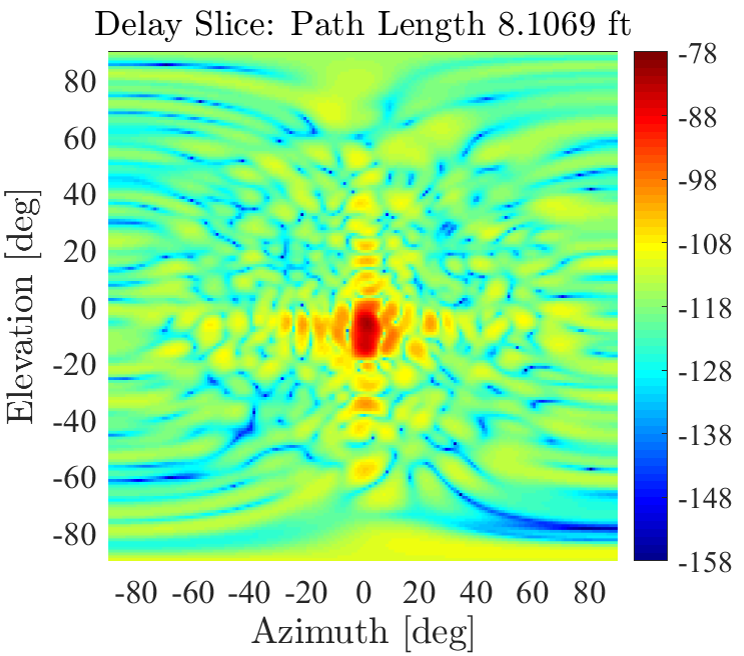}
  \caption{Wideband delay slice of measured cylinder constructed using Algorithm 2.}%\medskip
\label{fig:meas_delay_slice}
%\end{minipage}
\end{figure}

\begin{figure}
% \begin{minipage}[b]{1.0\linewidth}
  \centering
  \vspace{0.2cm}
  \includegraphics[width=7cm]{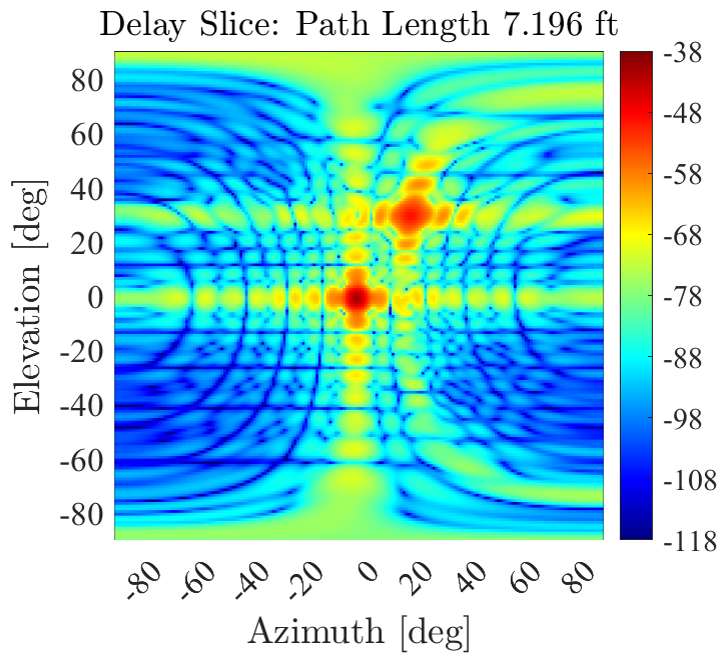}
  \caption{Simulated delay slice using spherical beamforming showing 2 signal sources at the same delay.}%\medskip
\label{fig:sim_delay_slice}
%\end{minipage}
\end{figure}

\begin{figure}
% \begin{minipage}[b]{1.0\linewidth}
  \centering
  \vspace{0.2cm}
  \includegraphics[width=7cm]{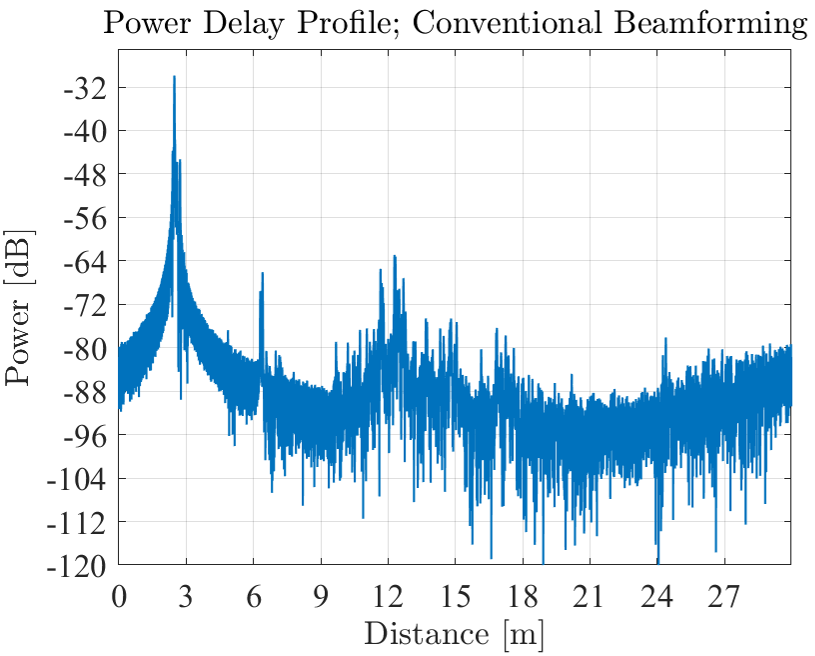}
%  \vspace{2.0cm}
  \caption{Conventional wideband PDP in direction of measured cylinder.}%\medskip
\label{fig:PDP_conv}
%\end{minipage}
\end{figure}

\begin{figure}
% \begin{minipage}[b]{1.0\linewidth}
  \centering
  \vspace{0.2cm}
  \includegraphics[width=7cm]{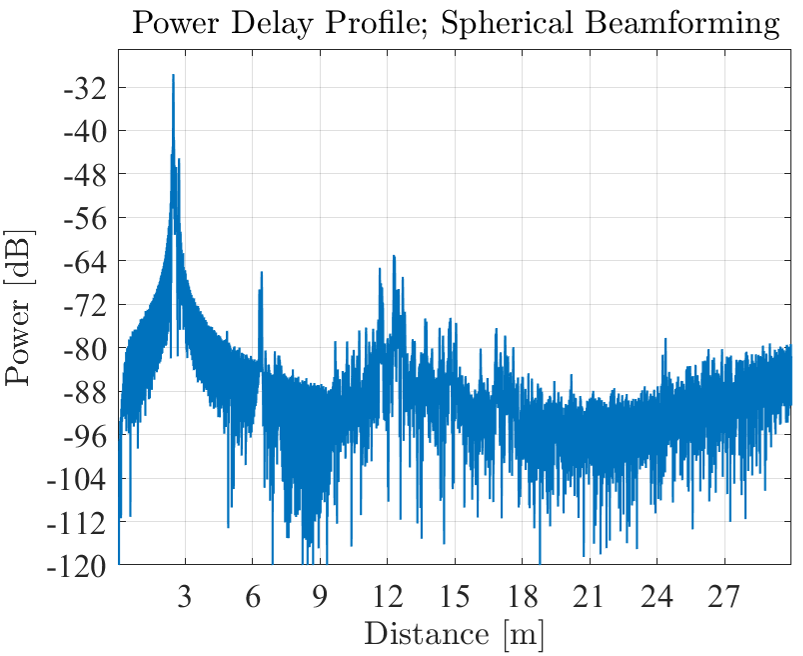}
%  \vspace{2.0cm}
  \caption{Spherical PDP in direction of measured cylinder.}%\medskip
\label{fig:PDP_sph}
%\end{minipage}
\end{figure}

\begin{figure}
% \begin{minipage}[b]{1.0\linewidth}
  \centering
  \vspace{0.2cm}
  \includegraphics[width=7cm]{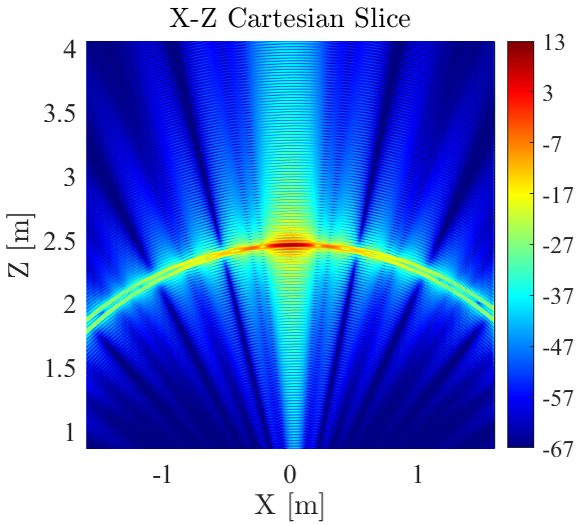}
%  \vspace{2.0cm}
  \caption{Simulated X-Z Cartesian slice showing sidelobes on a curved sinc function.}%\medskip
\label{fig:XZ_sim_BIG}
%\end{minipage}
\end{figure}

\begin{figure}
 \begin{minipage}[b]{1.0\linewidth}
  \centering
  \vspace{0.2cm}
  \includegraphics[width=7cm]{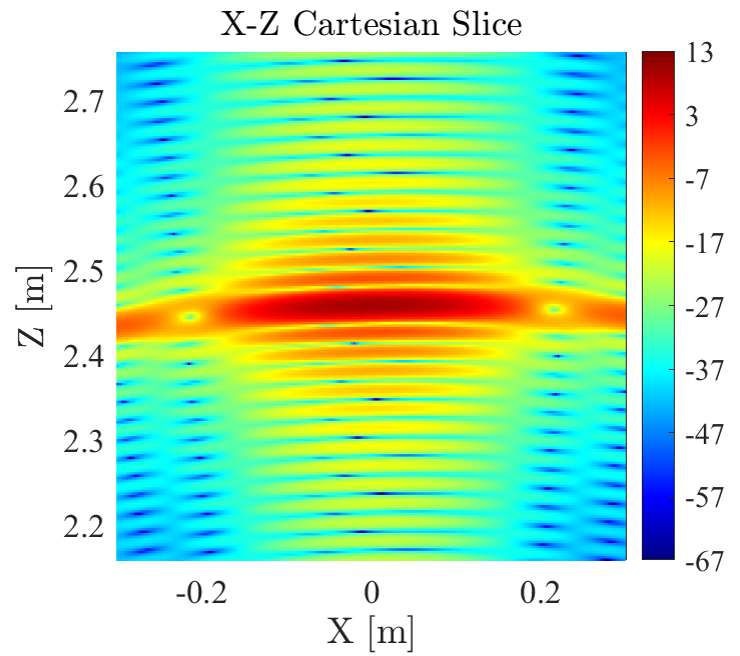}
%  \vspace{2.0cm}
  \caption{Simulated X-Z Cartesian slice; close-up view.}%\medskip
\label{fig:XZ_sim}
%\end{figure}
%\begin{figure}
% \begin{minipage}[b]{1.0\linewidth}
  \centering
  \vspace{0.2cm}
  \includegraphics[width=7cm]{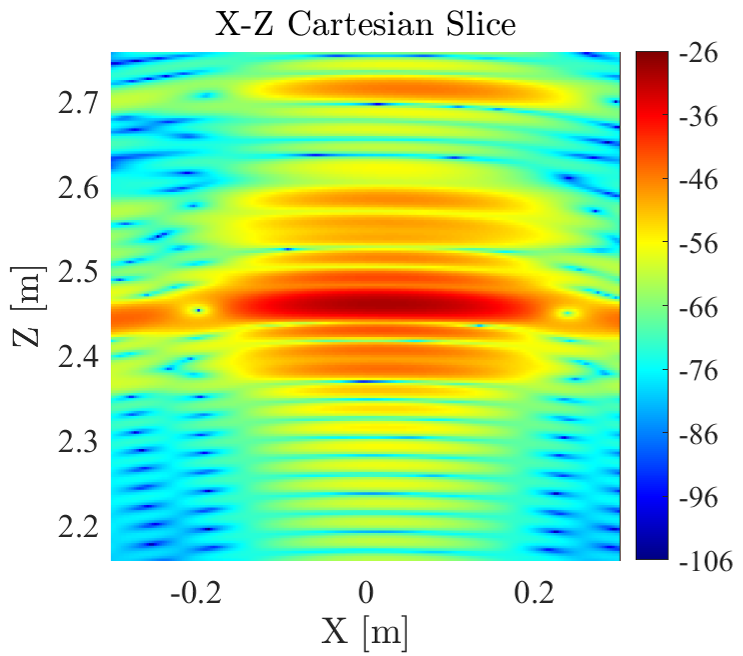}
%  \vspace{2.0cm}
  \caption{Measured X-Z Cartesian slice corresponding to top-down view of boresight cylinder; close-up look.}%\medskip
\label{fig:XZ_meas}
\end{minipage}
\end{figure}

Simulation and empirical results can be used to compare the differences between far-field plane-wave and near-field spherical beamforming.  In an experiment conducted at NIST, two aluminum cylinders were placed on an optical table and illuminated by a WR-28 horn antenna.  The back scattered energy was measured using a synthetic aperture as shown in Fig. \ref{fig:2_cylinders}.  A 35-by-35 planar sampling lattice with 3 mm spacing between points was used to create the synthetic aperture.  Fig. \ref{fig:meas_delay_slice} is a delay slice generated from measured data using Algorithm \ref{alg:alg_3}3 that clearly shows the boresight cylinder.  Fig. \ref{fig:sim_delay_slice} is a simulated delay slice for two cylinders at the same range with different angles of arrival.  Fig. \ref{fig:PDP_conv} is a conventional PDP computed using plane-wave beamforming and Algorithm \ref{alg:alg_1} while Fig. \ref{fig:PDP_sph} is a PDP computed using spherical beamforming and Algorithm \ref{alg:alg_3}3.  The PDPs are similar but differences are evident at the early delays corresponding to the near field region.  Note that the nominal far field of the aperture corresponding to ${2D^{2}/{\lambda}}$ begins at 2.77 m.  Algorithm \ref{alg:alg_3}3 can also be used to display the array response in ${(x,y,z)}$ spatial coordinates.  For instance, Fig. \ref{fig:XZ_sim_BIG} depicts the simulated power spectrum in the X-Z plane due to an isolated scatterer at boresight.  Note the sidelobe structure corresponds to a curved sinc function.  Fig. \ref{fig:XZ_sim} is a closer view in the vicinity of the simulated scatterer and Fig. \ref{fig:XZ_meas} is a top-down view in the X-Z Cartesian plane of the fields in the vicinity of the measured boresight cylinder.

\subsection{Optimized SA Probes}
With SA channel sounders, the receive antenna (also called a probe) mounted on a mechanical positioner can be optimized for wideband performance. For example, vector antennas are capable of measuring all six components of an impinging electromagnetic wave (i.e., ${E_{x}, E_{y}, E_{z}, H_{x}, H_{y}, H_{z}}$) in a Cartesian coordinate system. Recent results in \cite{Duplouy2019} describe a dual-polarized vector antenna with 7:1 bandwidth (or $1$-$7$ GHz). Due to its size, such an antenna would be difficult to embed into a hardware phased array, but it could serve as the receive probe for a SA. Quantum sensors that measure electric fields are also being investigated for use as probe antennas in SA systems, as described next.

\begin{figure}[t]
	\centering
	\includegraphics[width=1\linewidth]{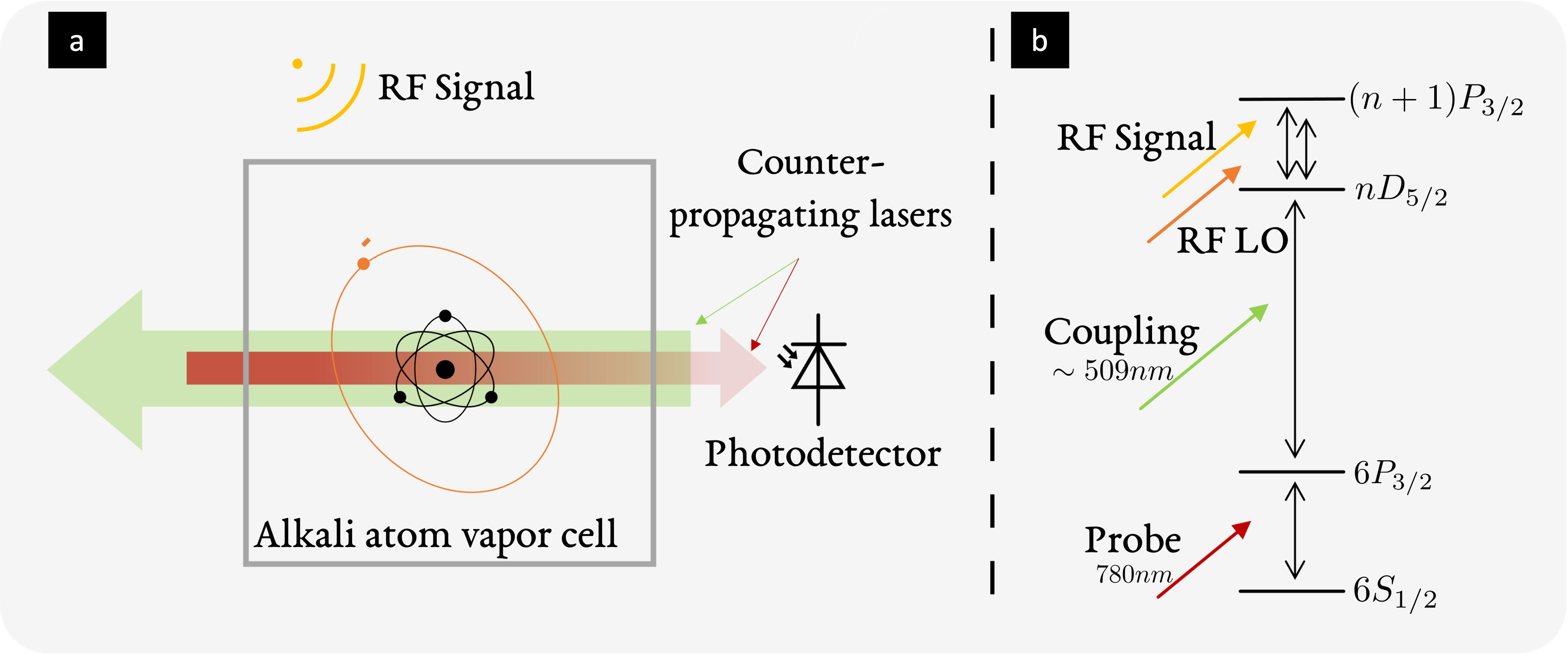}
	\caption{(a) Depiction of a Rydberg atom sensor receiving over the air RF signals. (b) Basic ladder diagram showing the excited states (in cesium, e.g.) coupled by two counter-propagating lasers (probe and coupling) - reaching the first high $n$ Rydberg state - and the RF signal coupling two Rydberg states. When an over the air RF local oscillator (LO) is on, the atoms act like a mixer where the signal output of the photodetector is the down-converted signal at the intermediate frequency between the RF signal and LO.
}
	\label{fig:Rydberg}
\end{figure}

Researchers at National Metrology Institutes~\cite{Holloway2014,Aksyuk2021,Holloway2021_patent}, academic labs~\cite{Sedlacek2012,Adams2019}, and in industry~\cite{RydbergTech2015,Amarloo2021_1,Amarloo2021_2,Amarloo2021_3,Anderson2021,Anderson2021_patent,Coldquanta2021,Darpa2021} are developing a new type of wideband receiver that detects the atomic response of room temperature vapor in the presence of a RF electric field. These quantum receivers use lasers to excite alkali atoms to a high principal quantum number, known as a Rydberg state, where the valence electron is weakly bound to the nucleus and, therefore, is highly sensitive to perturbations from an incident RF electric field; see Fig.~\ref{fig:Rydberg}. The operating frequency of this quantum receiver is defined by the frequency of the lasers that excite the atoms. Due to the fine tunability of these lasers, the quantum receivers the lasers drive are widely tunable, detecting incident fields from kilohertz to terahertz without any changes to the hardware~\cite{Holloway2014,Holloway2017,Meyer2020}. Meanwhile, the instantaneous bandwidth of these receivers is on the order of megahertz~\cite{Sapiro2020} due primarily to the state lifetime of the Rydberg atoms.  The receivers can also be used in a mixer configuration~\cite{Simons2019} wherein sub-Hertz frequency distinguishability is possible~\cite{Gordon2019}. 

Early investigations using these Rydberg atom-based quantum microwave receivers to resolve spatial variations of an incident electric field and to detect angles of arrival have been published in recent years. These probes have been shown to detect spatial variations in the strength $|E(x,y)|$ of a microwave field either within an atomic vapor cell~\cite{Holloway2017} or in the near-field of a transmitter~\cite{Simons2018,Cardman2020}. Using the mixer method~\cite{Simons2019} to determine phase of an incident RF electric field, Robinson, et al. demonstrated an angle of arrival measurement wherein two spatial locations inside the vapor cell were probed to determine the phase angle of the incident plane-wave field~\cite{Robinson2021}. This concept was then extended by scanning a single-point Rydberg atom receiver over a SA to determine the spatial distribution of phase in the plane of measurement and to extract a set of angles of arrival~\cite{Simons2021}. The SA measurement was an early study in understanding the use of these receivers in such measurements, ultimately to be used in channel sounding. Other than the intrinsic wideband tunability of these receivers, a benefit of using the Rydberg atom quantum receivers is that field strength measurements with these devices are directly traceable to Planck's constant ($h$), Eq.~\ref{eq:rydbergE} ~\cite{Holloway2017_June}, a fundamental unit in the new SI~\cite{Newell2018}, with a calculable scaling factor, the Rydberg transition dipole moment ($\wp_{ij}$) ~\cite{Holloway2017_June}, where $f_s$ is the splitting frequency of the narrow electromagnetically induced transparency (EIT) spectral line as the laser frequency is scanned (and can be measured very accurately), and $ij$ denotes the two Rydberg states coupled by the RF field,
\begin{equation}\label{eq:rydbergE}
    |E_{RF}| = \frac{f_s h}{\wp_{ij}}.
\end{equation}
When the atoms are used in the mixer method~\cite{Simons2019}, the RF field imposed on the atoms is 
\begin{align}
    E_{atoms} &= E_{res}E_{mod}, \\ \nonumber
    E_{res} &= \cos(2\pi f_{LO}+\phi_{LO}) \\ \nonumber
    E_{mod} &= \left(E_{LO}^2+E_{SIG}^2+2E_{LO}E_{SIG}\cos(2\pi\Delta f t+\Delta\phi)\right)^{1/2},
\end{align}
with a component of the field that is resonant with the pair of Rydberg states $E_{res}$ (see Fig.~\ref{fig:Rydberg}) and a component that causes a modulation of the EIT spectral line splitting (or a modulation of the transmitted power on the photodetector when the lasers are locked on resonance) at the intermediate frequency $\Delta f=f_{LO}-f_{SIG}$ between the over-the-air signal RF field and over-the-air LO, where $\Delta f \ll (f_{LO}+f_{SIG})/2$. The phase of this modulated voltage signal measured by the photodetector $\Delta\phi = \phi_{LO}-\phi_{SIG}$ contains information about the phase of the signal RF field, and, as long as the LO provides a constant phase reference, allows measurement of the signal field phase for angle of arrival and channel sounding applications described above.  What is more, this receiver is fully dielectric and scatters much less radiation, potentially leading to more reliable wireless channel characterizations.

\section{SAs in Optics}
\label{sec:optics}
Michelson and Pease published a seminal study on optical incoherent SA in 1921 which started the history of SA \cite{michelson1921measurement}. As a result, all of the interferometer systems in the radio and optical spectral regimes now provide high resolution images of astronomical objects through SA imaging \cite{merkle1988synthetic,bulbul2021super}. Essentially, these astronomical interferometers measure the statistical correlation between two electromagnetic signals originating in the object and passing through two telescopes spaced apart. Hence, all of these interferometers allow two signals to propagate simultaneously through two channels. Since the radio signals, including their amplitude and phase, are recorded in the radio antennas and transmitted electronically to the point of cross-correlation, this presents a less serious problem than in optics. In fact, only recording the amplitude of the signal leads to a problem called \textit{phase retrieval}. 
	
In the optical regime, however, an electrical detector cannot directly record the signal's phase without interfering with another telescope's wave. As a result, optical signals should be transferred by waveguides from the two telescopes far apart to the interference laboratory, with an optical path difference of about 100 micrometers between them, which is typical for optical sources. Thus, optical astronomical interferometers are heavy devices whose baseline (the distance between two telescopes) is limited to a few hundred meters. There is one exception to the two-wave interference problem: the intensity interferometer proposed by Hanbury Brown and Twiss \cite{brown20132}, in which the intensity, rather than the complex amplitude, is cross-correlated between the two telescopes. Although intensity interferometers are capable of estimating target sizes in addition to imaging, these interferometers are no longer used because of relatively low SNRs.
	
	\begin{figure}[t]
		\centering
		\includegraphics[width=0.45\textwidth]{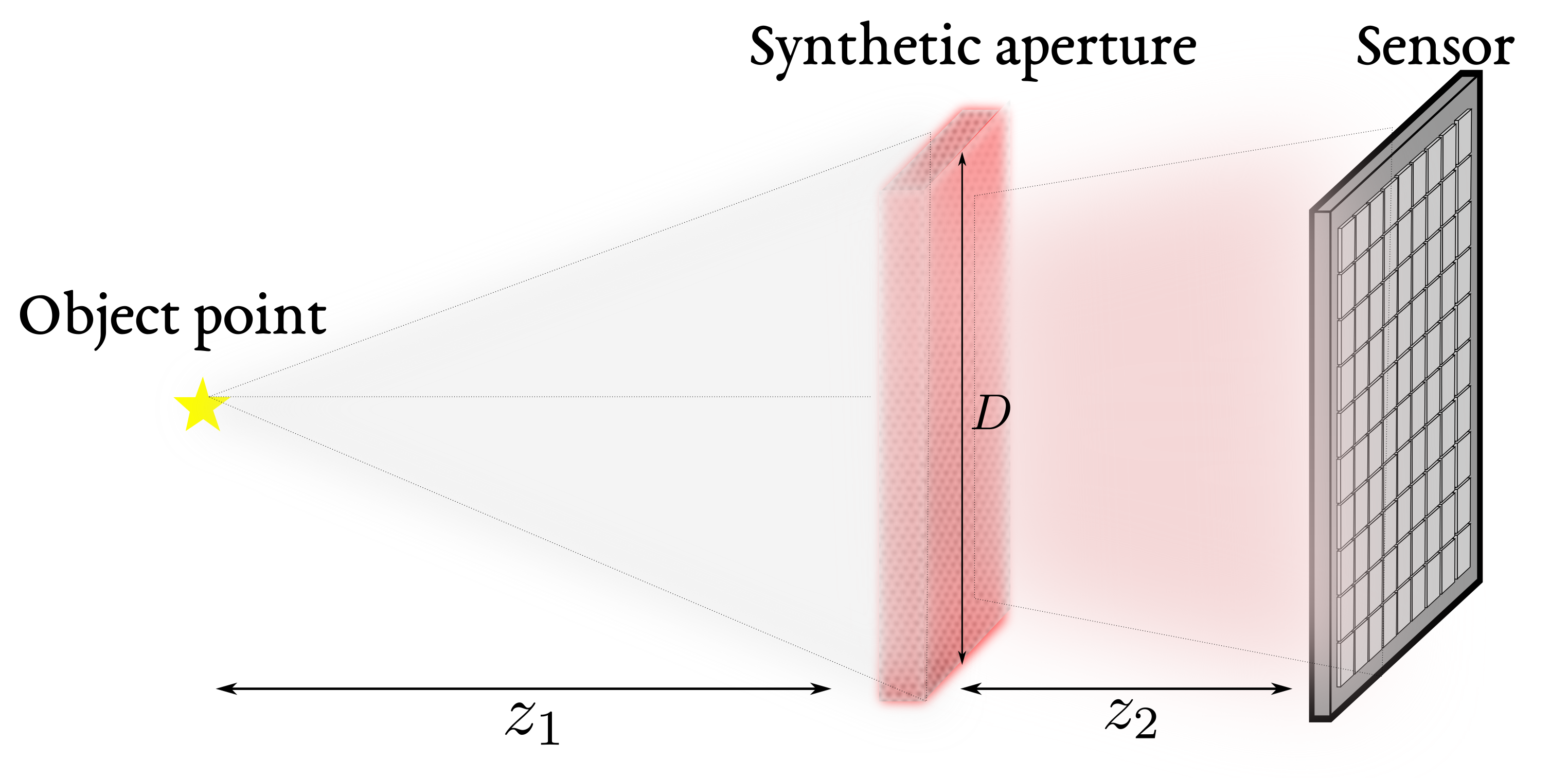}
		\caption{Scheme of an SA optical system. The incoming light, either coherent or incoherent, from the object propagated to a distance $z_{1}$ is modulated by the SA. The resultant encoded light which passes through a different aperture is then recorded at a distance $z_{2}$ by an intensity sensor.}
		\label{fig:system}
	\end{figure}
	
	In 2007, \textit{F}resnel \textit{in}coherent \textit{c}orrelation \textit{h}olography (FINCH) was introduced, opening up a number of new opportunities for incoherent SA imaging \cite{rosen2007digital}.  A typical setup is illustrated in Figure \ref{fig:system}. A FINCH hologram is created using objects emitting incoherent light. As a result of the recording system splitting each object's light into two waves, it is possible to make holograms. In the camera, where these waves interact as holograms, both waves are modulated differently. Rather than record a hologram all at once, it was proposed to record it piecemeal over a period of time \cite{katz2010super}. Nevertheless, this FINCH-based optical incoherent SA was not optimal in the sense that it resulted in relatively low image resolution. With FINCH \cite{katz2011could} configured optimally, the various parts of the hologram were formed by interference between two waves from two far-apart subapertures. The problem of processing simultaneously through two far-apart channels still exists even when the SA imaging is implemented with a different physical effect than the traditional statistical correlation.
	
	Three more systems in the history of optical incoherent SAs were introduced: \textit{co}ded \textit{a}perture \textit{c}orrelation \textit{h}olography (COACH) \cite{vijayakumar2016coded}, interference-less COACH \cite{vijayakumar2017interferenceless}, followed by an imaging system for partial illumination \cite{bulbul2017partial}. COACH is a generalized version of FINCH as it is also a self-interference method for recording incoherent holograms, but instead of a quadratic phase mask in FINCH it uses a chaotic phase mask for one of the waves. In contrast, the interference-less COACH is a degenerate version of the self-interference COACH. Due to the necessity of recording a wave interference from two far-apart subapertures at the same time, a high-resolution image is achievable only if the system operates in the mode of two far-apart channels at the same time. In situations in which the same rule of simultaneously using two channels exists in three different systems that each rely on different physical effects, the rule may be considered as a generic law of nature that cannot be changed.
	
	Researchers are also looking to quantum mechanics to find tools that may enable longer baselines between optical telescopes. One proposed method uses quantum repeater networks to create shared entangled states over arbitrarily long distances~\cite{Gottesman2012,Borregaard2020}. Rather than needing to preserve the astronomical photons, entangled photons from a source between distant telescopes are transmitted over the long distance, with help from the quantum repeaters, and interferograms between the laboratory entangled photons and the astronomical photons are generated at each telescope. Those interferograms are subsequently compared to generate an image of the celestial object. Alternatively, quantum hard drives have been proposed to entirely avoid the need for optical fiber links between distant telescopes~\cite{Bland-Hawthorn2021,Ma2021-2}. The concept here is the quantum state of the astronomical photons is preserved in quantum memory on a physical quantum hard drive. That hard drive is then transported to a common location where all hard drives from each of the optical telescopes in the array are combined and interference between the preserved photon states is used to extract the high resolution image of the celestial object. At this point, quantum repeaters are limited in range due to their complexity and losses while the best reported storage lifetime of a quantum hard drive is on the order of 1 hour~\cite{Ma2021-2}. Further advancements in these spaces is needed before a practical quantum enhanced long baseline optical interferometer is implemented.

\subsection{Phase retrieval algorithms}
Phase information characterizes the delay accrued by an electromagentic wave during propagation. This information is typically lost in the optical detection process, because light detectors measure intensity-only variations. The phase information is regained at the cost of greater experimental complexity, typically by requiring light interference with a known field, as in the process of holography.% An alternative means of measuring phase — without requiring interferometric configurations — is phase retrieval.
  Mathematically, the phaseless measurements $\{ y_{i} \}_{i=1}^{m}$ acquired in this problem are 
\begin{align}
 y_{i} = \lvert \langle \boldsymbol{a}_{i},\boldsymbol{x} \rangle \rvert^{2} + \eta_{i},
 \label{eq:basicPRProblem}
\end{align}
where $\boldsymbol{x}\in \mathbb{C}^{n}$ is the target unkown signal, $\boldsymbol{a}_{i}\in \mathbb{C}^{n}$ are the known sampling vectors, and $\eta_{i}$ models the noise.

Traditional algorithms to solve the phase retrieval problem are based on the error-reduction method \cite{fienup1982phase} proposed in 1970. However, this method does not have solid theoretical convergence guarantees \cite{candes2015phase,fienup1982phase}. Recently, a convex formulation was proposed in \cite{candes2014solving} via Phaselift, which consists in lifting up the original problem of vector recovery from a quadratic system into that of recovering a rank-1 matrix. In fact, this is possible because phaseless measurements in \eqref{eq:basicPRProblem} are equivalent to
\begin{align}
 y_{i} = \langle \boldsymbol{a}_{i}\boldsymbol{a}^{H}_{i},\boldsymbol{X} \rangle + \eta_{i},
\end{align}
where $\boldsymbol{X}=\boldsymbol{x}\boldsymbol{x}^{H}$. For this convex approach, large theoretical guarantees of convergence and recovery were provided, but its computational complexity becomes prohibitive when the signal dimension is large. 

More recent methods described in \cite{candWir} retrieve the phase by applying techniques such as matrix completion, and non-convex formulations \cite{pinilla2018phase,paulus2022reliable}. Specifically, one of the non-convex formulations, called the Wirtinger Flow (WF), is a gradient descent method based on the Wirtinger derivative, and has demonstrated it can attain exact recovery from the phaseless measurements \cite{candWir} up to a global unimodular constant. The WF method was improved by the truncated Wirtinger flow (TWF) algorithm proposed in \cite{chen2015solving}, which optimizes the Poisson likelihood and keeps the convergence by designing truncation thresholds to calculate the step gradient. Additionally, the WF and the TWF methods use the spectral initialization strategy to guarantee exact recovery of the true signal up to a global unimodular constant.
	
The reweighted gradient flow (RAF) \cite{wang2017solving}, stochastic truncated amplitude flow (STAF) \cite{wang2017scalable}, and the reshaped Wirtinger flow (RWF) \cite{zhang2016reshaped} algorithms are also gradient descent methods based on the Wirtinger derivative. These methods aim to solve
\begin{align}
 \minimize_{\boldsymbol{x}\in \mathbb{C}^{n}} \hspace{0.5em}\frac{1}{m}\sum_{i=1}^{m}\left( \sqrt{y_{i}} - \lvert \langle \boldsymbol{a}_{i},\boldsymbol{x} \rangle \rvert\right)^{2}.
\end{align}
Further, the RAF and RWF algorithms introduce different initializations, which attain a more accurate estimation of the true signal in comparison to the spectral initialization. In terms of the sample complexity and speed of convergence, the RAF and RWF methods exhibit a superior performance over the state-of-the-art algorithms. It is important to highlight that the functions optimized by the RAF and RWF methods are non-convex and non-smooth. In particular, in order to address the non-smoothness of the optimization cost function, TWF introduces truncation procedures to eliminate the erroneously estimated signs with high probability. 
\begin{figure*}[t]
		\centering
		\includegraphics[width=1.0\textwidth]{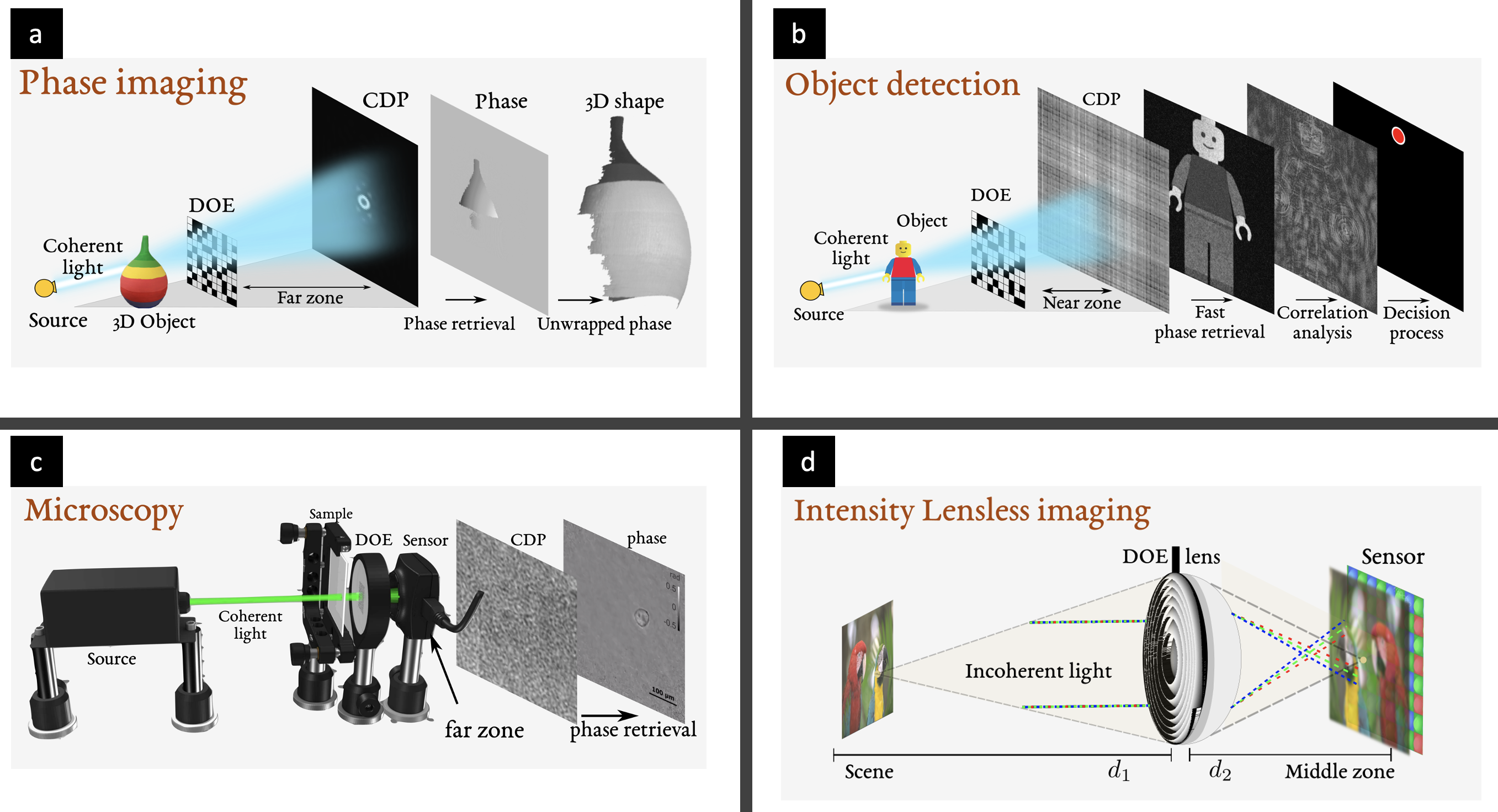}
		\caption{Common coded-based optical imaging applications: (a) phase imaging, (b) object detection, (c) microscopy, and (d) lensless imaging \cite{pinilla2022unfolding}.}
		\label{fig:cdp}
\end{figure*}
	
\subsection{Fourier Ptychography}
The trade-off between resolution and imaging field of view (FoV) is a long-standing problem in traditional optics. This trade-off implies an optical system can produce either an image of a small area with fine details, or an image of a large area with coarse details \cite{Zheng2021}. Fourier Ptychography (FP) was invented in 2013 \cite{zheng2013wide} and has proven to be an effective method of mitigating this trade-off. FP alleviates the physical constraints that limit resolution by integrating SA imaging and phase retrieval.

FP directly measures low-resolution intensity-only images and algorithmically creates a high-resolution complex image that contains the information carried by the light that interacted with a sample.  The information conveyed by an optical system can be quantified in terms of the space-bandwidth product (SBP).  The SBP describes the total number of independent pixel's over an optical system's FOV.  A high SBP is desirable in microscopy and can be obtained by making the lens larger.  However, by using FP the SBP can be increased without any hardware modifications.

\begin{figure}
		\centering
		\includegraphics[width=0.50\columnwidth]{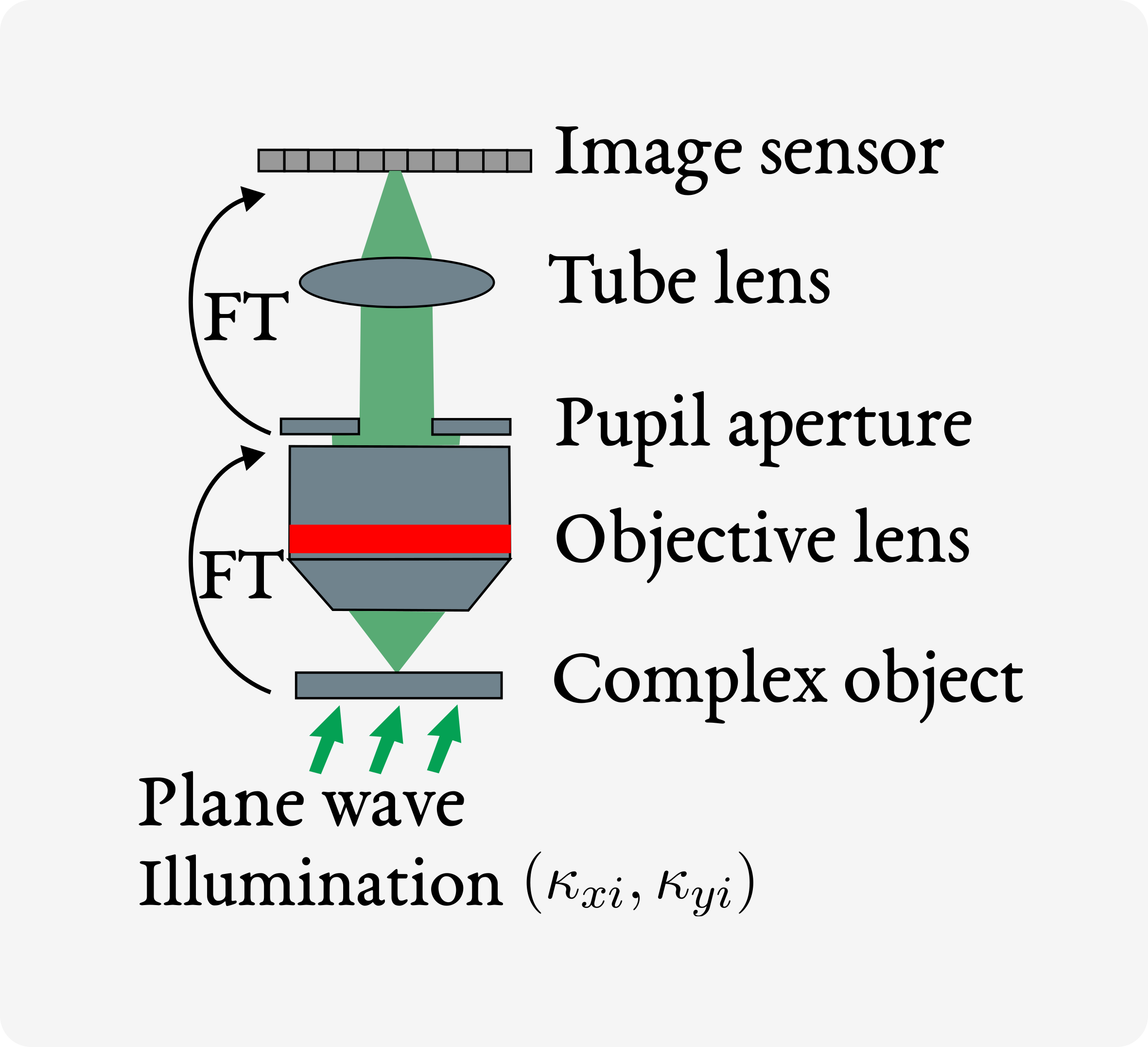}
		\caption{Configuration of a FP microscope \cite{Zheng2021}}
		\label{fig:FP_microscope}
\end{figure}
Fig. \ref{fig:FP_microscope} illustrates the configuration of a FP microscope.  The microscope achieves a higher SBP because an array of LEDs under the specimen varies the angle of incident light and illuminates different portions of Fourier space that are stitched together to form the image.  By activating one LED at a time the sample is illuminated sequentially by oblique plane waves carrying various spatial frequency vectors.  The illuminating plane wave from the ${k}$th LED is represented as
\begin{equation}
    \mathbf{\nu}_{k} = \frac{1}{\lambda}(\sin{\theta}_{xk},\sin{\theta}_{yk}).
\end{equation}
Here, ${\lambda}$ equals the wavelength of the LED probe and ${(\theta_{xk},\theta_{yk})}$ defines the inclination angle of the ${k}$th plane wave.  The field that intercepts the pupil aperture is effectively the Fourier transform (FT) of the field at the object plane.  The field at the imaging sensor is described by the FT of the field immediately after the pupil aperture.  The sequential low-resolution images of the specimen, ${I_k}$, can be expressed as
\begin{align}
    I_{k} &= \big\vert u(\mathbf{r})e^{\textrm{j}2{\pi}\mathbf{\nu}_{k}^{T}\mathbf{r}}{\ast}h(\mathbf{r}) \big\vert^2  \\
    &= \big\vert \text{FT}^{-1}[U(\mathbf{\nu} - \mathbf{\nu}_k)H(\mathbf{\nu})] \big\vert^2
\end{align}
where ${\mathbf{r} = [x \quad y]^{T}}$ are spatial coordinates, ${\ast}$ denotes the convolution operation, ${\text{FT}^{-1}}$ is the inverse Fourier transform, ${h(\mathbf{r})}$ represents the system point spread function (PSF), ${u(\mathbf{r})}$ is the complex amplitude of the specimen, ${U(\mathbf{\nu})}$ and ${H(\mathbf{\nu})}$ are the Fourier transforms of ${u}$ and ${h}$.

\begin{figure*}
		\centering
		\includegraphics[width=0.8\textwidth]{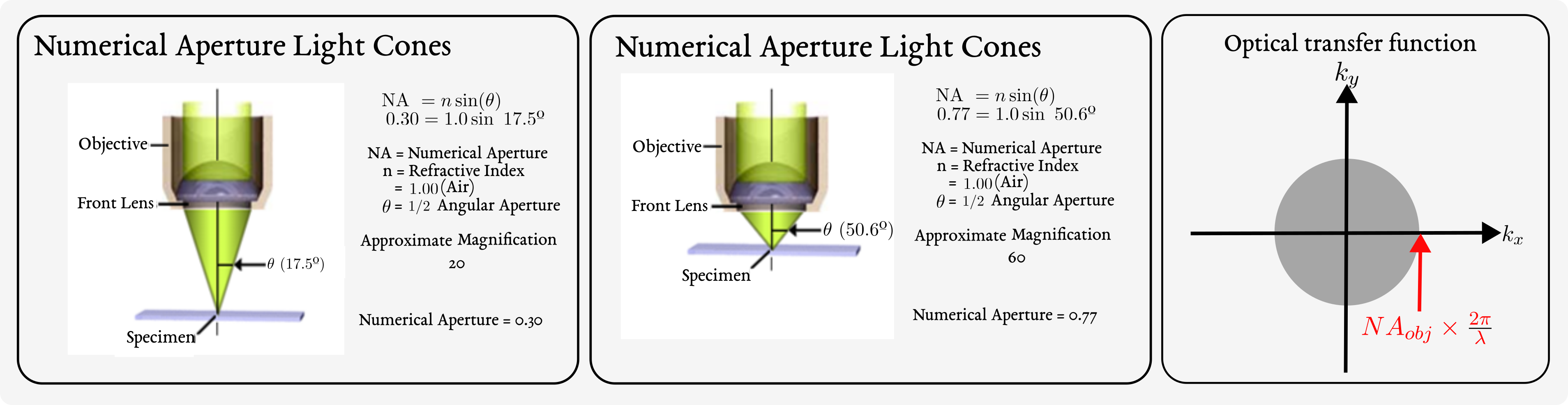}
		\caption{Optical transfer function of a microscope \cite{micro,Zheng2021}}
		\label{fig:NA}
\end{figure*}

In the ${k}$th algorithm iteration, the area within the high-resolution Fourier spectrum ${U(\mathbf{\nu})}$ centered at the spatial frequency ${(\mathbf{\nu}_{xk},\mathbf{\nu}_{yk})}$ is illuminated by the ${k}$th plane wave with an aperture set by the coherent transfer function of the optical system.  For a circularly symmetric pupil the transfer function ${H}$ is a circle in wavenumber space with a radius equal to the highest transmitted spatial frequency, shown as ${\text{NA}(2{\pi}/{\lambda})}$ in Fig. \ref{fig:NA}.  Numerical aperture (NA) describes the resolution of an optical system and is equal to ${n\sin{\theta}}$, where ${n}$ corresponds to the refractive index of the propagation medium (1 for air) and ${\theta}$ is the half-angle subtended by the imaging system from an axial object.

\begin{figure*}
		\centering
		\includegraphics[width=0.8\textwidth]{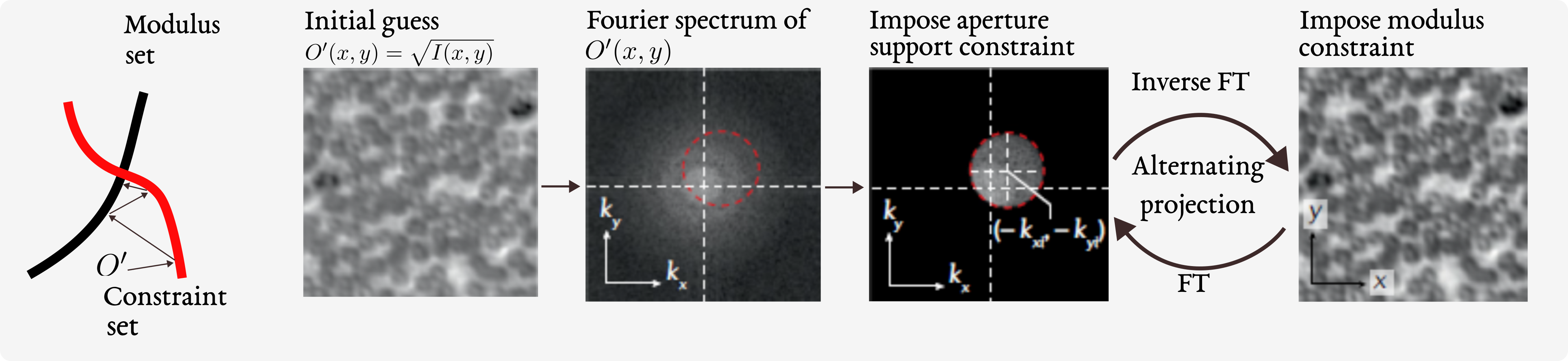}
		\caption{Phase retrieval using alternating projections \cite{Zheng2021}.}
		\label{fig:phase_retrieval}
\end{figure*}

The phase retrieval process recovers the complex object from the intensity measurement by alternating projections of the estimated object between the constraint set, defined as the signal support area of the pupil aperture in the Fourier domain, and the modulus set, which includes all objects in the spatial domain that are consistent with the intensity measurement.  The search for the intersection between the two sets is  shown in Fig. \ref{fig:phase_retrieval}.  The projection to the constraint set is accomplished in the Fourier domain by setting the spectrum values outside the shifted pupil aperture to zero. The shifted aperture offset ${(-\mathbf{\nu}_{xk},-\mathbf{\nu}_{yk})}$ is determined by the illumination wavevector.  Next, the projection onto the modulus set is performed in the spatial domain by computing an inverse FT and replacing the magnitude samples of the estimated object with measured values while keeping the phase unchanged.  The alternating projections are repeated until the estimated object converges.

% First, FP synthesizes the pupil aperture at the Fourier plane to bypass the resolution set by the objective lens. Mathematically, the acquired measurements in FP are given by
% \begin{align}
%     y_{i} = \lvert \langle \boldsymbol{L}\boldsymbol{f}_{i},\boldsymbol{x} \rangle \rvert^{2} + \eta_{i},
%     \label{eq:FP}
% \end{align}
% where $\boldsymbol{L}\in \mathbb{C}^{n\times n}$ is a diagonal matrix that models the effect of the pupil aperture, and $\boldsymbol{f}_{i}\in \mathbb{C}^{n}$ for $i=1,\dots,n$ are the rows of the inverse Fourier transform.

With FP, phase is not measured during the acquisition process, thereby eliminating the challenge of performing phase measurements via an interference field that exists in holography. Instead, FP recovers the missing phase from intensity measurements during the iterative phase retrieval process. FP also provides the ability to computationally correct optical aberrations post-measurement and solves the problems of phase loss, aberration-induced artifacts, shallow depth of field (DOF), and allows for higher resolution and a larger FOV simultaneously \cite{pan2019adaptive}. Current applications include digital pathology and quantitative phase imaging (QPI) with high precision \cite{zhang2017data}, high-throughput imaging \cite{pan2018subwavelength}, high-speed imaging \cite{tian2015computational}, three dimensional (3D) imaging \cite{horstmeyer2016diffraction}, and biomedical applications \cite{chung2019computational}. Combining reflective imaging, the authors in \cite{guo2015fourier}, \cite{dong2014aperture} reported a proof-of-concept study for active remote sensing using visible light with FP.
	
In comparison to microwaves, visible light provides higher resolution. As well as providing additional phase information, FP also greatly increases the feasibility of remote sensing. Thereafter, a long-distance subdiffraction-limited visible imaging technique based on FP was developed in \cite{holloway2016toward}, i.e., SAVI, which allows the imaging distance to be set freely according to system parameters and is applicable to 0.7 to 1.5 m imaging distances. In contrast, the SAVI system's imaging range is limited by commercially available products, which is comparable to a finite correction system. Moreover, due to the camera's scanning scheme, the FoV will change, resulting in a smaller overlapped FoV, leading to a higher cost for the array camera.
	
FP imaging with few photons is a challenge because stray light and noise can overwhelm the signal. Aidukas, et al., tackled this low SNR obstacle applicable to the imaging of delicate biological samples by leveraging quantum correlations. The team showed the ability to extract phase and intensity information of a microscopic object through correlations between the signal and idler of a parametric down-conversion illumination source in their experimental demonstration. 

\subsection{Coded Diffractive Imaging}
Coded diffraction imaging refers to acquisition of images through a setup that employs coherent/incoherent light and a coded aperture (also known as a diffractive optical element (DOE)) to modulate the scene. Typically this setup allows the acquisition of several snapshots of the scene by changing the spatial configuration of the coded aperture. This modulated data is experimentally acquired in three diffraction zones: near, middle, and far~\cite{poon2014introduction,guerrero}. For the coherent case, mathematically, assuming a diagonal matrix $\boldsymbol{D}_{\ell}\in \mathbb{C}^{n\times n}$ for modeling the DOE for the $\ell$-th snapshot, $\ell=1,\dots,L$, the coded measurements consist of quadratic equations of the form 
\begin{align}
    y_{i,\ell}^{k} = \lvert \langle \boldsymbol{D}_{\ell} \boldsymbol{a}_{i}^{k}, \boldsymbol{x} \rangle \rvert^{2} + \eta_{i}^{k},
\end{align}
where $\boldsymbol{a}_{i}^{k}\in\mathbb{C}^{n}$ are the known wavefront propagation vectors associated with the $k$-th diffraction for $i=1,\cdots,n$, $\boldsymbol{x}\in\mathbb{C}^{n}$ is the unknown scene of interest, and $k=1, 2, 3$ indexes the near, middle, and far zones, respectively.

By harnessing specific properties of each diffraction zone, several advances in imaging applications have been made, and Fig.~\ref{fig:cdp} summarizes common coded aperture (or DOE) applications. Specifically, phase imaging deals with the reconstruction of the three-dimensional (3-D) shape of an object via phase retrieval. The far zone scenario consists of estimating the optical phase of the object by low-pass-filtering the leading eigenvector of a carefully constructed matrix \cite{pinilla2020single}. In the case of object detection the optical phase is used to detect objects within a scene. Near zone coded data for rapid detection uses cross-correlation analysis to detect the target using its optical phase as a discriminant \cite{ajerez}. Moreover, the imaging task in microscopy is the reconstruction of the object wavefront. In \cite{kocsis2021ssr}, a novel approach is described for lens-less single-shot phase retrieval for pixel super-resolution phase imaging in the middle zone by suppressing the noise in a combination of sparse- and deep learning-based filters. Single-shot allows recording of dynamic scenes (frame rate limited only by the camera). And lastly, computational imaging with DOEs is a multidisciplinary research field at the intersection of optics, mathematics, and digital image processing \cite{rostami2021power}. Particularly, in \cite{rostami2021power} a DOE is effectively designed for all-in-focus intensity imaging where the diffraction patterns associated with the DOE were studied in the middle zone.

\subsection{Coded Imaging Setups }
The advantage of the coded aperture lies in its capability to successfully recover the phase without additional optical elements (such as lenses) leading to even more compact imaging devices \cite{kocsis2021ssr}. The eschewing of the lens makes the system not only light and cost-effective but also lens-aberration-free and with a larger FoV. 

Hyperspectral complex-domain imaging is a comparatively new development that deals with a phase delay of coherent light in transparent or reflective objects \cite{katkovnik2021admm}. Hyperspectral broadband phase imaging is more informative than the monochromatic technique. Conventionally, for the processing of hyperspectral images, 2-D spectral narrow-band images are stacked together and represented as 3-D cubes with two spatial coordinates $(x,y)$ and a third longitudinal spectral coordinate. In hyperspectral phase imaging, data in these 3-D cubes are complex-valued with spatially and spectrally varying amplitudes and phases. This makes phase image processing more complex than the hyperspectral intensity imaging, where the corresponding 3-D cubes are real-valued.

In certain coded diffractive imaging applications, the combination of blind deconvolution, super-resolution, and phase retrieval naturally manifests. While this is a severely ill-posed problem, it has been shown \cite{pinilla2021nonconvex} that an image-of-interest could be estimated in polynomial-time. The approach relies on previous results that established the DOE design to achieve high quality images \cite{bacca2019super} and partially analyzed the combined problem by solving the super-resolution phase retrieval problem \cite{katkovnik2017computational} from coded data. These designs are obtained by exploiting the model of the physical setup using machine learning methods where the DOE is modeled as a layer of a NN (data-driven model or unrolled) that is trained to act as an estimator of the true image \cite{rostami2021power}. This data-driven design has shown an outstanding image quality using a single snapshot as well as robustness against noise.

\section{SA Sonar}
\label{sec:sonar}
SA processing with sonar poses certain challenges that have delayed the development and applicability of SAS imaging methods compared to its radar counterpart.
The complexity of SA processing for underwater mapping applications stems mainly from: 1) the propagation speed of acoustic waves in water ($1.5\times 10^{3}$~m/s), which is at least 5 orders of magnitude smaller than the propagation speed of electromagnetic waves in air ($3\times 10^{8}$~m/s), and 2) the coherence loss of the received signal along the SA due to the random motion of the sonar platform, the instability of the medium and the multipath arrival pattern in shallow waters\cite{Cutrona1975}. 
The low propagation speed of acoustic waves requires a long acquisition time to achieve a practically useful imaging range, limiting the ping repetition rate.
Moreover, the spatial sampling requirement for unaliased imaging results in an inversely proportional relation between the speed of the platform carrying the sonar system and the maximum imaging range \cite{HansenBook}.
Therefore, the time required to form the SA for sonar is much longer than that for radar, making phase errors due to random platform motion and wave propagation critical for SAS imaging \cite{Williams1976}. 

It was soon demonstrated that the temporal and spatial instability of the underwater environment, e.g., due to turbulence or spatial inhomogeneity of the acoustic parameters, is not the main limiting factor for practical SAS \cite{Williams1976, Christoff1982}.
Nevertheless, environmental factors can degrade SAS imaging, e.g., due to refractive effects from internal waves \cite{Hansen2014}.
Introducing an array of multiple receivers instead of a single sensor provided a practically feasible pulse repetition rate for unaliased imaging and became the standard SAS configuration \cite{Lee1979}.
However, stable navigation of underwater vehicles and motion estimation and compensation with sub-wavelength accuracy is still a great operational challenge \cite{Hansen2011, Hunter2017}.
Multi-channel systems increase the complexity of platform motion estimation introducing ping-to-ping yaw errors, but they offer a refined relative position estimate by cross-correlating the signals of overlapping elements in the displaced phase center antenna (DPCA) taking into account the spatio-temporal coherence of homogeneous reverberation \cite{Doisy1998, Doisy2004, BellettiniDPCA}.
State-of-the-art SAS systems are equipped with inertial navigation systems (INS) for coarse motion estimation, combined with DPCA micronavigation to compensate for residual navigation errors \cite{Fossum2008, Bellettini2009, Larsen2010}.
For many years, the cost and complexity of SAS systems limited their scope to military applications, such as mine countermeasures and unexploded ordnance remediation \cite{Hayes2009, Hansen2013}. It is only for the last 20 years that SAS has become common and inexpensive enough to be used for commercial applications such as underwater archaeology, inspection of underwater pipelines and seafloor mapping for offshore windfarm installation and monitoring \cite{Hagen1999, Hansen2013}.

A comprehensive review of past work on SAS image reconstruction algorithms, platform motion estimation and compensation methods, interferometric SAS and SAS system configurations is presented in \cite{Hayes2009}.
In the following, we summarize the basic SAS model before we highlight recent developments in SAS and current research trends categorized into generic research focus areas.
We limit our review to methods that involve coherent signal processing, rather than incoherent image processing such as image segmentation and automatic target recognition \cite{Williams2014, Sledge2021}. \\

\subsection{SAS model}

\begin{figure}[t]
\centering
\includegraphics[width=1\linewidth]{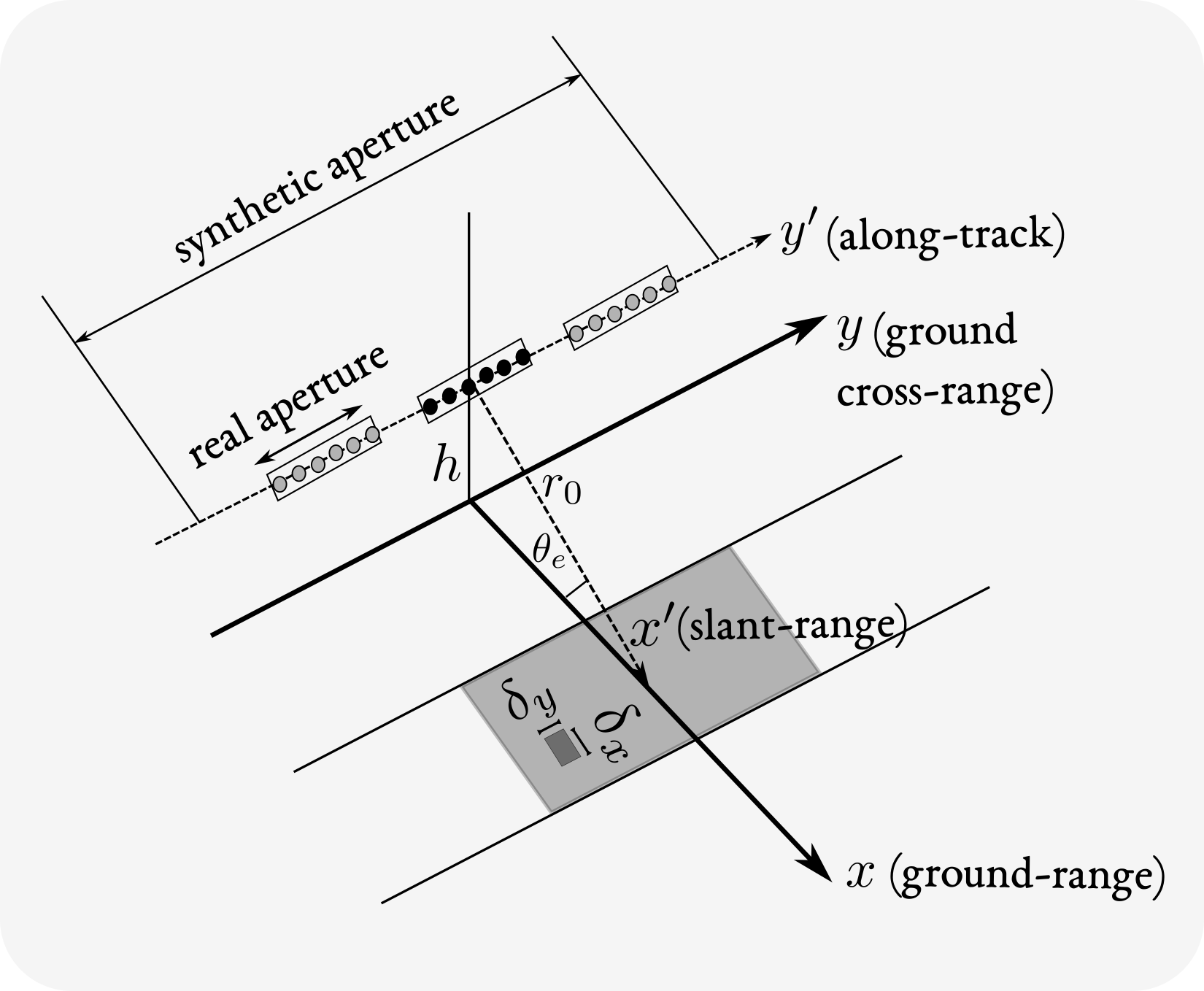}
\caption{SAS imaging geometry in strip-map mode.}
\label{fig:sas_stripmap}
\end{figure}

The SAS geometry in the simplest and most applied strip-map modality is depicted in Fig.~\ref{fig:sas_stripmap}. 
A platform carrying an active sonar, with an arrangement of transmitters and receivers, moves along a linear path, parallel to the seafloor plane.
In strip-map mode, the antenna is focused towards broadside, i.e., the central axis of the real-aperture beampattern is perpendicular to the platform path \cite{Gough1997}. 

The active sonar transmits a short pulse, referred to as a ping, and records the backscattered echoes repeatedly as the platform moves along the track. The stop-and-hop approximation postulates that the platform is stationary during each ping transmission and reception, before it jumps instantaneously to the next position \cite{Gough1997}. Hence, the position of the platform is discretized according to the ping number $p$ as,
\begin{equation}
y_{p} = p v_{p} \tau_{rec},
\label{eq:PlatformCoordinates}
\end{equation}
where $v_{p}$ is the constant speed of the platform and $\tau_{rec}$ is the duration of the recording, which defines the ping repetition period. The insonified area per ping is determined by the radiation pattern of the transmitting antenna. The total imaging area is determined by the SA length and the recording duration $\tau_{rec}$ allowing a maximum swath width of,
\begin{equation}
r_{\text{max}} = \frac{c\tau_{rec}}{2},
\label{eq:MaxRange}
\end{equation}
where $c$ is the speed of sound in water. Multi-element receiver arrays are employed to allow longer imaging ranges without violating the spatial sampling condition for a moderate platform speed (a few knots) \cite{HansenBook}. 

Consider a pulsed acoustic source $h(\mathbf{r}_{t}, f) = b_{T}(\mathbf{r}_{t})q(f)$, where $b_{T}(\mathbf{r}_{t})$ describes the transmitter aperture as a function of the spatial coordinates $\mathbf{r}_{t}$ and $q(f)$ is the transmitted waveform as a function of frequency $f$.
The sound pressure incident to a point located at $\mathbf{r}_{s}$ in the scattering medium due to the pulsed source centered at $\mathbf{r}_{t_{c}}$ is,
\begin{equation}
p_{i}(\mathbf{r}_{st}, f) = q(f) \int_{A_{T}}\!\! b_{T}(\mathbf{r}_{tt})\frac{e^{-\mathrm{j} \frac{2\pi f}{c}\lVert\mathbf{r}_{st} - \mathbf{r}_{tt}\rVert_{2}}}{4\pi\lVert\mathbf{r}_{st} - \mathbf{r}_{tt}\rVert_{2}} \mathrm{d}\mathbf{r}_{tt},
\label{eq:IncidentField}
\end{equation}
where the integration is over the transmitter aperture $A_{T}$ and the position vectors are defined relative to the center of the aperture, i.e., $\mathbf{r}_{st} = \mathbf{r}_{s} - \mathbf{r}_{t_{c}}$ and $\mathbf{r}_{tt} = \mathbf{r}_{t} - \mathbf{r}_{t_{c}}$. 
Equation~\eqref{eq:IncidentField} is the solution to the inhomogeneous Helmholtz equation and describes the wavefield at the far-field of a spatially distributed harmonic source in an unbounded medium as the result of the integrated contributions of the elementary point sources $b_{T}(\mathbf{r}_{tt})\mathrm{d}\mathbf{r}_{tt}$ constituting the source aperture \cite{MorseBook}.
With the Fraunhofer approximation for far field propagation,\cite[Chapter 4]{GoodmanBasicsBook} $\lVert\mathbf{r}_{st} - \mathbf{r}_{tt}\rVert_{2} \approx \lVert \mathbf{r}_{st}\rVert_{2} - \widehat{\mathbf{r}}_{st}\mathbf{r}_{tt}$, where $\widehat{\mathbf{r}}_{st}= \mathbf{r}_{st}/\lVert\mathbf{r}_{st}\rVert_{2}$ is the unit vector in the direction of $\mathbf{r}_{st}$, Eq.~\eqref{eq:IncidentField} is simplified to, 
\begin{equation}
\begin{aligned}
p_{i}(\mathbf{r}_{st}, f) &\approx q(f)\frac{e^{-\mathrm{j} \frac{2\pi f}{c}\lVert\mathbf{r}_{st}\rVert_{2}}}{4\pi\lVert\mathbf{r}_{st}\rVert_{2}}\!\!\int_{A_{T}}\!\! b_{T}(\mathbf{r}_{tt})e^{\mathrm{j} \frac{2\pi f}{c}\widehat{\mathbf{r}}_{st}\mathbf{r}_{tt}} \mathrm{d}\mathbf{r}_{tt}\\
&=q(f)\frac{e^{-\mathrm{j} \frac{2\pi f}{c}\lVert\mathbf{r}_{st}\rVert_{2}}}{4\pi\lVert\mathbf{r}_{st}\rVert_{2}}B_{T}(\mathbf{k}_{st}),
\end{aligned}
\label{eq:IncidentFieldFresnel}
\end{equation}
where $\mathbf{k}_{st}=(2\pi f/c)\widehat{\mathbf{r}}_{st}$ is the wavenumber vector in the direction of $\widehat{\mathbf{r}}_{st}$ and $B_{T}(\mathbf{k}_{st})$ denotes the beampattern of the transmitter as the spatial Fourier transform of its aperture function.
 
Assuming that the platform is stationary during transmission and reception, the backscattered signal at a receiver centered at $\mathbf{r}_{r_{c}}$ from the point scatterer at $\mathbf{r}_{s}$ with complex scattering amplitude $s$ is, 
\begin{equation}
\begin{aligned}
p(\mathbf{r}_{sr}, f) &= p_{i}(\mathbf{r}_{st}, f)s(\mathbf{r}_{sr}, f)\int_{A_{R}} \!\!b_{R}(\mathbf{r}_{rr}) \\
& \qquad\qquad\qquad\qquad\;\; \times \frac{e^{-\mathrm{j} \frac{2\pi f}{c}\lVert\mathbf{r}_{sr} - \mathbf{r}_{rr}\rVert_{2}}}{4\pi\lVert\mathbf{r}_{sr} - \mathbf{r}_{rr}\rVert_{2}} \mathrm{d}\mathbf{r}_{rr}\\
&\approx p_{i}(\mathbf{r}_{st}, f)s(\mathbf{r}_{sr}, f)\frac{e^{-\mathrm{j} \frac{2\pi f}{c}\lVert\mathbf{r}_{sr} \rVert_{2}}}{4\pi\lVert\mathbf{r}_{sr}\rVert_{2}}B_{R}(\mathbf{k}_{sr}),
\end{aligned}
\label{eq:BackscatteredField}
\end{equation}
where the integration is over the receiver aperture $A_{R}$ with beampattern $B_{R}(\mathbf{k}_{sr})$ and $\mathbf{r}_{sr} = \mathbf{r}_{s} - \mathbf{r}_{r_{c}}$, $\mathbf{r}_{rr} = \mathbf{r}_{r} - \mathbf{r}_{r_{c}}$.

In monostatic systems $\mathbf{r}_{t_{c}} = \mathbf{r}_{r_{c}} = \mathbf{r}_{\text{v}}$ by definition, whereas in multi-static configurations the phase center approximation (PCA) \cite{BellettiniDPCA} replaces each transmitter-receiver pair with a virtual element at $\mathbf{r}_{\text{v}} = (\mathbf{r}_{t_{c}} + \mathbf{r}_{r_{c}})/2$ such that $\lVert\mathbf{r}_{s}-\mathbf{r}_{t_{c}}\rVert_{2}\approx\lVert\mathbf{r}_{s}-\mathbf{r}_{r_{c}}\rVert_{2}\approx\lVert\mathbf{r}_{s}-\mathbf{r}_{\text{v}}\rVert_{2} = \lVert\mathbf{r}_{s\text{v}}\rVert_{2}$.
At any given time frame, the total backscattered field at $\mathbf{r}_{\text{v}}$ with the Born approximation \cite{MorseBook} is the superposition of the backscattered echoes from all scatterers within the corresponding isochronous insonified volume $A_{s}$,
\begin{equation}
p(\mathbf{r}_{\text{v}}, f) = \frac{q(f)}{(4\pi)^{2}}\!\!\int_{A_{s}} \!\!\! s(\mathbf{r}_{s\text{v}}, f)B(\mathbf{k}_{s})\frac{e^{-\mathrm{j} \frac{2\pi f}{c}2\lVert\mathbf{r}_{s\text{v}}\rVert_{2}}}{\lVert\mathbf{r}_{s\text{v}}\rVert_{2}^{2}}\mathrm{d}\mathbf{r}_{s},
\label{eq:TotalBackscatteredField}
\end{equation}
where $B(\mathbf{k}_{s}) = B_{T}(\mathbf{k}_{st})B_{R}(\mathbf{k}_{sr})$ is the combined beampattern of the transmitter and the receiver.
In the case that the receiver is much smaller than the transmitter, the receiver's beampattern can be considered omnidirectional, hence $B(\mathbf{k}_{s}) \approx B_{T}(\mathbf{k}_{st})$.

\begin{figure}[t]
\centering
\includegraphics[width=1\linewidth]{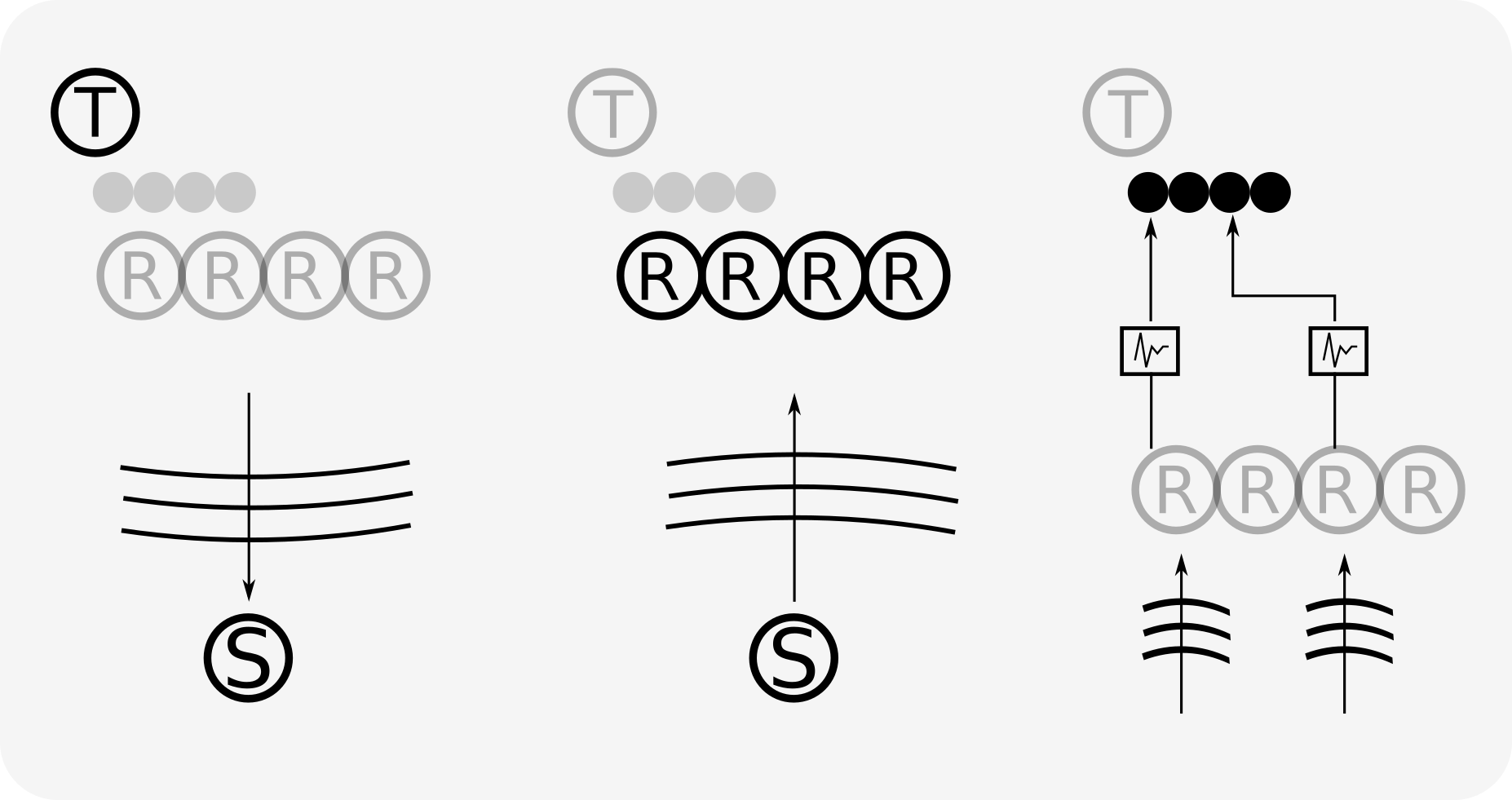}
\caption{Transmission~\eqref{eq:IncidentFieldFresnel}, reception~\eqref{eq:BackscatteredField} and matched filtering~\eqref{eq:TotalBackscatteredFieldMatchedFiltered} with a single-transmitter/multiple-receiver configuration. The PCA replaces the multistatic configuration with a virtual array of monostatic elements located at the middle of the distance between each transmitter-receiver pair.}
\label{fig:sas_pca}
\end{figure}

The recorded backscattered signal~\eqref{eq:TotalBackscatteredField} is compressed with matched filtering, i.e., by multiplication with the complex conjugate of the transmitted pulse $q^{*}(f)$,
\begin{equation}
\begin{aligned}
p_{\text{mf}}(\mathbf{r}_{\text{v}}, f) &\!=\! q^{*}(f)p(\mathbf{r}_{\text{v}}, f)\\
&\!=\! \frac{\lVert q(f)\rVert_{2}^{2}}{(4\pi)^{2}}\!\!\int_{A_{s}} \!\!\! s(\mathbf{r}_{s\text{v}}, f)B(\mathbf{k}_{s})\frac{e^{-\mathrm{j} \frac{2\pi f}{c}2\lVert\mathbf{r}_{s\text{v}}\rVert_{2}}}{\lVert\mathbf{r}_{s\text{v}}\rVert_{2}^{2}}\mathrm{d}\mathbf{r}_{s}.
\end{aligned}
\label{eq:TotalBackscatteredFieldMatchedFiltered}
\end{equation}
Figure~\ref{fig:sas_pca} shows a schematic of the transmission, reception and matched-filtering operations.

Discretizing the scattering field into an imaging grid of $N$ points, the data model~\eqref{eq:TotalBackscatteredFieldMatchedFiltered} can be written in a matrix-vector formulation,
\begin{equation}
\mathbf{d}(p, f)= \mathbf{A}(p, f)\mathbf{s}(p, f) + \mathbf{n}(p, f),
\label{eq:LinearModel}
\end{equation}
where $\mathbf{d}\in\mathbb{C}^{M}$ is the vector of the matched filtered measurements at frequency $f$ for all $M$ receivers comprising the real aperture at ping $p$, $\mathbf{s}\in\mathbb{C}^{N}$ is the unknown vector of the complex scattering values over a grid of $N$ points and $\mathbf{n}\in\mathbb{C}^{M}$ is the additive noise vector. The matrix $\mathbf{A}\in\mathbb{C}^{M\times N}$ maps the unknown scattering $\mathbf{s}$ to the observations $\mathbf{d}$ and has as columns the steering vectors,
\begin{equation}
\mathbf{a}(\mathbf{r}_{s}, p, f)= [e^{-\mathrm{j}\frac{2\pi f}{c} 2\lVert\mathbf{r}_{sv_{1}}\rVert_{2}}, \cdots,e^{-\mathrm{j}\frac{2\pi f}{c} 2\lVert\mathbf{r}_{sv_{M}}\rVert_{2}}]^{T},
\label{eq:SteeringVector}
\end{equation}
which describe the propagation delay from the $s$th scatterer to all the $M$ sensors on the real aperture at ping $p$.
Note that we have incorporated the gain factor, $\lVert q(f)\rVert_{2}^{2}B(\mathbf{k}_{s})/(4\pi\lVert\mathbf{r}_{s\text{v}}\rVert_{2})^{2}$, into the scattering vector $\mathbf{s}$ as it can be easily accounted for in a calibrated system. 

\begin{figure}[t]
\centering
\includegraphics[width=0.5\textwidth]{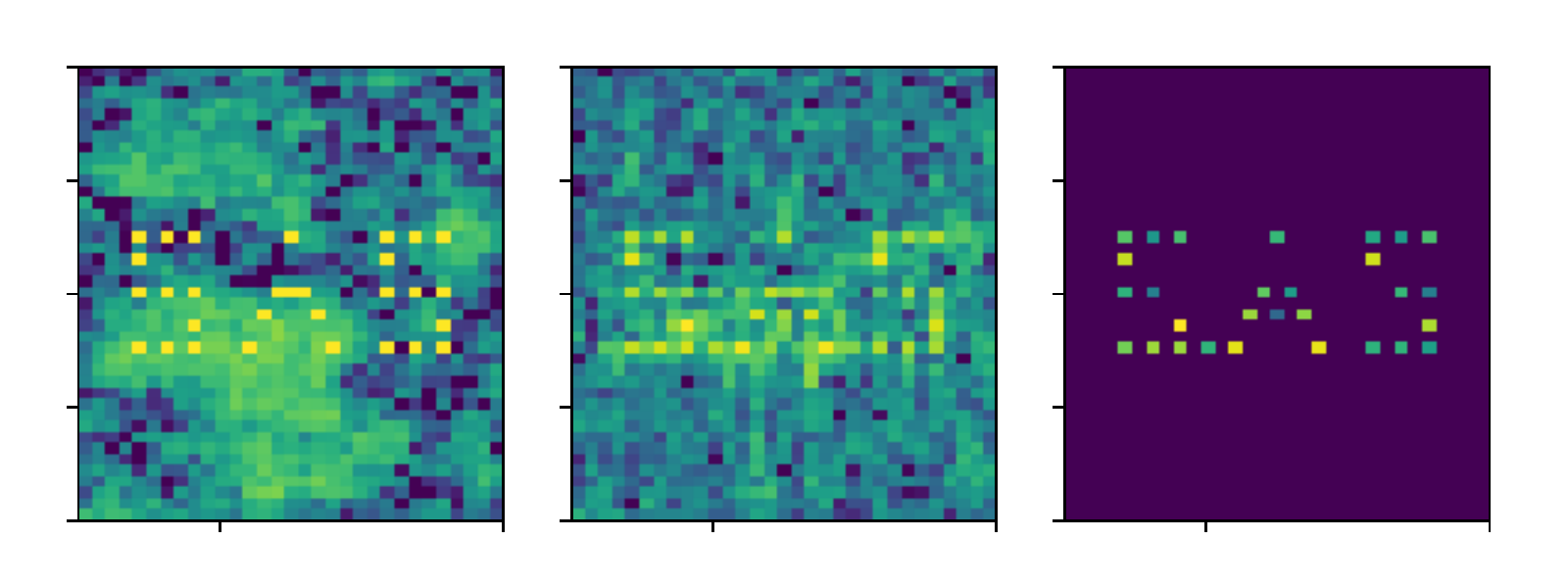}
\caption{Seafloor map with a configuration of strong scatterers in a weakly scattering background, and reconstruction with conventional and sparse SAS processing.}
\label{fig:sas_recosntruction}
\end{figure}

SAS imaging refers to the inverse problem of reconstructing the scattering field $\mathbf{s}$, given the sensing matrix $\mathbf{A}$ and a set of measurements $\mathbf{d}$ over a range of frequencies and pings.
Conventional (delay-and-sum) beamforming uses the steering vectors~\eqref{eq:SteeringVector} as spatial weights to combine the sensor outputs coherently, compensating for the geometrically induced spatial Doppler modulation. In SAS imaging, conventional beamforming provides the scattering estimate,
\begin{equation}
\widehat{\mathbf{s}}_{\text{CBF}} = \sum\limits_{i=1}^{P}\sum\limits_{j=1}^{F}\mathbf{A}^{H}(p_{i}, f_{j})\mathbf{d}(p_{i}, f_{j}),
\label{eq:sasCBF}
\end{equation}
by combining coherently the sensor outputs over $P$ pings and $F$ frequencies.
In the case that there are only a few strong scatterers in the scattering field ($K \ll N$), SAS imaging can be solved as a sparse model fitting problem,
\begin{equation}
\underset{\mathbf{s}(p_{i}, f_{j})}{\min} \; \frac{1}{2}\lVert\mathbf{A}(p_{i}, f_{j})\mathbf{s}(p_{i}, f_{j}) - \mathbf{d}(p_{i}, f_{j})\rVert_{2}^{2} + \mu\lVert\mathbf{s}(p_{i}, f_{j})\rVert_{1}.
\label{eq:sasLasso}
\end{equation}
where $\mu > 0$ is a regularization parameter which controls the relative importance between the quadratic data-fitting term and the sparsity promoting $\ell_{1}$-norm regularization term \cite{Xenaki2019}.
Figure~\ref{fig:sas_recosntruction} demonstrates SAS images of the reconstructed scattering field with conventional and sparse beamforming.
The reader is referred to \cite{Gough1997} for a comprehensive review of SAS imaging algorithms.

The resolution of a SAS system is defined in the range and cross-range directions, indicated in Fig.~\ref{fig:sas_stripmap} as $\delta_{x}$ and $\delta_{y}$ respectively. The range resolution, obtained with matched filtering, is determined by the bandwidth $\Delta f$ of the transmitted ping,
\begin{equation}
\delta_{x} \approx \frac{c}{2\Delta f}.
\label{eq:RangeResolution}
\end{equation}
The cross-range (angular) resolution depends on the apparent SA length $L_{\text{SA}}$, 
\begin{equation}
\delta_{y} \approx \frac{\lambda r_{0}}{2L_{\text{SA}}}.
\label{eq:CrossRangeResolution}
\end{equation}
For a given transducer size ${D}$, the corresponding SA length is proportional to the wavelength and the range of a point scatterer, $L_{\text{SA}}\approx \lambda r_{0}/D$, resulting in a cross-range resolution which is independent of frequency and range, $\delta_{y}\approx D/2$ \cite{HansenBook}.

\subsection{Wideband signal processing for SAS}

There is a progressive shift of interest towards \textit{wideband} and \textit{widebeam SAS} systems that can provide information about the frequency- and aspect-dependent properties of sea-bottom scattering.
For example, low frequencies can be partly transmitted through objects or penetrate the seafloor providing information about internal structure and buried objects \cite{Pinto2002, Lucifredi2006, Pailhas2010}, while multiple views provide information about the object's shape and dimensions \cite{Cook2001}.
As a result, several wideband SAS systems have been developed \cite{Chatillon1999, Shock2005, Larsen2010, Steele2019} and SAS processing algorithms have been adapted for wideband and widebeam applications \cite{Marston2010, Hunter2014, Synnes2017, Xenaki2019}.

Specifically, sub-band or sub-view processing for frequency- and aspect-dependent feature extraction reduces the resolution and the SNR compared to conventional SAS methods.
To alleviate the corresponding SAS image degradation, signal processing methods such as spatial filtering followed by deconvolution \cite{Marston2010} or wavenumber-domain filtering \cite{Synnes2017} have been proposed. 
Sparse signal reconstruction methods, such as feature selection through wavelet shrinkage \cite{Hunter2014} or distributed optimization \cite{Xenaki2019}, show great potential in wideband low-frequency SAS imaging.
In interferometric SAS, wideband methods allow direct estimation of the absolute phase difference, providing robust three-dimensional imaging even with complicated scenes \cite{Saebo2013}.

\subsection{Micronavigation}

Micronavigation refers to platform motion estimation methods, which use redundant recordings between pings in multi-channel systems to refine the coarse motion estimates from navigational instruments and achieve the subwavelength motion estimation accuracy required for SAS processing.
Current work focuses on achieving sub-sample localization accuracy of the peak correlation from spatially and temporally sampled correlation measurements. 

Methods that exploit the spatial correlation of overlapping measurements between consecutive pings for \textit{along-track micronavigation} propose maximum likelihood estimators \cite{Zhu2011}, analytical and numerical coherence models \cite{Blanford2019} and smoothing interpolation kernels \cite{Brown2019,Thomas2021} to improve the accuracy of the along-track, ping-to-ping translation estimate. 
Similarly, time delay estimation between signals recorded on overlapping along-track positions between pings provides an \textit{across-track micronavigation} estimate. 
Across-track micronavigation is particularly challenging in repeat-pass SAS processing for coherent change detection, where aggregated navigation errors between passes can result in baseline decorrelation not attributed to scene changes \cite{Hunter2016, Myers2020, Synnes2021}.
To achieve data co-registration in repeat-pass SAS, the authors in \cite{Hunter2016} introduce a repeat-pass SAS micronavigation algorithm that is a generalization of DPCA method, whereas Ref.~\cite{Thomas2020} proposes a phase unwrapping approach to best fit the temporal correlation function in the presence of noise.
The authors in Ref. \cite{Synnes2021} show that, by combining elements in the multi-channel system into larger effective elements, the along-track and across-track decorrelation baseline increases.
A machine learning approach based on variational inference has been recently proposed for robust data-driven estimation of the three-dimensional platform translation between pings from spatiotemporal coherence measurements of diffuse backscatter \cite{Xenaki2022}.

\subsection{System configuration}
With regard to system design, recent developments propose MIMO configurations \cite{Pailhas2016, Marston2019, Xenaki2020} and circular synthetic trajectories \cite{Marston2021}.
Specifically, \textit{MIMO SAS} systems use multiple channels not only on receive, but also on transmit \cite{Pailhas2016}.
For example, Ref. \cite{Marston2019} examines the use of spatially distributed transmitters in the across-track direction of a planar receiver array to increase the effective non-synthetic length of the array in that dimension and, consequently, improve the depth resolution of the resulting SAS system.
Ref. \cite{Xenaki2020}, instead, proposes a MIMO SAS configuration with spatially distributed transmitters in the along-track direction to improve the spatial sampling rate and increase the imaging range. The authors 
employ a sparse reconstruction algorithm to reduce the impact of the residual waveform correlation and produce high-quality SAS imaging.
\textit{Circular SAS (CSAS)} improves the resolution and reduces the speckle in SAS imaging. Ref. \cite{Marston2021} addresses the challenge of focusing CSAS data due to the non-linear trajectory.

\section{SA Radiometry}
\label{sec:radiometry}
%--------------------------------------------
\begin{figure}[t]
\centering
\includegraphics[width=1\linewidth]{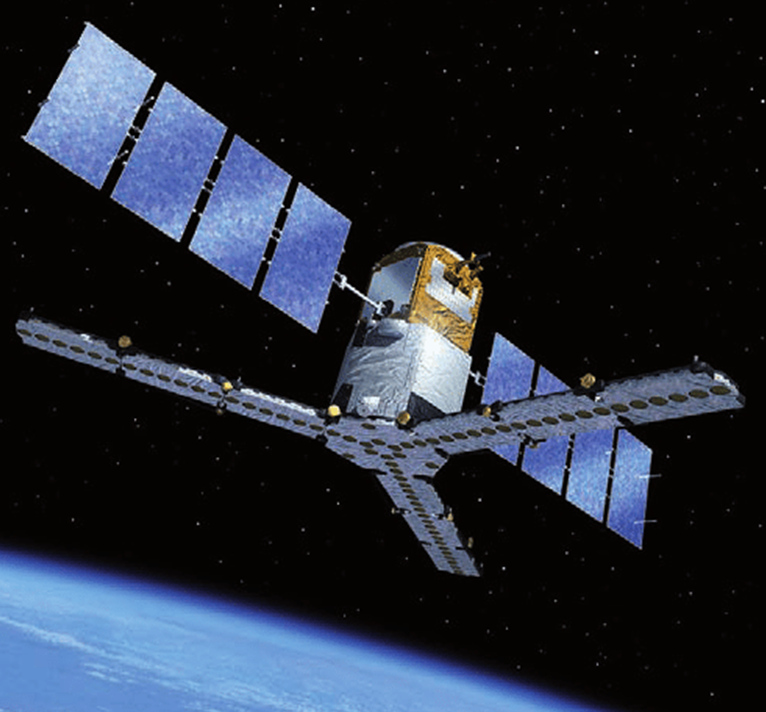}
\caption{The microwave imaging radiometer using aperture synthesis (MIRAS) \cite{LeVine1999} is a Y-shaped SA radiometer on-board the soil moisture and ocean salinity (SMOS) satellite. Photo credit: European Space Agency.}
\label{fig:Y_arm}
\end{figure}
%--------------------------------------------
\begin{figure}[t]
\centering
\includegraphics[width=1\linewidth]{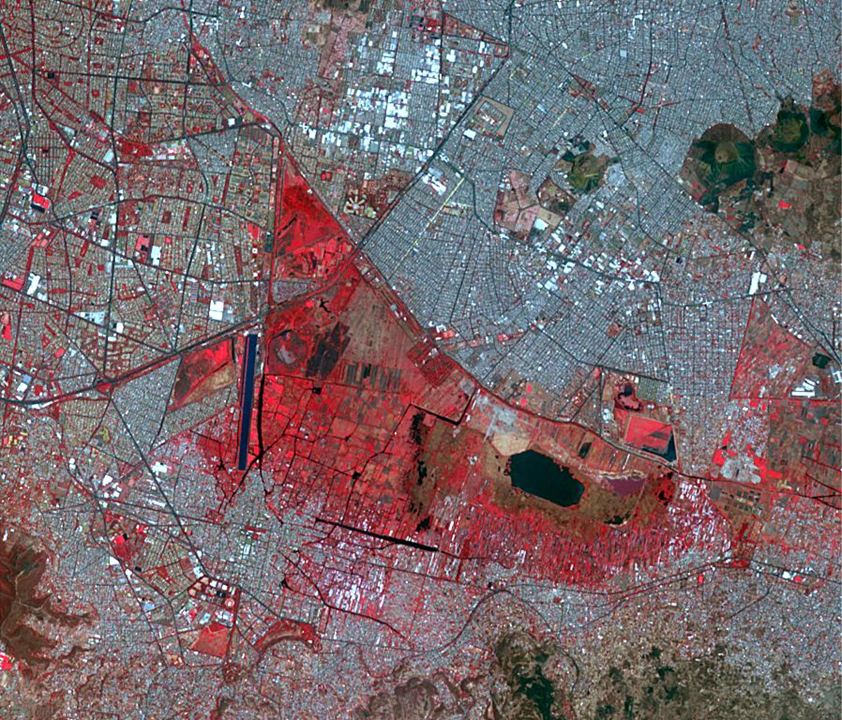}
\caption{Synthetic aperture radiometer image of the Floating Gardens of Xochimilco near Mexico City. Photo credit: NASA/METI/AIST/Japan Space Systems, and U.S./Japan ASTER Science Team.}
\label{fig:xochimilco}
\end{figure}
%--------------------------------------------
Passive microwave sensing at L-band (1.4 GHz) of the earth's surface provides information on important geophysical parameters including sea surface temperature, salinity, windspeed, soil moisture content, and arctic ice levels.  The measurement of these parameters from space requires resolution on the order of 10 km. To obtain resolution on this order of magnitude requires placing very large antennas in low-earth orbit. For example, to obtain an image resolution of 10 km at 1.4 GHz would require an aperture of more than 15 m flying at an altitude of ${800}$ km. The engineering problems associated with placing large antennas into orbit limit the deployment of passive sensors at this frequency ~\cite{LeVine1999}.

Synthetic aperture interferometry using thinned arrays has found widespread use in passive microwave remote sensing applications since it mitigates the technical challenges associated with placing large apertures in space. In aperture synthesis using a thinned array of radiometers, the coherent product (cross-correlation) of the complex signal from pairs of antennas is measured for different antenna-pair spacings (also called baselines). The product at each baseline yields a sample point in the Fourier transform of the brightness temperature map of the scene, and the scene itself can be reconstructed by inverting the sampled transform ~\cite{Thompson1986,LeVine1990,LeVine1994,Ruf1988,Swift1991,Piepmeier}. The resolution of the temperature image is determined by the largest baseline. The individual antennas determine the FOV on the ground surface. The compromise one makes in using aperture synthesis is a potential loss of radiometric sensitivity because small antennas imply a decrease in SNR for each measurement compared to a filled aperture.

Fig.~\ref{fig:Y_arm} illustrates a concept for employing aperture synthesis in both spatial dimensions. In the picture, small antennas are arranged along the arms of a ${Y}$, but other arrangements are also possible. The necessary baselines are obtained by making measurements between all independent pairs of antennas. One can show that this configuration has the same spatial resolution as a filled aperture with the dimensions of the arms. Aircraft instruments with antennas arranged in ${Y}$ and ${U}$ configurations have been built and satellite instruments in space have used the ${Y}$ configuration. Fig.~\ref{fig:xochimilco} shows a synthetic aperture radiometer image of an area near Mexico City known as Xochimilco.

Consider two separated antennas at a height ${h}$ over the earth's surface that have overlapping fields of view.  One antenna is centered at the origin of a Cartesian coordinate system such that the surface of the earth lies on the ${z=h}$ plane, using a flat-earth approximation.  The other antenna is located at the coordinates ${(x,y,z)}$ and at a distance ${(D_{x},D_{y})}$ from the origin.  The thermal emissions by the earth provide an incident flux density ${S_{i}(x,y,f,\theta,\phi)}$ measured in ${W/(m^2{\cdot}Hz{\cdot}sr)}$ at the antennas.  Here, ${f}$ is the frequency of the emissions, ${(\theta,\phi)}$ is the solid angle corresponding to the direction of the emissions, and ${(x,y)}$ is the antenna location in the ${xy}$ plane.  If the antennas are isotropic, then the power available at their output terminals per unit steradian is
\begin{equation}
    P_{i}(\theta,\phi) = 0.5S_{i}(\theta,\phi)A_{eff}B
\end{equation}
where ${P_{i}}$ is in units of ${W/sr}$, the factor of 0.5 accounts for the polarization of the antennas, ${A_{eff} = {\lambda}^2/4{\pi}}$ is the effective aperture area (${m^2}$) of an isotropic antenna, and ${B}$ is the signal bandwidth (${Hz}$).  The corresponding incident brightness temperature per steradian ${T_{i}(\theta,\phi)}$ is defined by
\begin{equation}
    T_{i}(\theta,\phi) = \frac{1}{kB}P_{i}(\theta,\phi)
\end{equation}
where ${k}$ is Boltzmann's constant.  For directional, non-isotropic antennas, the received brightness temperature ${T_r}$ per steradian includes the antenna gain as in
\begin{equation}
    T_{r}(\theta,\phi) = T_{i}(\theta,\phi)G(\theta,\phi)
\end{equation}
where ${G(\theta,\phi)}$ is the power gain pattern of the antenna.  The measured brightness temperature ${T_{m}}$ is obtained by integrating ${T_{r}}$ over all solid angles,
\begin{equation}
    T_{m} = \displaystyle\int_{0}^{2\pi}\displaystyle\int_{0}^{\pi}T_{r}(\theta,\phi)\sin{\theta}d\theta{d\phi}.
\end{equation}
The quantity ${T_{m}}$ is directly measurable  while ${T_{r}}$ is derived by dividing ${T_{m}}$ by the appropriate solid angle.

The cross-correlation of the signals produced by the pair of separated antennas viewing the same brightness temperature distribution ${T_{r}(\theta,\phi)}$ produces a sample of the visibility function given by \cite{Ruf1988}
\begin{equation}
V(u,v) = \displaystyle\int_{0}^{2\pi}\displaystyle\int_{0}^{\pi}T_{r}(\theta,\phi) e^{\textrm{j}2{\pi}(u{\sin}{\theta}{\cos}{\phi} + v{\sin\theta{\sin\phi}})} \sin{\theta}d\theta{d\phi}
\end{equation}
where ${V}$ is in units of ${K}$, ${u = D_{x}/{\lambda}}$, ${v = D_{y}/{\lambda}}$, and ${\lambda}$ is the wavelength of the incident radiation at the antennas.  The final brightness temperature image ${\widehat{T}(\theta,\phi)}$ is formed by an inverse FT of the measurements \cite{Ruf1988},
\begin{equation}
    \widehat{T}(\theta,\phi) = \displaystyle\int_{-\infty}^{\infty}\displaystyle\int_{-\infty}^{\infty}V(u,v) e^{-\textrm{j}2{\pi}(u{\sin}{\theta}{\cos}{\phi} + v{\sin\theta{\sin\phi}})} du{dv}.
\end{equation}

In some system designs, the antennas are placed in a configuration known as a minimally redundant linear array (MRLA) \cite{Moffett68}.  The spacing of antennas in a MRLA is at integer multiples of ${\lambda/2}$ chosen such that the cross-correlation operation will generate every missing half-wavelength spacing with as few redundancies as possible \cite{Moffett68}.  The redundancies do not improve the spatial resolution of the image but they do affect the noise floor.

\section{Intelligent SA Systems}
\label{sec:cognitive}
\begin{figure}[t]
\centering
\includegraphics[width=1\linewidth]{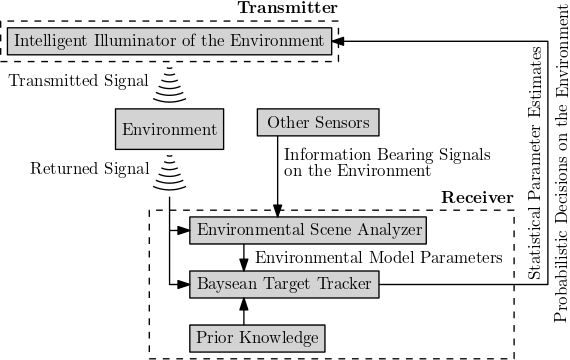}
\caption{The basic cognitive transmit/receive sensing system architecture proposed by Haykin, from Fig.~1 in \cite{haykin2006cognitive}. This architecture assumes a feedback loop from receiver to transmitter, which may not exist (e.g., passive systems).}
\label{fig:CognitiveRadar}
\end{figure}

With the arrival of computing systems with sufficient memory and clock rates, machine learning has grown massively in recent years as researchers have tried to apply it to myriad applications, including SAs. One of the first forays into the concept of intelligent control of sensing systems was Haykin's seminal article on cognitive radar \cite{haykin2006cognitive}, which defined the general architecture to support agile control of radar systems. Such systems, as posited by Haykin, share three defining features. First, they use intelligent signal processing, which builds on learning from the results of the radar's interactions with its environment. Second, they provide some type of channel for the receiver to provide feedback to the transmitter, which allows the transmitter to adapt to its environment in an intelligent way. Third, the system has a means for preserving the information content of radar returns. From this general architecture, depicted in Fig.~\ref{fig:CognitiveRadar}, it is possible to apply machine learning to multiple elements of the system. Wideband systems present a set of unique challenges relative to narrrowband systems, such as greater impact of frequency selective fading, as well as larger volumes of data, that invite the use of machine learning to efficiently develop system improvements.

Several current lines of machine learning work focus on SA front-end functions, such as beamforming. In addition, a vast literature has developed around SA image classification, but this considerable body of work lies outside the scope of this paper. As an example of research efforts in recent years that have explored using machine learning to do beamforming for SA, Luchies and Byram \cite{Luchies2018Deep} proposed a deep feed-forward architecture for reducing sidelobes in ultrasound applications, and noted that SA techniques could provide further improvements by expanding the depth of field. Peretz and Feuer discuss a beamforming algorithm for SA ultrasound in \cite{peretz2020deep}. The Delay and Sum (DAS) technique in SA ultrasound, allows for a higher frame rate but at the cost of lower Signal to Noise Ratio (SNR) and reduced contrast resolution. Using Synthetic Transmit Aperture (STA) ultrasound results in higher SNR, but requires more memory use because all the elements in the array are receiving and storing the returns from a single transmitting element. Peretz and Feuer introduce Deep NN Beamforming (DNNB) for SA, STA, and phased array, which involves a training component that compares large aperture and small aperture data, and a utilization component that uses the trained deep NN to process the small-aperture data to achieve large-aperture type results.

Yonei et al.~\cite{Yonel2018Deep} used deep NNs for image reconstruction in passive SA radar applications, where the user deploys receivers that operate opportunistically, i.e., they use transmitters in the environment, with which they do not coordinate operations, and measure the transmitter's signal and its reflection from a target of interest. The opportunistic nature of passive SAR means that the transmitter's location, waveform, and beam pattern may not be known to the receiver as it moves across the aperture. The authors trained a recurrent NN (RNN) model to obtain estimates of the parameters associated with both forward projection and backprojection and filtering. Using the Born approximation, the authors model the forward projection operation in a sampled aperture as the linear transformation $\boldsymbol{d} = \mathbf{F} \boldsymbol{\rho}$, where $\boldsymbol{\rho}$ is a vector of reflectivity values of sample points in the environment, $\boldsymbol{d}$ is the vector of corresponding received signal values, and $\mathbf{F}$ is the forward scattering matrix. They implement the backprojection and filtering operation using $\mathbf{F}$ and a filtering matrix $\mathbf{Q}$ that they employ at each stage of the RNN, and use unsupervised learning to train the RNNs to estimate the weights. 

Yonel et al. built on their deep learning concept in a subsequent work \cite{Yonel2019Deep}, in which they apply a decoder to the estimated reflectivity vector at the RNN output to map it back to the data space and thus produce an auto-encoder. By incorporating feedback that adaptively minimizes the error between received SAR signals and those generated by the auto-encoder, the authors demonstrate that they are able to estimate the SAR waveform. In addition, other researchers have applied cognitive methods to SAR waveform design, such as the work by Xu et al. \cite{Xu2019Cognitive}, which develops a joint optimization algorithm for designing a SAR waveform that maximizes resolution performance and the Signal to Clutter and Noise Ratio (SCNR).

 Another major area of work involving machine learning is beamforming for ultrasound applications, including those that use either transmitted plane waves (PWs) or transmitted Diverging Waves (DWs). Early work includes the investigation by Gasse et al.~\cite{Gasse2017HighQuality}, on PW beamforming using multiple transmissions to produce a compound image. Because PW transmissions do not use a focused beam, the image resolution for a single PW transmission is poor; a resulting strategy is to combine multiple PW images, but this oversampling reduces the ultrasound image frame rate. Gasse et al. developed a Convolutional NN (CNN) approach that would increase frame rate by learning channel parameters so that good quality images could be obtained by compounding fewer PW transmissions. The authors trained a six-layer CNN and were able to use it to match the performance metrics obtained with about 20~PW transmissions using only 3~PW transmissions.
 
 More recent work by Ghani et al.~in \cite{Ghani2019High} addresses issues with using Diverging Waves (DWs) to illuminate targets. The authors also used a six-layer CNN with relatively low computational complexity by having only the first two CNN layers be 3D convolutional layers, with the subsequent layers being 2D. Ghani et al. also trained using invidual pixels rather than entire images, which prevents the CNN from learning features unique to ultrasound environments. The authors also used a compound loss function to train their CNN, rather than using a simple MMSE criterion. This produced improved performance due to the relationships between the various elements of the compound loss function. 

Recent work has explored the use of machine learning to solve the forward and reverse scattering problems. This approach uses a linear approximation such as Eq.~\eqref{eq:LinearModel}, which arises from sampling the scattering field. The approach employed by \cite{DeGuchy2020Machine} uses a NN architecture that consists of a single layer of neurons, with no activation function, to find the sensing matrix, $\mathbf{A}$, or its pseudo-inverse $\mathbf{B} = (\mathbf{A}^H \mathbf{A})^{-1}\mathbf{A}^H$, where $H$ is the Hermetian (complex transpose) matrix operator. The authors note that knowing $\mathbf{A}$ is sufficient to solve the scattering problem when the observation space sampling is chosen so that $\mathbf{A}^H \mathbf{A}$ is approximately a diagonal matrix; however, their experimental results using the CIFAR-10 image dataset \cite{Krizhevsky2009learningmultiple} show better performance using the estimated pseudo-inverse.

\section{Future Outlook and Summary}
\label{sec:summary}
%We have shown that ... \cite{amari2000methods}
This paper has provided an overview of the broad utility of SA techniques to a wide range of imaging applications, including radar, channel sounding, optics, radiometry, and sonar. The overarching advantage of a SA is that the available angular resolution can be increased beyond the limits imposed by the physical size of the antenna. This is accomplished by using a mechanical positioner to move the receive antenna through space while it collects signal samples. Provided the samples are phase coherent then they can be combined in post-processing as if they were measured by a physical aperture with the same size as the SA. In the temporal domain, synthetic techniques can also be applied to increase the available delay resolution. For example, in frequency domain channel sounding, a wide measurement bandwidth is synthesized by measuring the frequency response of a wireless channel using many narrowband frequency tones spaced over a wide frequency grid.

Out of the synthetic apertures considered, SAR and SAS have the most in common.  Both platforms are capable of measuring the temporal and spatial phase of propagating waves, although the wave medium and the propagation velocity are very different.  In fact many of the concepts being pursued today for digital arrays were first developed for digital phased array sonars \cite{Trider72,Knight81}.  Both SAR and SAS also suffer degraded performance if there is uncertainty in the position of the sensor.  Beamforming is the workhorse algorithm for SAR and SAS and if the exact locations of the sensor when spatial samples are taken is not known then the array model errors will affect the beamformed outputs.

Fourier ptychography describes an optical synthetic aperture that indirectly estimates spatial phase to image a microscope specimen.  A synthetic aperture channel sounder also measures spatial phase to image the scattering characteristics of a wireless channel.  Synthetic aperture radiometry does not measure phase but estimates the time difference of arrival by cross-correlating signals across different separated antennas.  In all cases, the motivation for using a synthetic aperture is to achieve greater spatial and angular resolution than the physical system allows.

Many recent innovations have improved the performance of SAs with potential for even more significant advances in the future. Primary among these is the use of self-calibrating quantum sensors based on Rydberg atoms that measure electric field strength to extraordinary precision. The quantum states of atoms are fundamental constants of nature that never drift or change and don't need to be calibrated. Therefore, they provide traceable measurements that are hugely important in many metrology applications.

New technology for micronavigation and the precision geolocation of ocean-going vessels has had a dramatic impact on SAS. By knowing precisely the location and orientation of a sonar vessel it is possible to significantly improve the detection and imaging performance of sonar, especially in severe weather. Advances in micronavigation have leveraged machine learning, which also underpins many other advances in SA imaging. New beamforming techniques and the development of intelligent systems that can adapt system parameters to optimize performance based on the operational environment are just a few examples of the advanced capabilities now possible via machine learning approaches.

\section*{Acknowledgements}
K. V. M. acknowledges helpful discussions with Lam Nguyen of United States DEVCOM Army Research Laboratory. A. A.-G. thanks Matthew L. Simons of NIST for helpful discussions and reference paper recommendations. P. V. acknowledges measurements and data generated through experiments conducted by the SAMURAI team in the NIST Channel Sounding section. P. V. also acknowledges contributions to the simulations of sparse arrays made by Mohamed Hany, Sudantha Perera, Katherine Remley, Carnot Nogueria, and Rick Candell of NIST.

Official contribution of the U.S. government; not subject to copyright in the United States.

%% References:
%\clearpage
%\balance
\bibliographystyle{IEEEtran}
\bibliography{main}

\end{document}